\newif\ifTechReport
\TechReporttrue

\ifTechReport
  \documentclass[acmsmall,nonacm,authorversion,screen,timestamp=true]{acmart}
\else
  \documentclass[acmsmall]{acmart}
\fi

\author{Arshavir Ter-Gabrielyan}
\affiliation{
  \department{Department of Computer Science}
  \institution{ETH Zurich}
  \country{Switzerland}
}
\email{ter-gabrielyan@inf.ethz.ch}

\author{Alexander J. Summers}
\affiliation{
  \department{Department of Computer Science}
  \institution{ETH Zurich}
  \country{Switzerland}
}
\email{alexander.summers@inf.ethz.ch}

\author{Peter M{\"u}ller}
\affiliation{
  \department{Department of Computer Science}
  \institution{ETH Zurich}
  \country{Switzerland}
}
\email{peter.mueller@inf.ethz.ch}

\ifTechReport
  \makeatletter
  \let\@authorsaddresses\@empty
  \makeatother
\else
  \acmJournal{PACMPL}
  \acmVolume{3}
  \acmNumber{OOPSLA} 
  \acmArticle{1}
  \acmYear{2019}
  \acmMonth{10}
  \acmDOI{} 
  \startPage{1}
\fi

\setcopyright{none}

\usepackage{booktabs}   
\usepackage{subcaption} 

\usepackage[T1]{fontenc}
\usepackage[title]{appendix}
\usepackage{array}
\usepackage{multirow}
\usepackage{amsmath}
\usepackage{amsfonts}
\usepackage{enumerate}
\usepackage{scalerel}
\newcommand{\RN}[1]{%
  \scriptscriptstyle{\textup{\expandafter{\romannumeral#1}}}%
}

\usepackage{mathtools}

\newtheorem{chuckwalladef}{Definition}

\newcommand*{\longeq}{\mathrel{\vcenter{\baselineskip0.5ex \lineskiplimit0pt
                      \hbox{.}\hbox{.}}}%
                      \Longleftrightarrow}
\usepackage[utf8]{inputenc}
\usepackage[english]{babel}

\usepackage{scalerel,stackengine}
\stackMath
\newcommand\longhat[1]{%
\savestack{\tmpbox}{\stretchto{%
  \scaleto{%
    \scalerel*[\widthof{\ensuremath{#1}}]{\kern.1pt\mathchar"0362\kern.1pt}%
    {\rule{0ex}{\textheight}}
  }{\textheight}%
}{2.4ex}}%
\stackon[-6.9pt]{#1}{\tmpbox}%
}

\usepackage[font=bf]{caption}
\usepackage{pdflscape}
\usepackage{graphicx}
\graphicspath{ {./figures/} }
\usepackage[]{hyperref}
\hypersetup{
  colorlinks=true,
}

\usepackage{listings}
\usepackage[scaled=0.81]{beramono}
\usepackage{tikz}

\usetikzlibrary{trees,shapes,decorations,decorations.pathmorphing}
\usetikzlibrary{shapes.geometric, arrows, positioning, calc, matrix, fit}

\newcommand{\defp}[1]{\tikz[remember picture]\node[minimum size=0](#1){};}
\tikzstyle{heap}=[densely dotted,rounded corners=3pt]
\tikzstyle{frame}=[heap,color=blue]
\tikzstyle{footprint}=[heap,color=red]
\tikzstyle{vertex}=[draw, circle,inner sep=0pt,minimum width=4pt]
\tikzstyle{hvertex}=[vertex,color=red]
\tikzstyle{fvertex}=[vertex,color=blue]
\tikzstyle{edge}=[->,shorten >=1pt]
\tikzstyle{hedge}=[edge,color=red]
\tikzstyle{fedge}=[edge,color=blue]
\tikzstyle{path}=[edge,decorate,decoration={snake,amplitude=0.75pt,segment length=6pt,pre length=0pt,post length=5pt}]
\tikzstyle{hpath}=[path,color=red]
\tikzstyle{fpath}=[path,color=blue]
\tikzstyle{lbl}=[label={[label distance=-1pt,text depth=0pt,font=\small]above:{#1}}]
\tikzstyle{lbld}=[label={[label distance=-1pt,text depth=0pt,font=\small]below:{#1}}]
\tikzstyle{lbll}=[label={[label distance=-1pt,text depth=0pt,font=\small]left:{#1}}]
\tikzstyle{lblr}=[label={[label distance=-1pt,text depth=0pt,font=\small]right:{#1}}]
\tikzstyle{lbl45}=[label={[label distance=-1pt,text depth=0pt,font=\small]45:{#1}}]
\tikzstyle{lbl135}=[label={[label distance=-1pt,text depth=0pt,font=\small]135:{#1}}]
\tikzstyle{heaplbl}=[font=\small,text depth=0pt]
\tikzstyle{framelbl}=[heaplbl,color=blue]
\tikzstyle{footprintlbl}=[heaplbl,color=red]

\makeatletter
\def\pgfutil@firstofmany#1#2\pgf@stop{#1}
\def\pgfutil@secondofmany#1#2\pgf@stop{#2}
\def\tikz@lib@place@of@#1#2#3{%
  \def\pgf@tempa{fit bounding box}%
  \edef\pgf@temp{\expandafter\pgfutil@firstofmany#2\pgf@stop}
  \if\pgf@temp(%
    \tikz@lib@place@fit@scan{#2}{0}%
  \else\if\pgf@temp|
      \expandafter\tikz@lib@place@fit@scan\expandafter{\pgfutil@secondofmany#2\pgf@stop}{1}%
    \else\ifx\pgf@temp\tikz@activebar
        \expandafter\tikz@lib@place@fit@scan\expandafter{\pgfutil@secondofmany#2\pgf@stop}{1}%
      \else\if\pgf@temp-
          \expandafter\tikz@lib@place@fit@scan\expandafter{\pgfutil@secondofmany#2\pgf@stop}{2}%
        \else\if\pgf@temp+
            \expandafter\tikz@lib@place@fit@scan\expandafter{\pgfutil@secondofmany#2\pgf@stop}{3}%
          \else
            \def\pgf@tempa{#2}%
          \fi
        \fi
      \fi
    \fi
  \fi
  \expandafter\tikz@scan@one@point\expandafter\tikz@lib@place@remember\expandafter(\pgf@tempa)%
  \iftikz@shapeborder%
    \iftikz@lib@ignore@size%
      \edef\tikz@node@at{\noexpand\pgfpointanchor{\tikz@shapeborder@name}{center}}%
      \def\tikz@anchor{center}%
    \else%
      \edef\tikz@node@at{\noexpand\pgfpointanchor{\tikz@shapeborder@name}{#3}}%
    \fi%
  \fi%
  \edef\tikz@lib@place@nums{#1}%
}
\def\tikz@lib@place@fit@scan#1#2{
  \pgf@xb=-16000pt\relax%
  \pgf@xa=16000pt\relax%
  \pgf@yb=-16000pt\relax%
  \pgf@ya=16000pt\relax%
  \if\pgfutil@firstofmany#1\pgf@stop(%
    \tikz@lib@fit@scan#1\pgf@stop%
  \else
    \tikz@lib@fit@scan(#1)\pgf@stop
  \fi
  \ifdim\pgf@xa>\pgf@xa
  \else
     \expandafter\def\csname pgf@sh@ns@fit bounding box\endcsname{rectangle}%
     \expandafter\edef\csname pgf@sh@np@fit bounding box\endcsname{%
       \def\noexpand\southwest{\noexpand\pgfqpoint{\the\pgf@xa}{\the\pgf@ya}}%
       \def\noexpand\northeast{\noexpand\pgfqpoint{\the\pgf@xb}{\the\pgf@yb}}%
     }%
     \expandafter\def\csname pgf@sh@nt@fit bounding box\endcsname{{1}{0}{0}{1}{0pt}{0pt}}%
     \expandafter\def\csname pgf@sh@pi@fit bounding box\endcsname{\pgfpictureid}%
     \ifcase#2\relax
     \or 
       \pgf@y=\pgf@yb%
       \advance\pgf@y by-\pgf@ya%
       \edef\pgf@marshal{\noexpand\tikzset{minimum height={\the\pgf@y-2*(\noexpand\pgfkeysvalueof{/pgf/outer ysep})}}}%
       \pgf@marshal
     \or 
       \pgf@x=\pgf@xb%
       \advance\pgf@x by-\pgf@xa%
       \edef\pgf@marshal{\noexpand\tikzset{minimum width={\the\pgf@x-2*(\noexpand\pgfkeysvalueof{/pgf/outer xsep})}}}%
       \pgf@marshal
     \or 
       \pgf@y=\pgf@yb%
       \advance\pgf@y by-\pgf@ya%
       \pgf@x=\pgf@xb%
       \advance\pgf@x by-\pgf@xa%
       \edef\pgf@marshal{\noexpand\tikzset{minimum height={\the\pgf@y-2*(\noexpand\pgfkeysvalueof{/pgf/outer ysep})},minimum width={\the\pgf@x-2*(\noexpand\pgfkeysvalueof{/pgf/outer xsep})}}}%
       \pgf@marshal
     \fi
  \fi
}
\tikzset{
  fit bounding box/.code={\tikz@lib@place@fit@scan{#1}{0}},
  span vertical/.code={\tikz@lib@place@fit@scan{#1}{1}},
  span horizontal/.code={\tikz@lib@place@fit@scan{#1}{2}},
  span/.code={\tikz@lib@place@fit@scan{#1}{3}}}

\makeatother

\DeclareRobustCommand{\outDegreeSymb}{%
\parbox{8pt}{%
  \tikz[baseline=-3pt,scale=0.66]{%
  \node[circle, draw,scale=0.3](down){};%
  \foreach \X in {45,-45}%
  {\draw[->,>=stealth] (down.\X) to ++ (\X:7.5pt);}  }%
}}

\DeclareRobustCommand{\inDegreeSymb}{%
\parbox{8pt}{%
  \tikz[baseline=-3pt,scale=0.66]{%
  \node[circle, draw,scale=0.3](down){};%
  \foreach \X in {225,135}%
  {\draw[<-,>=stealth] (down.\X) to ++ (\X:7.5pt);}  }%
}}

\tikzstyle{linkbg}=[ -, thick, white, solid, line width = 5pt ]
\tikzstyle{link}=[ ->, >=stealth, thick, gray, dashed, line width = 1pt ]
\tikzstyle{block}=[ -, >=stealth, thick, gray, line width = 1pt ]
\newcommand\chkwToViper[4]{
    \path[link] (#1) edge node[ sloped, fill=white, anchor=center, pos=0.5, font=\small] {#2} ([xshift=#4-2pt]#3);
}

\newcommand\viperCodeBlock[4]{
    \draw[block] ([xshift=#4]#1 |- #2) -- ++(0,4pt) -- ++(3pt,0);
    \draw[block] ([xshift=#4]#1 |- #2) -- ([xshift=#4]#1 |- #3);
    \draw[block] ([xshift=#4]#1 |- #3) -- ++(0,-4pt) -- ++(3pt,0);
}
\newcommand\chkwCodeBlock[3]{
    \draw[block] (#1 |- #2) -- ++(0,4pt) -- ++(-3pt,0);
    \draw[block] (#1 |- #2) -- (#1 |- #3);
    \draw[block] (#1 |- #3) -- ++(0,-4pt) -- ++(-3pt,0);
}

\newcommand\macroDefinition[4]{
  \node[thick, gray, line width = 1pt, anchor=west, font=\small] at ([xshift=#4]#1 |- #2) {//#3};   
}

\usepackage{letltxmacro}
\newcommand*{\SavedLstInline}{}
\LetLtxMacro\SavedLstInline\lstinline
\DeclareRobustCommand*{\lstinline}{%
  \ifmmode
    \let\SavedBGroup\bgroup
    \def\bgroup{%
      \let\bgroup\SavedBGroup
      \hbox\bgroup
    }%
  \fi
  \SavedLstInline
}

\newcommand*{\thead}[1]{\bfseries #1}
\newcommand{\acyclic}{DAG}

\newcommand{\zeroOnePG}{ZOPG}
\newcommand{\tick}{\ensuremath{\checkmark}}

\newcommand{\ditto}{\hspace{10mm}---~\raisebox{-0.5ex}{''}~---}

\newcommand*{\figref}[1]{Fig.~\ref{fig:#1}}
\newcommand*{\tabref}[1]{Tab.~\ref{tab:#1}}
\newcommand*{\secref}[1]{Sec.~\ref{sec:#1}}
\newcommand*{\appref}[1]{%
\ifTechReport%
App.~\ref{sec:#1}%
\else%
\cite[\begin{NoHyper}App.~\ref{sec:#1}\end{NoHyper}]{techreport}%
\fi%
}

\let\oldeqref\eqref
\renewcommand*{\eqref}[1]{\oldeqref{eq:#1}}

\newcommand*{\defref}[1]{Def.~\ref{def:#1}}

\newcommand*{\coderef}[1]{\hyperref[lst:#1]{\fontseries{m}\texttt{#1}}}
\newcommand*{\appcoderef}[1]{%
\ifTechReport%
\hyperref[lst:#1]{\fontseries{m}\texttt{#1}}%
\else%
\hyperlink{citeTechReport}{\fontseries{m}\texttt{#1}}%
\fi%
}

\usepackage{etoolbox}
\newcommand{\todo}[2]{\noindent %
  \textit{\ifstrequal{\string #1}{\string ATG}{\color{purple}}{\ifstrequal{\string #1}{\string PM}{\color{blue}}{\color{red}}}%
  \textbf{#1---}#2}%
}

\usepackage[normalem]{ulem}

\usepackage{arydshln}
\newcommand{\etal}{{et al.\@}} 
\newcommand{\eg}{{e.g.,\@}} 
\newcommand{\ie}{{i.e.,\@}} 
\newcommand{\cf}{{cf.\@}} 
\newcommand{\WLOG}{{w.l.o.g.\@}}

\newcommand{\mo}[1]{\ensuremath{\mathtt{#1}}} 
\newcommand{\code}[1]{\fontseries{m}\lstinline!#1!}
\newcommand{\CONVEXIN}{\ensuremath{\prec}}
\newcommand{\DISJUNION}{\ensuremath{\uplus}} 

\newcommand{\NULLLIT}{\texttt{null}}
\newcommand{\ACCESS}[1]{\mo{acc(#1)}}
\newcommand{\OLDEXP}[1]{\mo{old(#1)}}

\newcommand{\RELCLOSED}[2]{\ensuremath{\mathtt{CLOSED}_{#2}(#1)}}
\newcommand{\ACYCLIC}[1]{\ensuremath{\mathtt{ACYCLIC(#1)}}}

\newcommand{\RESULT}[1][]{\ensuremath{\mathfrak{r}_{#1}}}
\newcommand{\GRAPH}[1][]{\ensuremath{\mathfrak{g}_{#1}}}
\newcommand{\HEAP}[1][]{\ensuremath{\mathfrak{h}_{#1}}}

\newcommand{\FRAME}[1][]{\ensuremath{\mathfrak{f}_{#1}}}

\newcommand{\EDGE}[4]{\mbox{\ensuremath{\mo{E}_{#1}(#2, #3, #4)}}}
\newcommand{\PATH}[4]{\mbox{\ensuremath{\mo{P}_{#1}(#2, #3, #4)}}}
\newcommand{\DEP}[6]{\mbox{\ensuremath{\mo{DEP}_{#1}(#2, #3, #4, #5, #6)}}}
\newcommand{\BEDGE}[2]{\mbox{\ensuremath{\mo{E}(#1, #2)}}}
\newcommand{\BPATH}[2]{\mbox{\ensuremath{\mo{P}(#1, #2)}}}
\newcommand{\BDEP}[4]{\mbox{\ensuremath{\mo{DEP}(#1, #2, #3, #4)}}}

\newcommand{\EDGEF}[5]{\mbox{\ensuremath{\mo{E}_{#1}^{#5}(#2, #3, #4)}}}
\newcommand{\PATHF}[5]{\mbox{\ensuremath{\mo{P}_{#1}^{#5}(#2, #3, #4)}}}
\newcommand{\DEPF}[7]{\mbox{\ensuremath{\mo{DEP}_{#1}^{#7}(#2, #3, #4, #5, #6)}}}

\newcommand{\PATHREL}[2]{\mbox{\ensuremath{\mo{P}_{#1}(#2)}}}

\newcommand{\EDGESYMB}[1]{\mbox{\ensuremath{\mo{E}_{#1}}}}
\newcommand{\PATHSYMB}[1]{\mbox{\ensuremath{\mo{P}_{#1}}}}
\newcommand{\DEPSYMB}[1]{\mbox{\ensuremath{\mo{DEP}_{#1}}}}

\newcommand{\HoareTriple}[3]{
  \begin{displaymath}
    \left\{\begin{array}{cc}#1\end{array}\right\}
    \begin{array}{c}#2\end{array}
    \left\{\begin{array}{cc}#3\end{array}\right\}
  \end{displaymath}
}

\begin{document}

\title{Modular Verification of Heap Reachability Properties in Separation Logic}


\begin{abstract}
The correctness of many algorithms and data structures depends on reachability properties, that is, on the existence of chains of references between objects in the heap. Reasoning about reachability is difficult for two main reasons. First, any heap modification may affect an unbounded number of reference chains, which complicates modular verification, in particular, framing. Second, general graph reachability is not supported by SMT solvers, which impedes automatic verification.

In this paper, we present a modular specification and verification technique for reachability properties in separation logic. For each method, we specify reachability only locally within the fragment of the heap on which the method operates.
A novel form of reachability framing for relatively convex subheaps allows one to extend reachability properties from the heap fragment of a callee to the larger fragment of its caller, enabling precise procedure-modular reasoning.
Our technique supports practically important heap structures, namely acyclic graphs with a bounded outdegree as well as (potentially cyclic) graphs with at most one path (modulo cycles) between each pair of nodes. The integration into separation logic allows us to reason about reachability and other properties in a uniform way, to verify concurrent programs, and to automate our technique via existing separation logic verifiers. We demonstrate that our verification technique is amenable to SMT-based verification by encoding a number of benchmark examples into the Viper verification infrastructure.
\end{abstract}

\maketitle

\section{Introduction}
\label{sec:intro}
Separation logic~\citep{reynolds2002} has greatly simplified the verification of basic heap data structures such as lists and trees by leveraging the disjointness of sub-heaps to reason about the effects of heap modifications. However, verifying data structures that permit unbounded sharing remains challenging. Their correctness often depends on heap reachability properties, that is, the existence of paths of references between objects. For instance, the path compression of union-find needs to preserve the reachability of the root object, the termination of heap traversals might rely on the absence of cyclic paths, and the invariant of a garbage collector may prescribe that each object is reachable from the list of allocated objects or the free-list, but not from both.

Reasoning about reachability properties is difficult for two main reasons. (1)~Modularity: reachability is inherently a non-local property. Any heap modification may affect an unbounded number of heap paths, which complicates framing, that is, proving modularly that a heap update or method call does not affect a given reachability property. (2)~Automation: general graph reachability is not supported by SMT solvers, which power most automatic verification tools.

Existing work addresses these challenges typically by supporting only certain kinds of reachability properties or certain classes of data structures. For instance, \citet{ItzhakyPOPL14} present a modular verification technique for reachability properties of a broad class of linked-list programs, but do not support structures that can have more than one outgoing reference per object. Such structures may contain an arbitrary number of alternative paths between two objects, and maintaining reachability information via first-order formulas becomes infeasible. 
Flows~\cite{Krishna2017GoWT} is a technique providing local reasoning for updates to subgraphs which \emph{preserve} properties such as reachability, \eg{}~changes to a subgraph which neither add nor remove paths between nodes in its boundary. However, flows do not provide analogous means for local reasoning about methods which are \emph{intended} to change these paths (\eg{}~a function which connects two subgraphs).
%

This paper presents a modular verification technique for general heap reachability properties that supports both acyclic data structures with a bounded number of outgoing references per object (for instance, DAG structures such as BDDs~\cite{Akers78}) and (potentially cyclic) 0--1-path graphs, that is, graphs that contain at most one path (modulo cycles) between each pair of objects (such as a ring buffer). Our technique is integrated into separation logic, which allows us to reason about reachability and other properties in a uniform way, to verify concurrent programs, and to automate our technique via existing separation logic verifiers. Our technique enables modular reasoning by specifying reachability properties locally within the memory footprint of a method rather than in the entire heap. A novel form of reachability framing allows one to extend the reachability properties guaranteed by a callee method to the (larger) footprint of its caller. As a result, each method can be verified modularly, without considering the heap outside its footprint, the implementations of other methods, or other threads.

\paragraph{Contributions.} Our paper makes the following technical contributions:

\begin{itemize}

\item \textit{Specification:} We introduce a specification technique for reachability properties in the context of separation logic. It enables modular verification, even for concurrent programs (\secref{methodology}).

\item \textit{Verification:} We present a novel verification technique for reachability properties. In particular, we identify relative convexity of method footprints as a property that enables precise reachability framing and procedure-modular reasoning. Our technique goes beyond prior work~\cite{ItzhakyPOPL14} by supporting all acyclic graphs (\secref{DAGs}).

\item \textit{Cyclic graphs:} We extend our verification technique to cyclic 0--1-path graphs. While reachability framing carries over from the acyclic case, cyclic graphs require a more elaborate machinery to handle reference field updates (\secref{ZOPs}).

\item \textit{Automation:} We demonstrate that our verification technique is amenable to SMT-based verification by encoding a number of benchmark examples into the Viper verification infrastructure~\cite{mueller-2015-viper} (\secref{evaluation}).

\end{itemize}

\section{Specification Technique}
\label{sec:methodology}
In this section, we illustrate our technique using a DAG data structure with node type \code{Node} and fields \code{left} and \code{right}. Method \coderef{merge} in Fig.~\ref{fig:runningExample} takes as arguments references \code{l} and \code{r} to two nodes of disjoint DAGs and attaches \code{r} as descendant of \code{l}. It returns \code{link}, a node of the first DAG to which \code{r} was attached. 
The postcondition ensures that exactly one connection was created (via an edge from \code{link} to the root of the second DAG, \code{r}), and that heap paths exist either if they existed in the pre-state or were connected by the new edge, $(\texttt{link}, \texttt{r})$. We explain the specification of \coderef{merge} in full detail in the remainder of this section.

\begin{figure}[t]
\phantomsection\label{lst:merge}
\begin{lstlisting}[escapeinside={(*@}{@*)}]
method merge(l: Node, r: Node,
             $\GRAPH$: Graph, ldag: Graph, rdag: Graph)$\defp{ghostParamsComment}$
  returns link: Node$\defp{updatedNodeComment}$
  requires $\GRAPH =\texttt{ldag}\DISJUNION\texttt{rdag} \land \texttt{l} \in \texttt{ldag} \land \texttt{r} \in \texttt{rdag}$
           $\forall x,y \in \GRAPH \>\bullet\> \lnot \EDGE{}{\GRAPH}{x}{y} \lor \lnot \PATH{}{\GRAPH}{y}{x}\defp{preAcyclicComment}$
           $\forall n \> \bullet \> n\in\texttt{ldag} \Leftrightarrow \PATH{}{\GRAPH}{\texttt{l}}{n}$
           $\forall n \> \bullet \> n\in\texttt{rdag} \Leftrightarrow \PATH{}{\GRAPH}{\texttt{r}}{n}$
  ensures $\texttt{link} \in \texttt{ldag}$
          $\forall x,y \in \GRAPH \>\bullet\> \lnot \EDGE{}{\GRAPH}{x}{y} \lor \lnot \PATH{}{\GRAPH}{y}{x}\defp{postAcyclicComment}$
          $\forall x,y \> \bullet \> \EDGE{}{\GRAPH}{x}{y} \iff \EDGE{0}{\GRAPH}{x}{y} \lor x=\texttt{link} \land y=\texttt{r}$
          $\forall x,y \> \bullet \> \PATH{}{\GRAPH}{x}{y} \iff \PATH{0}{\GRAPH}{x}{y} \lor \PATH{0}{\GRAPH}{x}{\texttt{link}} \land \PATH{0}{\GRAPH}{\texttt{r}}{y}$
{
    if (l.right != $\NULLLIT$) {
        var nldag := $\coderef{sub}$($\GRAPH$, ldag, l.right)$\defp{newGhostParamComment}$
        link := $\coderef{merge}$(l.right, r, nldag$~\DISJUNION~$rdag, nldag, rdag)
    } else {
        l.right := r
        link := l
    }
}
\end{lstlisting}
\begin{tikzpicture}[ remember picture, overlay ]
  \path let \p0=(newGhostParamComment),
            \p1=(ghostParamsComment),
            \p2=(updatedNodeComment),
            \p3=(preAcyclicComment),
            \p4=(postAcyclicComment)
            in coordinate (Q2) at (\x1, {max(\y0,\y1,\y2,\y3,\y4)});
  \def\minArrowLen{0pt}
  \macroDefinition{Q2}{ghostParamsComment}{ ghost parameters }{\minArrowLen}
  \macroDefinition{Q2}{updatedNodeComment}{ updated node }{\minArrowLen}
  \macroDefinition{Q2}{newGhostParamComment}{ define new ghost parameter }{\minArrowLen}
  \macroDefinition{Q2}{preAcyclicComment}{ acyclic invariant }{\minArrowLen}
  \macroDefinition{Q2}{postAcyclicComment}{ acyclic invariant }{\minArrowLen}
\end{tikzpicture}
\vspace{-4mm}
\caption{An example program and specification. Method \coderef{merge} attaches the DAG rooted in \code{r} to a node of the DAG rooted in \code{l}, and returns that node. We use the edge predicate \EDGESYMB{} and the path predicate \PATHSYMB{} to specify reachability properties, within a set of objects $\GRAPH$. Each specification line is a separate conjunct. The footprint $\GRAPH$ is closed due to the equivalences in the last two preconditions.}
\label{fig:runningExample}
\end{figure}

\subsection{Footprints}
\label{sec:footprints}

Separation logics associate an \emph{access permission} with each memory location. Access permissions are held by method executions and may be transferred between methods upon calls and returns; they can be thought of as additional program state used for reasoning (\emph{ghost} state). A heap location can be accessed only while the corresponding permission is held. The set of locations that a method may access is called its \emph{footprint}. Due to (de)allocation or concurrency, the footprint of a method may change during its execution.

A method's precondition specifies which permissions to transfer on calling the method. The initial footprint of a method contains exactly the locations for which its precondition requires permission. Conversely, the method postcondition specifies which permissions to return to the caller when the method terminates.

The footprint of any method operating on linked heap structures, \eg{} lists and DAGs, contains a statically unknown number of memory locations. To provide a convenient way to refer to a method's footprint, we equip each method with a distinct ghost parameter \code{$\GRAPH$: Graph} to denote its footprint. For simplicity, instead of specifying the footprint as a set of object-field pairs, we let \code{Graph} denote sets of non-null objects and keep the fields implicit when they are clear from the context. The set stored in $\GRAPH$ is updated whenever the footprint changes, for instance, due to allocation. In order to be able to refer to the final footprint of a method execution in its postcondition, we make $\GRAPH$ an in-out parameter. For simplicity, we assume in the following that the footprint of a method remains unchanged, s.t.~the value of $\GRAPH$ is constant; an extension is straightforward.

We equip each method with implicit pre- and postconditions to require and ensure permissions to all locations in the footprint:
\begin{lstlisting}[mathescape,escapeinside={(*@}{@*)}]
  requires $\forall$n $\in \GRAPH \>\bullet\> \ACCESS{n.left} \>*\> \ACCESS{n.right}$
  ensures  $\forall$n $\in \GRAPH \>\bullet\> \ACCESS{n.left} \>*\> \ACCESS{n.right}$
\end{lstlisting}

\noindent
Here, \ACCESS{\textit{x}.\textit{f}\>} denotes an access permission to the \emph{memory location} for field $f$ of object $x$ (like $x.f$$\mapsto$$\_$ in traditional separation logic \cite{reynolds2002}), $\>*\>$ denotes separating conjunction, and the universal quantifier is an iterated separating conjunction~\cite{reynolds2002,QuantifiedPermissions}, which (here) denotes permissions to all field locations of objects in the footprint $\GRAPH$. In contrast to using recursive definitions to specify unbounded heap structures (\eg{}~separation logic predicates~\cite{YangPhD,Parkinson05}), iterated separating conjunction permits arbitrary sharing within the set $\GRAPH$ (many field \emph{values} may alias the same node) and does not prescribe a traversal order within the data structure.
We assume for simplicity that a method specification expresses \emph{all} required and returned permissions via these implicit contracts with respect to $\GRAPH$, but it is easy to also support other permission specifications, \eg{}~points-to predicates and recursive predicates.

In our example, we use two additional ghost parameters \code{ldag} and \code{rdag} to allow our specification to simply denote the sets of objects constituting the first and second DAG, respectively. The first precondition expresses that the method footprint is the disjoint union of these two DAGs.

\subsection{Local Reachability}
\label{sec:localReachability}

Reasoning in a separation logic has the key advantage that one can modularly verify properties of a method, and reuse this verification for all calling contexts (and concurrently-running threads).
Enforcing that properties verified for the method depend \emph{only} on its footprint, guarantees that they hold independently of the context; we refer to these as the \emph{local} properties of the footprint. However, classical reachability in the heap is \emph{not} a local property of this form. Hence, combining reachability and separation logic requires us to refine the notion of reachability to one that is local, as we explain next.

Our technique provides two predicates to express reachability properties in specifications. We generalize classical reachability by adding an extra footprint parameter, $\GRAPH$ to make the property local. The \emph{edge predicate} $\EDGEF{}{\GRAPH}{x}{y}{F}$ expresses that object $x$ is in the set $\GRAPH$ and has a field from the set of fields $F$ storing a non-null object $y$ (which need not be in $\GRAPH$). The \emph{path predicate} \PATHSYMB{} denotes, for a fixed $\GRAPH$ and $F$, the reflexive, transitive closure of \EDGESYMB{}, that is, $\PATHF{}{\GRAPH}{x}{y}{F}$ expresses that either $x = y$, or there is a path of field references from $x$ to $y$ s.t.~all objects on the path (except possibly $y$) are in $\GRAPH$ and all fields are in $F$; in particular $\PATHSYMB{}$ may denote reachability via multiple fields. We omit the parameter $F$ when the set of fields is clear from the context; for instance, in our example, $F$ consists of the (only) reference-typed fields \code{left} and \code{right}. We say that a path $x \ldots y$ is \emph{$\GRAPH$-local} if $\PATH{}{\GRAPH}{x}{y}$ holds. Both our edge and path predicates are defined over a \emph{mathematical abstraction} of the current heap graph (\cf{}~\secref{snapshots}), and are \emph{pure} in the separation logic sense, allowing us to freely repeat them in specifications. 

Our edge and path predicates enable rich reachability specifications within a method's footprint. The preconditions of \coderef{merge} express that the method footprint is acyclic and closed under the edge relation (due to the second and the last two preconditions), and that \code{ldag} and \code{rdag} contain exactly the objects reachable from \code{l} and \code{r}, resp. In general, method specifications are checked to only employ edge and path predicates whose first parameter is the method's footprint or a subset thereof. 

Method postconditions typically express how reachability \emph{changes} within this footprint. In our example, the first postcondition specifies that the result \code{link} is part of the first DAG and its \code{right}-field was initially null. The \code{old}-expression allows postconditions to refer to pre-state values; we write $\EDGESYMB{0}(\ldots)$ to abbreviate \code{old($\EDGESYMB{}(\ldots)$)}, and analogously for $\PATHSYMB{}$. We can freely mix reachability specifications with specifications in terms of the program heap (\eg{}~the \code{link.right} expression). The other postconditions illustrate how we can specify the new edge and path relations in terms of their originals, summarizing the method's effect. In particular, the last postcondition expresses that an object $x$ reaches an object $y$ in the post-state \emph{iff} it reached $y$ already in the pre-state, or if $x$ reaches \code{link} in the first DAG and $y$ is in the second DAG\@. Our method specification leaves \code{link} underspecified, whereas the implementation chooses the rightmost node in the first DAG\@. We could easily provide a less abstract specification by using path predicates over (only) the \code{right}-field.

\begin{figure}[t]
\begin{center}
	\begin{tikzpicture}[>=latex,xscale=0.5,yscale=0.5]
	\def\xa{ 0.0}
	\def\xb{ 1.5}
	\def\xc{ 4.5}
	\def\xd{ 6.0}
	\def\xe{ 7.5}
	\def\xf{10.5}
	\def\xg{12.0}
	
	\def\ya{4.5}
	\def\yb{3.0}
	\def\yc{1.5}
	\def\yd{0.0}
	
	\def\p{1.2}              
	\def\d{0.15}              
	\def\l{{(\xa-\p)}}       
	\def\r{{(\xg+\p)}}       
	\def\t{{(\ya+\p)}}       
	\def\b{{(\yd-\p)}}       
	\def\ml{(0.5*(\xb+\xc))} 
	\def\mr{(0.5*(\xe+\xf))} 
	
	\draw [frame] (\l,\b) rectangle ({\ml-\d},\t);
	\draw [footprint] ({\ml+\d},\b) rectangle (\r,\t);
	\draw [footprint,densely dashed] ({\mr},\b) -- ({\mr},\t);
	
	\node [framelbl,anchor=south] at ({(\l+\ml-\d)*0.5},\b) {frame};
	\node [footprintlbl,anchor=south] at ({(\ml+\d+\mr)*0.5},\b) {\texttt{nldag}};
	\node [footprintlbl,anchor=south] at ({(\r+\mr)*0.5},\b) {\texttt{rdag}};
	
	\node [vertex,lbl=\texttt{l}] (a1) at (\xa,\ya) {};
	\node [vertex]         (a2) at (\xa,\yb) {};
	\node [vertex]         (a3) at (\xa,\yc) {};
	\node [vertex]         (a4) at (\xa,\yd) {};
	\node [vertex]         (a5) at (\xb,\yb) {};
	\node [vertex]         (a6) at (\xb,\yc) {};
	\node [vertex]         (a7) at (\xb,\yd) {};
	\node [vertex,lbl=\texttt{l.right}]         (b1) at (\xc,\ya) {};
	\node [vertex]         (b2) at (\xc,\yb) {};
	\node [vertex]         (b3) at (\xc,\yc) {};
	\node [vertex]         (b4) at (\xd,\ya) {};
	\node [vertex]         (b5) at (\xd,\yb) {};
	\node [vertex]         (b6) at (\xd,\yc) {};
	\node [vertex,lbl=\texttt{link}]         (b7) at (\xe,\ya) {};
	\node [vertex]         (b8) at (\xe,\yb) {};
	\node [vertex]         (b9) at (\xe,\yc) {};
	\node [vertex,lbl=\texttt{r}] (c1) at (\xf,\ya) {};
	\node [vertex]         (c2) at (\xf,\yb) {};
	\node [vertex]         (c3) at (\xg,\ya) {};
	\node [vertex]         (c4) at (\xg,\yb) {};
	
	\draw [edge] (a1) -- (a2);
	\draw [edge] (a2) -- (a3);
	\draw [edge] (a2) -- (a5);
	\draw [edge] (a3) -- (a4);
	\draw [edge] (a3) -- (a6);
	\draw [edge] (a4) -- (a7);
	\draw [edge] (a6) -- (a7);
	
	\draw [edge] (a1) -- (b1);
	\draw [edge] (a5) -- (b2);
	\draw [edge] (a5) -- (b3);
	\draw [edge] (a6) -- (b3);
	\draw [edge] (a7) -- (b3);
	
	\draw [edge] (b1) -- (b2);
	\draw [edge] (b1) -- (b4);
	\draw [edge] (b2) -- (b3);
	\draw [edge] (b2) -- (b5);
	\draw [edge] (b4) -- (b5);
	\draw [edge] (b4) -- (b7); 
	\draw [edge] (b5) -- (b6);
	\draw [edge] (b5) -- (b8);
	\draw [edge] (b6) -- (b9);
	\draw [edge] (b8) -- (b9);
	
	\draw [edge,densely dashed] (b7) -- (c1);
	
	\draw [edge] (c1) -- (c2);
	\draw [edge] (c1) -- (c3);
	\draw [edge] (c2) -- (c4);
	\draw [edge] (c3) -- (c4);
	\end{tikzpicture}
\end{center}
	\caption{An example scenario of running \coderef{merge} on two DAGs rooted in \code{l} and \code{r}. Small circles correspond to heap objects; solid arrows represent fields initialized in the pre-state that are  unchanged; the dashed arrow represents the new heap edge (created in the post-state by initializing a field). The frame of the recursive call is surrounded with blue; the footprint is surrounded with red.}
	\label{fig:merge}
\end{figure}
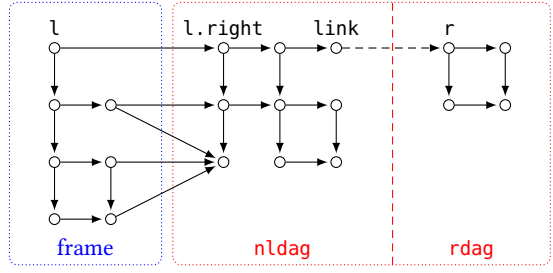

The recursive call in method \coderef{merge} needs to supply values for the three ghost parameters. We construct these values using a predefined function \code{$\coderef{sub}$($\GRAPH$:Graph,$~$$\HEAP$:Graph,$~$root:Node)}, which yields the subset of $\HEAP$ reachable from the node \code{root} via $\GRAPH$-local paths. The properties known for the resulting set are summarized by the following \emph{heap-dependent function}~\cite{mueller-2015-viper} declaration:\footnote{Unlike methods, functions in our language are guaranteed to be side-effect free. Hence, we do not distinguish between $\PATHSYMB{}$ and $\PATHSYMB{0}$ in the postcondition of \coderef{sub} (similar for $\EDGESYMB{}$ and $\EDGESYMB{0}$).}
\phantomsection\label{lst:sub}
\begin{lstlisting}[mathescape,escapeinside={(*@}{@*)}]
function sub($\GRAPH$: Graph, $\HEAP$: Graph, root: Node): Graph
  requires $\texttt{root} \in \HEAP \land \HEAP \subseteq \GRAPH$
  ensures $\texttt{\textbf{result}} \subseteq \HEAP \land \texttt{root} \in \texttt{\textbf{result}} \land \RELCLOSED{\texttt{\textbf{result}}}{\HEAP} \> \land$
          $\forall n \>\bullet\> n \in \texttt{\textbf{result}} \Leftrightarrow \PATH{}{\GRAPH}{\texttt{root}}{n}$
\end{lstlisting}
where \code{result} refers to the result value of the function; $\RELCLOSED{\RESULT}{\HEAP}$ denotes that an edge that exits must not end in $\HEAP$: 
\begin{equation}
\RELCLOSED{\RESULT}{\HEAP} \>\>\longeq\>\> \forall x \in \RESULT, y \>\bullet\> \EDGE{}{\RESULT}{x}{y} \implies y \notin \HEAP{\setminus}\RESULT
\label{eq:RelativeClosedness}
\end{equation}

Note that $\RELCLOSED{\RESULT}{\HEAP}$ is permissive enough to allow selecting new footprints for method calls even if the current footprint is \emph{open}, \ie{}~if there exist edges that exits the current footprint. To specify that a subheap is closed \emph{in the global heap}, we would use a stronger condition:
\begin{equation}
\mo{CLOSED}(\GRAPH) \>\>\longeq\>\> \forall x\in\GRAPH, y\notin\GRAPH \>\>\bullet\>\> \lnot \EDGE{}{\GRAPH}{x}{y}
\label{eq:closed}
\end{equation}

\subsection{Verification Challenges}

The specification ingredients presented above allow us to combine separation logic specification with reachability. However, practical verification of these specifications requires the solution of three challenges.
Firstly, we must handle direct updates to the program heap, and model their effects on our $\EDGESYMB{}$ and $\PATHSYMB{}$ predicates. SMT solvers cannot efficiently automate reasoning about a direct definition of $\PATHSYMB{}$ as transitive closure, but it has been shown that a first-order approximation technique can be efficiently used for this purpose~\cite{Dong1995IncrementalAD,SagivLMCS09}.
Secondly, and most challenging, we require a technique to deduce reachability for a method \emph{caller} from what is known about its callee's footprint: a problem we call \emph{reachability framing}. This is necessary, for example, when reasoning about the recursive call to \coderef{merge} in our example; we must relate local reachability information in the caller's footprint to that of its callee.
Finally, we require a modeling of our verification technique in an automated tool; we aim for proof obligations ultimately amenable to first-order SMT solvers, which necessitates effective quantifier instantiation strategies.


\section{Reachability in Acyclic Structures}
\label{sec:DAGs}
In this section, we explain the core ingredients of our verification technique for combining reachability information with separation logic style reasoning. Reasoning about a method starts with assuming its precondition. The precondition provides permissions to access the objects (\ie{}~nodes) in its footprint and the reachability constraints that guarantee the existence or the absence of heap paths connecting some objects from the footprint. As the program performs modifications to some parts of the heap, our goal is to determine a precise way of checking any (local) reachability query (\eg{}~in the method's postcondition) after these changes. Hence it is important to identify the paths that were unchanged and those that were created or destroyed by each operation. 

Heap modifications are performed either directly by field updates or indirectly through method calls. In the former case, the reachability properties known to hold before the update need to be adjusted to reflect the change of heap references (\secref{fieldupdates}).
For a field update, the local reachability properties before and after the update can be expressed within the same (enclosing method's) footprint. The situation is more complex for method calls (\secref{convexframe}). To determine the reachability properties after a call (the reachability framing problem), one needs to combine reachability properties before the call that are known to be outside of the call's footprint (hence, unaffected by the call) with reachability properties guaranteed by the callee method (as expressed in the callee's postcondition). These two sets of properties are expressed within the footprints of the client and the callee, respectively. If these footprints are not equal, then the reachability properties guaranteed by the callee need to be re-interpreted in the client's footprint.

We present our techniques for tackling these challenges in the remainder of the paper. We discuss how our reachability reasoning technique is integrated with separation logic in~\secref{snapshots}. The technique for direct field updates discussed in~\secref{fieldupdates} requires the current method's footprint to be acyclic; note that we generally permit arbitrary structures, including those with heap cycles, \emph{outside} of the footprint. However, our technique for method calls, and all of the formulas that we present in~\secref{convexframe}, \emph{do not} require acyclicity. Instead, we require and exploit \hyperref[def:convexity]{\emph{relative convexity}} of method footprints, a novel restriction that is strong enough to reduce the reachability framing problem to first-order formulas tractable for SMT solvers, but permissive enough to embrace a broad spectrum of challenging data structures. \secref{ZOPs} explains how our technique can be extended to  potentially \emph{cyclic} 0--1-path graphs. 

\subsection{Encoding of Edge and Path Predicates}
\label{sec:snapshots}

Our specification technique supports reachability via the edge predicate $\EDGESYMB{}$ and the path predicate $\PATHSYMB{}$. In order to verify such specifications, we encode them into a flavor of separation logic and use an existing verification tool to construct proofs in that logic. We use Implicit Dynamic Frames~\cite{ImpDynFrames} for this purpose, a variation of separation logic~\cite{ParkinsonSummers12} that separates specifications of access permissions for memory locations from specifications of the values stored in these locations. For instance, separation logic's points-to predicate $x.f \mapsto v$ is specified in implicit dynamic frames as a conjunction of the access permission and the field content: \code{acc($x$.$f$) * $\ x$.$f = v$}. This separation of permissions and value properties allows us to conveniently express additional value properties, \eg{}~sortedness, in addition to reachability properties, without having to define a new graph-abstraction that exposes the values of interest. 

Our edge predicates could be defined directly, \eg{}~as $(x\mo{.}f_1 = v \lor x\mo{.}f_2 = v)$ for two fields $f_1$ and $f_2$; conceptually, $\EDGESYMB{}$ is a first-order abstraction over this property, which may, in particular, be used in the \emph{syntactic triggering patterns}~\cite{Simplify,MoskalEtAl,SMTLIB,Z3solver} that the SMT solver requires to control quantifier instantiations (and which cannot include logical operations such as $\lor$ above).

Unlike the edge predicate $\EDGESYMB{}$, directly defining the path predicate $\PATHSYMB{}$ would compromise automation. A definition would involve transitive closure, which is notoriously difficult to handle for SMT solvers. Therefore, we take a different approach here. We leave the path predicate undefined and axiomatize its essential properties, for instance, how it is affected by heap updates. We specify these axioms over mathematical graphs and not directly over the heap-dependent edge and path predicates. Therefore, our encoding first abstracts the heap within a footprint to a set of edges (ordered pairs of nodes) and then expresses reachability over those. This abstraction is defined by a predefined function called \coderef{snapshot}. For simplicity, we define \coderef{snapshot} using the notation of our source language, but it is only used internally by our encoding. In particular, the function implicitly depends on the heap and requires permissions to all objects in its footprint $\GRAPH$:

\phantomsection\label{lst:snapshot}
\vspace{-2mm}
\begin{lstlisting}[mathescape,escapeinside={(*@}{@*)}]
function snapshot$^F$($\GRAPH$: Graph): Edgeset
  ensures  $\forall x,y \>\bullet\> x \in \GRAPH \land y \neq \NULLLIT \land
           (x.f_1 = y \vee \ldots \vee x.f_n=y) \iff (x,y) \in$ result
\end{lstlisting}

\noindent
Here, \code{Edgeset} is the type of sets of pairs of nodes, and  $F = \{f_1,\ldots, f_n\}$; we omit this parameter when it is clear from the context. The postcondition can be thought of as an axiom over an uninterpreted function that defines its semantics. Note that \coderef{snapshot} also collapses edges between two objects for different field names (duplicate edges are not needed to keep track of reachability).

This abstraction function lets us define the edge predicate in a straightforward way:
\begin{equation}\label{eq:edge}
\EDGEF{}{\GRAPH}{x}{y}{F} \>\>\longeq\>\>
(x,y) \in \textrm{\coderef{snapshot}}^F(\GRAPH)
\end{equation}

\noindent
To avoid the issues with transitive closure mentioned above, we do not define the path predicate directly, but axiomatize the properties we need for verification. In fact, we define the path relation in terms of a function $\hat{P}$ over graphs and then axiomatize the latter, state-independent function:
\begin{equation}
\PATHF{}{\GRAPH}{x}{y}{F} \>\>\longeq\>\>
\hat{P}(\textrm{\coderef{snapshot}}^F(\GRAPH), x, y)
\end{equation}
For this axiomatization, we carefully control the quantifier instantiation performed by SMT solvers to avoid diverging proof search. For instance, we include the axiom below, but let the solver instantiate it only to a fixed depth of unrolling $\hat{P}$~\cite{limitedfunctions}.
\begin{equation}\label{eq:partialDefinitionOfPathPredicate}
  \hat{P}(G,x,y) \>\>\longeq\>\>
  x = y \vee
  \exists z \>\bullet\> (x,z) \in G \land \hat{P}(G,z,y)
\end{equation}

\subsection{Field Updates}
\label{sec:fieldupdates}

A field update \code{x.f := v} may affect reachability properties in the heap and, thus, both edge and path predicates. Since our encoding contains a precise definition of the edge predicate in terms of the underlying heap (via \eqref{edge} and the definition of \coderef{snapshot}), the verifier can determine which edge predicates hold after a field update.

However, determining the effect of a single field update on the \emph{path} relation is more intricate as its the axiomatization is not sufficient to determine which predicates hold after a field assignment (\eg{}~because this reasoning step would require induction proofs, which SMT solvers cannot find automatically). We solve this problem by adapting an existing approach: for \emph{acyclic} graphs (which we focus on in this section), one can provide first-order \emph{update formulas} that express precisely how adding or deleting a single edge affects reachability~\cite{Dong1995IncrementalAD,SagivLMCS09}. For example, the following update formula characterizes the effect of adding an edge between nodes $a$ and $b$ (\eg{}~by initializing a field of $a$):
\begin{equation}\label{eq:update}
\forall x,y \>\>\bullet\>\> \PATHF{}{\GRAPH}{x}{y}{F} \iff \PATHF{0}{\GRAPH}{x}{y}{F} \>\vee\> \PATHF{0}{\GRAPH}{x}{a}{F} \wedge \PATHF{0}{\GRAPH}{b}{y}{F}
\end{equation}
\noindent
where $\PATHSYMB{}$ and $\PATHSYMB{0}$ denote the path predicate in the states before and after the update.

\begin{figure}[t]
\begin{center}
\begin{tikzpicture}[>=latex,xscale=0.75,yscale=0.5]

\def\xa{0.0}
\def\xab{1.5}
\def\xb{2.0}
\def\xc{3.0}
\def\xcd{3.5}
\def\xd{5.0}

\def\yo{3.0}
\def\ya{2.0}
\def\yb{1.0}
\def\yc{0.0}

\def\dta{0.5}

\node [vertex,lbl=$x$] (x) at (\xa,\yb) {};
\node [vertex,lbld=$u$] (u) at (\xb,\ya) {};
\node [vertex,lbld=$v$] (v) at (\xc,\ya) {};
\node [vertex,lbl=$y$] (y) at (\xd,\yb) {};

\draw [path] (x) -- (u);
\draw [edge] (u) -- node[draw, midway, -, black, sloped, cross out, line width=.2ex, minimum width=1.5ex, minimum height=1ex, anchor=center]{} (v);
\draw [path] (v) -- (y);
\draw [path] (x) to (\xab,\yc) to (\xcd,\yc) to (y);


\end{tikzpicture}
\end{center}
\caption{The reachability update problem in presence of alternative paths. Concrete heap edges are represented by straight arrows, while (possibly, zero-length) heap paths are represented by wavy arrows. The upper path $x \ldots y$ \emph{depends} on the edge $(u,v)$; removing this edge would destroy the path, but $x$ may still reach $y$ after deleting $(u,v)$, here, via the lower path. Alternative paths may occur in our setting because we permit multiple reference fields per object.}
\label{fig:atomicUpdateDiagram}
\end{figure}
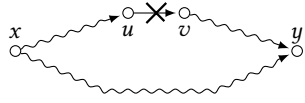

The update formula for removing an edge is more complex. Since we allow for an arbitrary out-degree of nodes (via multiple reference fields), it is possible for there to exist multiple paths between two different nodes (\figref{atomicUpdateDiagram}). When adding an edge between two nodes, the new \PATHSYMB{} relation can be updated relatively simply, \eg{}~via \eqref{update}; no paths have been lost, and only paths connected by this new edge are created. On \emph{removal} of an edge, no paths are created, but, for node pairs previously connected by a path using this edge, it is unclear whether or not they belong to the new \PATHSYMB{} relation, due to the possibility of \emph{alternative paths}. This entails a more-complex update formula for the edge-removal case (due to~\citet{Dong1995IncrementalAD}); see \appref{acyclicTCupdates} for details. A general field update entails removing and then adding an edge, as we demonstrate for our \coderef{merge} example in~\figref{viperencoding}. 

Our verification technique rewrites each field update \code{x.$f$ := v} with a method call to an internal \code{update} method with the same footprint $\GRAPH$ as for the current method. The postconditions of \code{update} make the reachability update formulas available to the SMT solver. This way, we assume the update formulas for each field set $F$ that is used in the current method specification and that contains the updated field $f$ (reachability for other field sets is not affected by the update).

The else-branch in the example from \figref{runningExample} modifies the heap through a single field update. The second postcondition describes the effect on the edge relation; it follows directly from the definition of the edge predicate. The third postcondition, about the path relation, is exactly the update formula~\eqref{update}, with \code{link} and \code{r} for $a$ and $b$, resp.

\subsection{Method Calls and Relatively Convex Footprints}
\label{sec:convexframe}

Update formulas allow us to precisely capture the effect of adding or removing \emph{individual} edges, which is sufficient to reason about field updates. However, reasoning modularly about method calls requires us to determine the effect of \emph{multiple} heap updates. According to the callee's specification, we can partition the footprint $\GRAPH$ of the client into the footprint $\HEAP$ of the callee  and the remainder $\FRAME$ ($\GRAPH = \FRAME \DISJUNION \HEAP$). This remainder is the \emph{frame} of the call and cannot be modified by the callee method. The postcondition of a callee method provides a specification of reachability information \emph{within} its footprint $\HEAP$. The challenge is to determine the effect of the call on reachability within the (generally larger) footprint $\GRAPH$ of its client. For the edge relation, this extrapolation is straightforward:
\begin{equation}
  \forall x \in \FRAME \DISJUNION \HEAP, y \>\bullet\> \EDGE{}{\FRAME \DISJUNION \HEAP}{x}{y} \iff \EDGE{}{\FRAME}{x}{y} \lor \EDGE{}{\HEAP}{x}{y}
  \label{eq:edgePartitioning}
\end{equation}
\noindent In separation logic, a method may modify any heap edges that \emph{originate} in its footprint; hence, the predicate $\EDGE{}{\HEAP}{x}{y}$ implies that $x$ is in the footprint $\HEAP$ and $\EDGE{}{\FRAME}{x}{y}$ implies that $x$ is in the frame $\FRAME$. We refer to edges that cross the boundary of the footprint as \emph{cut points}: if $x\in\HEAP, y\notin\HEAP$, then $(x,y)$ is an \emph{exit point} of the footprint, and if $x\notin\HEAP, y\in\HEAP$, then $(x,y)$ is an \emph{entry point} into the footprint.

Unfortunately, a simple rule such as \eqref{edgePartitioning} does not exist for relating \emph{paths} in $\FRAME \DISJUNION \HEAP$ to those in $\FRAME$ and $\HEAP$. A path can span fields from both heap partitions, and, in general, could cross the boundary between the two unboundedly many times. It is known that, in full generality, a first-order reachability framing formula for our path predicate cannot exist (see \eg{}~\cite{ItzhakyPOPL14}). The key insight behind our technique for handling method calls is that this intractable situation becomes tractable if the footprint of the callee is \emph{relatively convex} in the composed heap.

\begin{chuckwalladef}[Relatively Convex Footprints] \label{def:convexity}
In a given program state and for a given set of reference fields $F$, footprint $\HEAP$ defines a \emph{relatively convex sub-footprint} of footprint $\GRAPH$ (written $\HEAP \CONVEXIN \GRAPH$) iff $\GRAPH = \FRAME \DISJUNION \HEAP$ for some footprint $\FRAME$, and no paths within $\GRAPH$ leave $\HEAP$ and then return:
\[
  \forall x,y \in \HEAP, u \in \FRAME \>\bullet\> \lnot \PATHF{}{\GRAPH}{x}{u}{F} \lor \lnot \PATHF{}{\GRAPH}{u}{y}{F}
\]
\end{chuckwalladef}

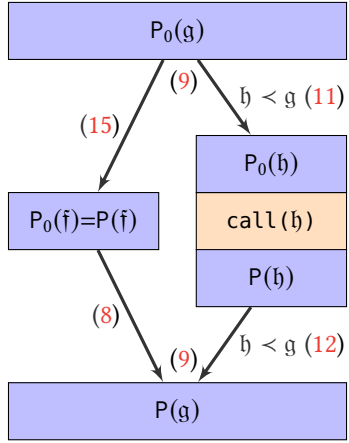
\begin{figure}[t]
\tikzstyle{process} = [rectangle, minimum width=2cm, minimum height=0.75cm, text centered, draw=black, fill=orange!25]
\tikzstyle{block} = [rectangle, minimum width=2cm, minimum height=0.75cm, text centered, draw=black, fill=blue!25]
\tikzstyle{line} = [draw, very thick, color=black!80, -latex']
\begin{center}
	
\begin{tikzpicture}[node distance=5pt,every node/.style={draw,minimum width=50pt}]

	\node (F) [process, node distance=1.0cm] {\code{call($\HEAP$)}};
	\node (C) [block, above of=F, node distance=0.75cm] {$\PATHREL{0}{\HEAP}$};
	\node (D) [block, below of=F, node distance=0.75cm] {$\PATHREL{}{\HEAP}$};
	\node (B) [block, left of=F, node distance=2.5cm] {$\PATHREL{0}{\FRAME}{=} \PATHREL{}{\FRAME}$};
	
	\node (A) [block, span horizontal=(B)(C), above of=C, node distance=1.75cm, xshift=-1.25cm, label=below:{~~\eqref{pathPartitioning}}] {$\PATHREL{0}{\GRAPH}$};
	\node (E) [block, span horizontal=(B)(C), below of=D, node distance=1.75cm, xshift=-1.25cm, label=above:{~~\eqref{pathPartitioning}}] {$\PATHREL{}{\GRAPH}$};

	\path [line] (A) -- (C) node[midway,right,draw=none] {$\HEAP \CONVEXIN \GRAPH$ \eqref{checkConvexityBeforeMethodCall}};
	
	\path [line] (D) -- (E) node[midway,right,draw=none] {$\HEAP \CONVEXIN \GRAPH$ \eqref{checkConvexityAfterMethodCall}}; 

	\path [line] (A) -- node[draw=none, midway, -, minimum width=1.5ex, minimum height=1ex, anchor=east]{\eqref{frameLocalization}} (B); 
	
	\path [line] (B) -- node[draw=none, midway, -, minimum width=1.5ex, minimum height=1ex, anchor=east]{\eqref{preservedFrameReachability}} (E); 

\end{tikzpicture}
	
\end{center}
\caption{The flow of reachability information in the presence of method calls. Reachability information in the client's footprint $\GRAPH = \FRAME \DISJUNION \HEAP$ is split into reachability information within $\HEAP$ and $\FRAME$ before the call, the effect of the call on $\HEAP$ is accounted for, and then the information is recombined to paths in $\GRAPH$. The numbers in parentheses indicate ingredients of our technique explained in this section.}
\label{fig:infFlow}
\end{figure}

We show, in the remainder of this section, how we exploit this property to enable precise, first-order, and modular reasoning about reachability in presence of method calls. In particular, we are able to make tractable the problem of framing reachability information when a method footprint $\HEAP$ is relatively convex in its client's footprint $\GRAPH$. This requirement is checked by our technique at the call site, but is typically naturally the case. For example, any method operating on a recursively-defined data type, its sub-structures or portions thereof (such as linked list segments), a strongly connected component of a potentially-cyclic structure, or combinations of these will have a relatively convex footprint. DAG traversals also have relatively convex footprints.
For instance, the recursive call to \coderef{merge} in our running example of \figref{runningExample} and the corresponding illustration in \figref{merge} demonstrate a method call with a relatively convex footprint. Note that both acyclicity and relative convexity are defined relatively to a field set $F$. Therefore, even operations on data structures with back-pointers (such as parent-pointers in a tree) typically have relatively convex footprints as long as the path predicates are defined in terms of only forward-references or only back-pointers.

\subsubsection*{Method call overview} A high-level overview of our solution is illustrated in \figref{infFlow}. We use $\PATHSYMB{0}$ to represent the paths before the call, and $\PATHSYMB{}$ for those afterwards. According to standard separation logic reasoning, method calls are only allowed if the callee's footprint $\HEAP$ is a subset of the client's ($\GRAPH = \FRAME \DISJUNION \HEAP$, for some frame $\FRAME$). Under the additional requirement that $\HEAP \CONVEXIN \GRAPH$, the technique we present in this section shows how  to decompose reachability information before the call (\ie{}~expressed in terms of $\PATHSYMB{0}(\GRAPH,\ldots)$) into paths in the callee's footprint ($\PATHSYMB{0}(\HEAP,\ldots)$) and paths in the frame ($\PATHSYMB{0}(\FRAME,\ldots)$). The callee's specification is responsible for relating $\PATHSYMB{}(\HEAP,\ldots)$ information to $\PATHSYMB{0}(\HEAP,\ldots)$ information, \ie{}~specifying how reachability changes \emph{within} the callee's footprint. Conversely, reachability purely in the \emph{frame} $\FRAME$ cannot be changed by a call, since it does not have the permissions to do so.
Indeed, based on consideration of the permissions \emph{not passed} to the method call, we know that the following formula holds (which we call \emph{separation-logic framing}):%
\begin{equation}
\forall x \in \FRAME, y \>\>\bullet\>\> \PATH{}{\FRAME}{x}{y} \iff \PATH{0}{\FRAME}{x}{y}
\label{eq:preservedFrameReachability}
\end{equation}
Our technique then provides means of reconstructing reachability in the client's footprint ($\PATHSYMB{}(\GRAPH,\ldots)$) from the information we have after the call in terms of $\PATHSYMB{}(\HEAP,\ldots)$ and $\PATHSYMB{}(\FRAME,\ldots)$.

\subsubsection*{Path partitioning} 
\begin{figure}[t]
\begin{center}
	\begin{tikzpicture}[>=latex,xscale=0.5,yscale=0.5]
		\def\xa{ 0.0}
		\def\xb{ 1.5}
		\def\xc{ 3.5}
		\def\xd{ 5.0}
		\def\xe{ 6.0}
		\def\xf{ 7.5}
		\def\xg{ 9.5}
		\def\xh{11.0}
	
		\def\ya{ 3.5}
		\def\yb{ 2.5}
		\def\yc{ 1.5}
		\def\yd{ 0.5}
		
		\def\px{2.85}              
		\def\py{1.2}              
		\def\d{0.15}             
		\def\l{{(\xa-\px)}}       
		\def\r{{(\xh+\px)}}       
		\def\t{{(\ya+\py)}}       
		\def\b{{(\yd-\py*0.5)}}       
		\def\ml{(0.5*(\xb+\xc))} 
		\def\mr{(0.5*(\xf+\xg))} 
		\def\mv{(0.5*(\yc+\yd))} 
		
		\draw [frame] (\l,\b) -- (\r,\b) -- (\r,\t) -- ({\mr+\d},\t) -- ({\mr+\d},{\mv-\d})
		                      -- ({\ml-\d},{\mv-\d}) -- ({\ml-\d},\t) -- (\l,\t) -- cycle;
		\draw [footprint] ({\ml+\d},{\mv+\d}) rectangle ({\mr-\d},\t);
		
		\node [framelbl,anchor=north] at ({\xh},\t) {frame (\FRAME)};
		\node [footprintlbl,anchor=north] at ({(\xc+\xf)*0.5},\t) {footprint (\HEAP)};
		
		\node [hvertex,lbl=$x_{\RN1}$] (x1) at (\xc,\ya) {};
		\node [fvertex,lbl=$x_{\RN2}$] (x2) at (\xa,\yb) {};
		\node [hvertex,lbl=$x_{\RN3}$] (x3) at (\xe,\yb) {};
		\node [fvertex,lbl=$x_{\RN4}$] (x4) at (\xa,\yc) {};
		\node [fvertex,lbl=$x_{\RN5}$] (x5) at (\xa,\yd) {};
		\node [hvertex,lbl=$y_{\RN1}$] (y1) at (\xf,\ya) {};
		\node [hvertex,lbl=$y_{\RN2}$] (y2) at (\xd,\yb) {};
		\node [fvertex,lbl=$y_{\RN3}$] (y3) at (\xh,\yb) {};
		\node [fvertex,lbl=$y_{\RN4}$] (y4) at (\xh,\yc) {};
		\node [fvertex,lbl=$y_{\RN5}$] (y5) at (\xh,\yd) {};
		\node [hvertex,lbl=$a_{\RN2}$] (a2) at (\xc,\yb) {};
		\node [hvertex,lbl=$a_{\RN4}$] (a4) at (\xc,\yc) {};
		\node [fvertex,lbl=$b_{\RN3}$] (b3) at (\xg,\yb) {};
		\node [fvertex,lbl=$b_{\RN4}$] (b4) at (\xg,\yc) {};
		
		\draw [hpath] (x1) -- (y1);
		\draw [fpath] (x2) -- (a2);
		\draw [hpath] (a2) -- (y2);
		\draw [hpath] (x3) -- (b3);
		\draw [fpath] (b3) -- (y3);
		\draw [fpath] (x4) -- (a4);
		\draw [hpath] (a4) -- (b4);
		\draw [fpath] (b4) -- (y4);
		\draw [fpath] (x5) -- (y5);
	\end{tikzpicture}
\end{center}
\caption{No paths originating and ending inside a relatively convex footprint may go though nodes of its frame. Therefore, the paths that originate in the frame may enter and exit the footprint at most once. This gives five possibilities for a path to interact with a relatively convex footprint. 
\label{fig:pathDistribution}
}
\end{figure}
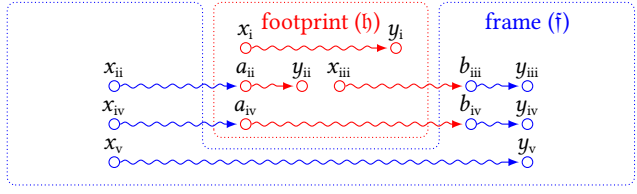

The first key step of our solution is \emph{path partitioning}. We exploit relative convexity of the callee's footprint to define a set of formulas for soundly and precisely relating reachability in a caller's footprint to reachability in the callee and its frame, and vice versa.
\figref{pathDistribution} illustrates the possibilities for a path in the client's footprint $\GRAPH$ to interact with a relatively convex footprint $\HEAP$. We proceed by analyzing
\figref{pathDistribution} by cases, deriving formulas one of which must hold in each possible case.


Crucially, our relative convexity assumption $\HEAP \CONVEXIN \GRAPH$ guarantees that no paths  $x \ldots y$ in $\GRAPH = \FRAME \DISJUNION \HEAP$ enter or leave $\HEAP$ more than once.
We summarize the five cases for paths from $x$ to $y$ based on the distribution of these nodes between the footprint, $\HEAP$, and the frame of the call, $\FRAME$:
~(i) $x,y\in\HEAP$ as is the whole path,
~(ii) $x\in\FRAME, y\in\HEAP$; the path crosses the boundary \emph{once},
~(iii) $x\in\HEAP, y\in\FRAME$ again crossing \emph{once},
~(iv) $x,y\in\FRAME$ with a path entering and leaving $\HEAP$ \emph{once}, and
~(v) $x,y \in \FRAME$ with a path entirely in $\FRAME$.
Note that these cases are exhaustive for a path between $x, y \in \GRAPH$, due to our convexity restriction.

These five cases translate to the following formulas, allowing us to relate reachability in $\FRAME \DISJUNION \HEAP$ and reachability in the two subheaps $\FRAME$ and $\HEAP$ individually, which we call \emph{path partitioning formulas}:
\begin{equation}
\begin{alignedat}{5}
\text{(i)}            \quad &\forall x\in \HEAP,  &&\>y\in \HEAP   &&\>\>\bullet\>\>&& \PATH{}{\FRAME\DISJUNION \HEAP}{x}{y} \iff \PATH{}{\HEAP}{x}{y} \\
\text{(ii)}           \quad &\forall x\in \FRAME, &&\>y\in \HEAP   &&\>\>\bullet\>\>&& \PATH{}{\FRAME\DISJUNION \HEAP}{x}{y} \iff \exists a\in \HEAP \>\bullet\> \PATH{}{\FRAME}{x}{a} \wedge \PATH{}{\HEAP}{a}{y} \\
\text{(iii)}          \quad &\forall x\in \HEAP,  &&\>y\in \FRAME  &&\>\>\bullet\>\>&& \PATH{}{\FRAME\DISJUNION \HEAP}{x}{y} \iff \exists b\in \FRAME \>\bullet\> \PATH{}{\HEAP}{x}{b} \wedge \PATH{}{\FRAME}{b}{y} \\
\text{(iv)--(v)}\quad &\forall x\in \FRAME, &&\>y\in \FRAME  &&\>\>\bullet\>\>&& \PATH{}{\FRAME\DISJUNION \HEAP}{x}{y} \iff \PATH{}{\FRAME}{x}{y} \>\vee \\
 ~                          & ~                   && ~             && ~             && \exists a\in \HEAP, b\in \FRAME \>\bullet\> \PATH{}{\FRAME}{x}{a} \wedge \PATH{}{\HEAP}{a}{b} \wedge \PATH{}{\FRAME}{b}{y}
\end{alignedat}
\label{eq:pathPartitioning}
\end{equation}

These formulas can be used left-to-right or right-to-left. In the former case, we obtain a canonical means of \emph{decomposing} information about paths in a composed footprint of the client into information about paths in the callee's footprint and paths in the frame. In the latter case, we obtain means of \emph{reassembling}
reachability information in the composed footprint from that in the constituent parts. In practice, we add separate \code{assume} statements for both directions of each formula, so that we can clearly specify to the underlying SMT solver when to instantiate the formula in which direction.

It is due to our relative convexity assumption that there exist simple first-order path partitioning formulas~\eqref{pathPartitioning}. Without this assumption, the number of cut points of $\HEAP$ could be \emph{arbitrary}, and localization of reachability information would require either considering an unbounded set of cases or higher-order reasoning, preventing automatic verification.

For simplicity, formulas~\eqref{pathPartitioning} cover only the cases $x, y \in \GRAPH$: however, paths that are local to a particular footprint may leave that footprint by a single edge (and so case~(i) above, for example, does not provide information about such paths in $\HEAP$). The following formulas reduce the case $y \notin \GRAPH$ to the cases already covered by introducing a node $u\in \GRAPH$ with an edge to $y$:
\begin{equation}
\begin{array}{c}
  \forall x\in\HEAP, y\notin\HEAP \>\>\bullet\>\> \PATH{}{\HEAP}{x}{y} \iff \exists u\in\HEAP \>\bullet\> \PATH{}{\HEAP}{x}{u} \land \EDGE{}{\HEAP}{u}{y}\\
  \forall x\in\FRAME, y\notin\FRAME \>\>\bullet\>\> \PATH{}{\FRAME}{x}{y} \iff \exists u\in\FRAME \>\bullet\> \PATH{}{\FRAME}{x}{u} \land \EDGE{}{\FRAME}{u}{y}\\
\end{array}
\label{eq:crossBorderPaths}
\end{equation}


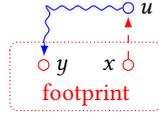
\begin{figure}[t]
\begin{center}
\begin{tikzpicture}[>=latex,xscale=0.75,yscale=0.5]

\def\xa{0.0}
\def\xb{1.5}

\def\ya{1.5}
\def\yb{0.0}

\def\dta{0.6}

\node [fvertex,lblr=$u$] (u1) at (\xb,\ya) {};
\node [hvertex,lblr=$y$] (y) at (\xa,\yb) {};
\node [hvertex,lbll=$x$] (x) at (\xb,\yb) {};

\draw [footprint] ({\xa-\dta},{\yb+\dta}) rectangle ({\xb+\dta},{\yb-2*\dta});
\node [footprintlbl,anchor=north] at (\xa+0.75,\yb-0.25) {footprint};

\draw [red,edge,dashed] (x) -- (u1);
\draw [fpath] (u1) -- (\xa+0.044,\ya) -- (y);


\end{tikzpicture}
\end{center}
\caption{The footprint (surrounded with red) is relatively convex \emph{before} a method call, satisfying~\eqref{checkConvexityBeforeMethodCall}. The heap edge created by the call is represented by a dashed arrow. The new edge $(x,u)$ is an \emph{exit point} of the footprint into the frame; since its end node $u$ reaches the footprint via some path $u \ldots y$, adding $(x,u)$ violates the relative convexity property of the footprint~\eqref{checkConvexityAfterMethodCall}.}
\label{fig:nonconvex}
\end{figure}

\subsubsection*{Checking relative convexity of footprints.} In terms of reasoning about calls, we emit \code{assume} statements for our path partitioning formulas \emph{both} before and after a method call (to decompose paths into those matching the callee's footprint and frame before the call, and to reconstruct information from these sources back to the client's footprint, afterwards; \cf{} \figref{infFlow}). In both cases, before assuming our path-partitioning formulas, we first check that the footprint is relatively convex (since this property justifies their soundness); as we show in~\figref{nonconvex}, a method's footprint could be relatively convex before the call but non-convex in the client's footprint afterwards. The two checks employed by our technique must be expressed in slightly different terms. Before the call (and without yet emitting our path-partitioning formulas) the $\GRAPH$-local reachability information is available, and we directly use the formula from \defref{convexity}:
\begin{equation}
\forall x,y \in \HEAP, u \in \FRAME \>\bullet\> \lnot \PATH{}{\GRAPH}{x}{u} \lor \lnot \PATH{}{\GRAPH}{u}{y}
\label{eq:checkConvexityBeforeMethodCall}
\end{equation}

However, after the call we obtain the $\HEAP$-local reachability from the postcondition of the callee, while the $\FRAME$-local reachability is preserved. We cannot use $\GRAPH$-local reachability in the post-state of the call; the aim of our path-partitioning formulas is to \emph{deduce} information in this form, and these are only justified \emph{after} making the convexity check. Therefore, after the method call, we use the following alternative formulation:
\begin{equation}
\forall x,y \in \HEAP, u \in \FRAME \>\bullet\> \lnot \PATH{}{\HEAP}{x}{u} \lor \lnot \PATH{}{\FRAME}{u}{y}
\label{eq:checkConvexityAfterMethodCall}
\end{equation}

\begin{landscape}
\begin{figure}[H]
\vspace*{-0.6cm}
\centering%
\hspace*{-5.1cm}%
\noindent\centering%
\begin{subfigure}[t!]{.5\textwidth}
\begin{lstlisting}
method merge(l: Node, r: Node, $\defp{Args-begin}$
    $\GRAPH$:Graph, ldag:Graph, rdag:Graph) $\defp{Args-end}$
  returns link: Node                 $\defp{Hpos}$
  requires ...    $\defp{Contracts-begin}$
  ensures  ...    $\defp{Contracts-end}$
{
  if (l.right != $\NULLLIT$) {
    var nldag := $\coderef{sub}$($\GRAPH$, ldag, l.right) $\defp{SubFootprint}$
    link := $\coderef{merge}$(l.right, r, $\defp{Mcall-begin}$ 
      nldag$\DISJUNION$rdag, nldag, rdag) $\defp{Mcall-end}$
  } else {
    l.right := r $\defp{FiledUpdate}$
    link := l
  }
}
\end{lstlisting}
\end{subfigure}
\hspace{0.045\linewidth}
\centering%
\begin{subfigure}{.5\textwidth}
\begin{lstlisting}
$\defp{Vhpos}$
method merge(l: Ref, r: Ref, $\defp{Vargs-begin}$
    g: Set[Ref], ldag: Set[Ref], rdag: Set[Ref])$\defp{Vargs-end}$
  returns link: Ref 
  requires ...$\defp{Vcontracts-begin}$
  ensures  ...$\defp{Vcontracts-end}$
{
  if (l.right != null) {
    var nldag: Set[Ref] := $\coderef{sub}$(g, ldag, l.right)$\defp{VsubFootprint-begin}$
    var g1: Set[Ref] := nldag union rdag$\defp{Vcall-begin}$
    DeduceRelationshipBetweenSubHeaps(g1, g)$\defp{DeduceRelationshipBetweenSubHeaps}$
    var frame: Set[Ref] := g setminus g1
    EnableFocusOnConvexSubHeap(g, g1)$\defp{EnableFocusOnConvexSubHeapA}$
    EnableFocusOnFrame(g1, g, frame)$\defp{EnableFocusOnFrame}$
    label l1$\defp{preState}$
    link := $\coderef{merge}$(l.right, r, 
      nldag union rdag, nldag, rdag)
    label l2$\defp{postState}$
    EnableFocusOnConvexSubHeap(g, g1)$\defp{EnableFocusOnConvexSubHeapB}$
    EnableFocus(g, frame)$\defp{EnableFocus}$
    ApplyConvexTCFraming(l1, l2, g1, g, frame)$\defp{Vcall-end}$
  } else {
    if (r != l.right) {$\defp{VfiledUpdate-begin}$
      if (l.right != null) $\appcoderef{unlinkDAG}^{\{\texttt{left}, \texttt{right}\}}_{\texttt{right}}$(g, l)$\defp{UnlinkOper}$
      if (r != null) $\appcoderef{linkDAG}^{\{\texttt{left}, \texttt{right}\}}_{\texttt{right}}$(g, l, r) }$\defp{VfiledUpdate-end}$ 
    link := l }
}
\end{lstlisting}
\end{subfigure}

\begin{tikzpicture}[ remember picture, overlay ]
    \node (Args-center) at ($(Args-begin)!0.5!(Args-end)$) {};
    \node (Contracts-center) at ($(Contracts-begin)!0.5!(Contracts-end)$) {};
    \node (Mcall-center) at ($(Mcall-end)!0.5!(Mcall-begin)$) {}; 
    
    \node (Vargs-center) at ($(Vargs-begin)!0.5!(Vargs-end)$) {};
    \node (Vcontracts-center) at ($(Vcontracts-begin)!0.5!(Vcontracts-end)$) {};
    \node (Vcall-center) at ($(Vcall-begin)!0.5!(Vcall-end)$) {};
    \node (VfiledUpdate-center) at ($(VfiledUpdate-begin)!0.5!(VfiledUpdate-end)$) {}; 
    
    \def\translationArrowXlen{-8pt}
    \chkwToViper{Hpos |- Args-center}{ parameters }{Vhpos |- Vargs-center}{\translationArrowXlen}
    \chkwToViper{Hpos |- Contracts-center}{ contracts }{Vhpos |- Vcontracts-center}{\translationArrowXlen}
    \chkwToViper{Hpos |- Mcall-center}{ call }{Vhpos |- Vcall-center}{\translationArrowXlen}
    \chkwToViper{Hpos |- FiledUpdate}{ field update }{Vhpos |- VfiledUpdate-center}{\translationArrowXlen}

    
    \chkwCodeBlock{Hpos}{Args-begin}{Args-end}
    \chkwCodeBlock{Hpos}{Contracts-begin}{Contracts-end}
    \chkwCodeBlock{Hpos}{Mcall-begin}{Mcall-end}
    \chkwCodeBlock{Hpos}{FiledUpdate}{FiledUpdate}
    
    \def\codeBlockMargin{-10pt}
    \viperCodeBlock{Vhpos}{Vargs-begin}{Vargs-end}{\codeBlockMargin}
    \viperCodeBlock{Vhpos}{Vcontracts-begin}{Vcontracts-end}{\codeBlockMargin}
    \viperCodeBlock{Vhpos}{Vcall-begin}{Vcall-end}{\codeBlockMargin}
    \viperCodeBlock{Vhpos}{VfiledUpdate-begin}{VfiledUpdate-end}{\codeBlockMargin}

    \path let \p1=(DeduceRelationshipBetweenSubHeaps),
              \p2=(EnableFocusOnConvexSubHeapA),
              \p3=(EnableFocusOnFrame),
              \p4=(preState),
              \p5=(postState),
              \p6=(EnableFocusOnConvexSubHeapB),
              \p7=(EnableFocus), 
              \p8=(Vcall-end),
              \p9=(UnlinkOper),
              \p0=(VfiledUpdate-end) in coordinate (Q1) at (\x1, {max(\y1,\y2,\y3,\y4,\y5,\y6,\y7,\y8,\y9,\y0)});


    \def\minArrowLen{32pt}
    \macroDefinition{Q1}{DeduceRelationshipBetweenSubHeaps}{ Convert \eqref{RelativeClosedness}$\rightleftarrows$\eqref{closed} }{\minArrowLen}
    \macroDefinition{Q1}{EnableFocusOnConvexSubHeapA}{ \eqref{crossBorderPaths}, case i of~\eqref{pathPartitioning} }{\minArrowLen}
    \macroDefinition{Q1}{EnableFocusOnFrame}{ \eqref{crossBorderPaths}, \eqref{frameLocalization} }{\minArrowLen}
    \macroDefinition{Q1}{preState}{ Pre-state of the call }{\minArrowLen}
    \macroDefinition{Q1}{postState}{ Post-state of the call }{\minArrowLen}
    \macroDefinition{Q1}{EnableFocusOnConvexSubHeapB}{ \eqref{crossBorderPaths}, case i of~\eqref{pathPartitioning} }{\minArrowLen}
    \macroDefinition{Q1}{EnableFocus}{ \eqref{RelativeClosedness}, \eqref{closed} }{\minArrowLen}
    \macroDefinition{Q1}{Vcall-end}{ \eqref{edgePartitioning}, \eqref{preservedFrameReachability}, cases ii--iv of~\eqref{pathPartitioning} }{\minArrowLen}
    \macroDefinition{Q1}{UnlinkOper}{ \eqref{update} }{\minArrowLen}
    \macroDefinition{Q1}{VfiledUpdate-end}{ \secref{fieldupdates} }{\minArrowLen}
    
\end{tikzpicture}
\vspace*{-0.3cm}
\caption{Encoding \coderef{merge} in Viper. Types are translated directly. The specifications (\secref{footprints}) are omitted for brevity. The reference field update is translated via \code{unlinkDAG}, \code{linkDAG} (\secref{fieldupdates}). The method call is augmented with \emph{local assumptions} in the form of macros (lines starting with capital letters), constraining the states \code{l1}, \code{l2}. The macros with infix \code{Convex} also check relative convexity of corresponding footprints (\secref{convexframe}). The complete encoding is part of the publicly available artifact~\cite{ArtifactCitation}.}
\label{fig:viperencoding}
\end{figure}
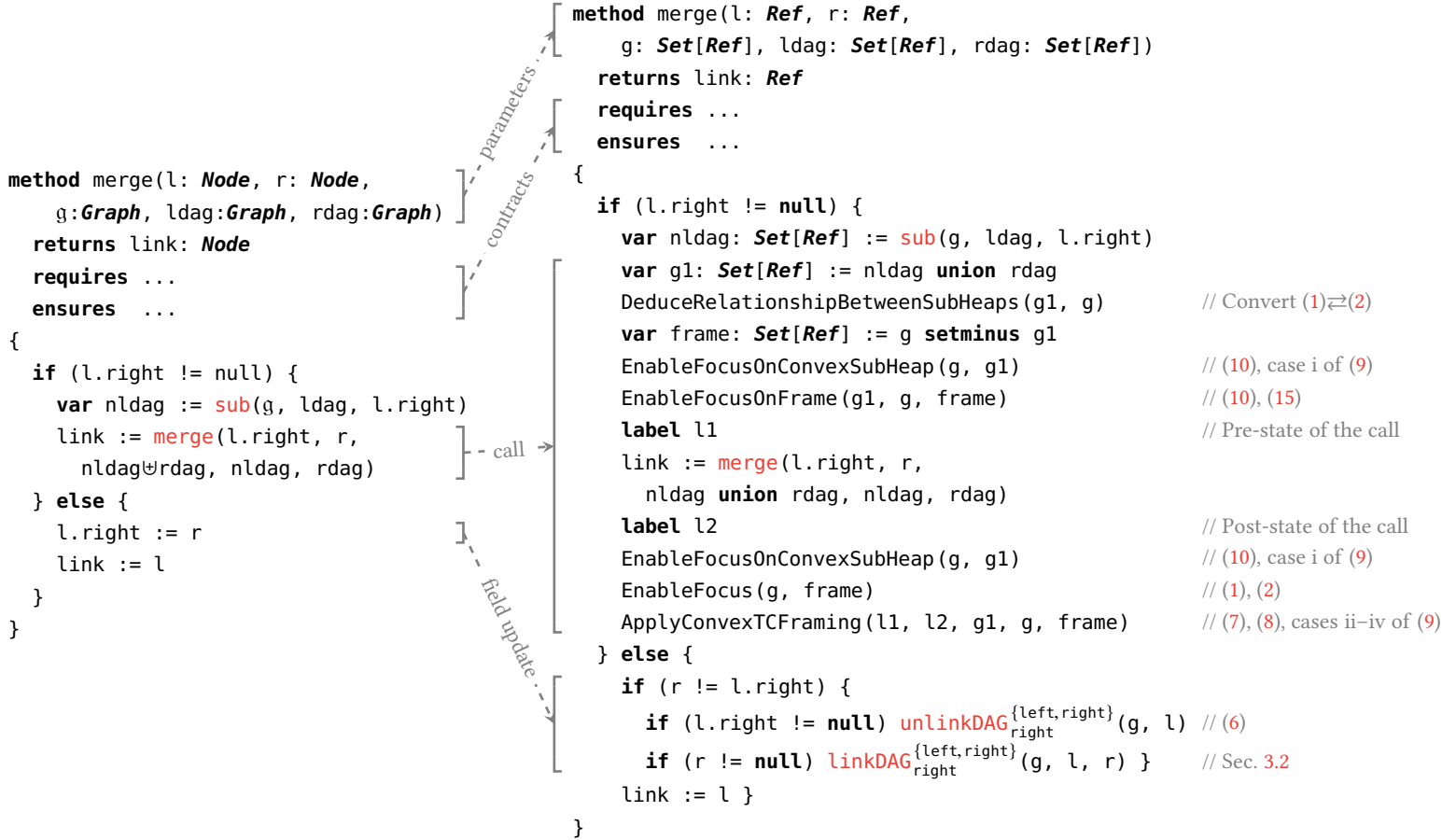
\end{landscape}  

\subsubsection*{Relating reachability information before and after a method call.}

Based solely on our assumption of relatively convex footprints, we now have a rich set of formulas available for precisely relating reachability information before and after a method call. To illustrate how our formulas can be used in practice, we consider
one of the verification conditions needed for verifying the postcondition in \figref{runningExample} after the recursive call to \coderef{merge}. Concretely, we consider the following Hoare triple: 
\HoareTriple{%
  \code{l.right} \neq \NULLLIT \land \PATH{0}{\GRAPH}{\code r}{n}
}{%
  \code{link := merge(l.right, r,$\>\>\HEAP$, ...)}
}{%
  \PATH{}{\GRAPH}{\code l}{n}
}
\noindent Here, $\GRAPH$ and $\HEAP$ are the footprints of the client and the callee, resp., \code{l} and \code{r} are the roots of the left and right DAGs, resp. (\figref{merge}), and $n$ is \emph{some} node reachable from \code{r} in the pre-state; we omit the rest of the arguments of \coderef{merge} for brevity. The condition $\code{l.right} \neq \NULLLIT$ comes from the \code{if} statement in \figref{runningExample} and holds before (and after) the recursive call; we enter this branch \emph{iff} we have not yet found the \code{link} node and must keep recursively traversing the current structure. We proceed with a proof sketch for the postcondition of this Hoare triple. Other checks needed to verify the Hoare triple include relative convexity checks~\eqref{checkConvexityBeforeMethodCall} and \eqref{checkConvexityAfterMethodCall}, and the precondition check before the recursive call; these require similar reasoning steps and are omitted for brevity.

The postcondition $\PATH{}{\GRAPH}{\code l}{n}$ expresses the existence of a path $\mo{l} \ldots n$. We justify this postcondition by showing the existence of a (single-edge) frame-local path $\code{l} \ldots \code{l.right}$ and an $\HEAP$-local path $\code{l.right} \ldots n$ (where $\HEAP$ is the footprint of the call). The former sub-path starts in $\code{l} \notin \HEAP$ and ends in $\code{l.right} \in \HEAP$ (since $\HEAP = \code{nldag} \DISJUNION \code{rdag}$, where $\code{nldag}$ was constructed via \coderef{sub}), and the latter sub-path starts in $\code{l.right} \in \HEAP$ and ends in $n \in \HEAP$ ($n \in \code{rdag}$ follows from the precondition $\PATH{0}{\GRAPH}{\code r}{n}$ of the Hoare triple and the last precondition of \coderef{merge}, while $\code{rdag} \subseteq \HEAP$ because $\HEAP = \code{nldag} \DISJUNION \code{rdag}$). The distribution of the starting and ending nodes of these sub-paths enables an instantiation of (ii)~from~\eqref{pathPartitioning} with $\code{l}$, $\code{l.right}$, $n$ for $x$, $a$, $y$, resp., reducing the overall proof goal to the two predicates $\PATH{}{\FRAME}{\code l}{\code{l.right}}$ and $\PATH{}{\HEAP}{\code{l.right}}{n}$. First, since $\code{l.right} \neq \NULLLIT$, the former predicate can be justified by \eqref{partialDefinitionOfPathPredicate} and the postcondition of \coderef{snapshot}. Second, we instantiate the last postcondition of \coderef{merge} with $\code{l.right}$, $n$ for $x$, $y$, resp. in order to reduce the latter predicate to $\PATH{0}{\HEAP}{\code{l.right}}{\code{link}}$ and $\PATH{0}{\HEAP}{\code r}{n}$. Note that, since the path $\code{l.right} \ldots \code{link}$ starts in the root of \code{nldag}, the former predicate is implied by the last postcondition of \coderef{merge}. We can justify the latter predicate, $\PATH{0}{\HEAP}{\code r}{n}$, with an instantiation of (i)~from~\eqref{pathPartitioning} with $\code{r}$, $n$ for $x$, $y$, resp., since (as we argued above) $\code{r},n\in\HEAP$.
\qed

\subsection{Frame-Localized Reachability}
\label{sec:nonconvexframe}

The ingredients presented thus far form the core of our solution for handling method calls, but are not yet sufficient to preserve reachability information in all cases, as we explain next. In some cases, we need to be able to localize reachability information in the \emph{frame} of the call (\cf{} the left branch in~\figref{infFlow}). But precise frame-local reachability information cannot be obtained the same way as footprint-local reachability because, unlike method footprints, our technique permits the frame to be \emph{non-convex} in the client's footprint. For example, consider a call to a method that operates on an acyclic list segment; the footprint of this call must be convex, while the frame would generally be non-convex in the entire list. Since the issue is subtle, we illustrate how information can be lost with a concrete example, and then show how to plug the gap.  

\subsubsection*{The problematic scenario.} The program in~\figref{nonConvexFrameCode} consists of two methods: the client, \coderef{joinAndModify}, and the callee, \coderef{disconnectAll}. This program concerns a particular shape of DAG structure, which we call a \emph{hammock} between two nodes. We say that a (closed) DAG $\HEAP$ is a hammock between two (distinct) nodes $s$ and $t$ \emph{iff} it consists of all nodes reachable from its node $s$ (called the \emph{source}) that reach its distinct node $t$ (called the \emph{sink}):
\begin{equation}
\begin{aligned}
\mo{HAMMOCK_{\GRAPH}(\HEAP, \mathit{s}, \mathit{t})} \>\> \longeq \>\> &s \in \HEAP \land t \in \HEAP \land \mo{CLOSED(\HEAP)} \land \mo{ACYCLIC_{\GRAPH}(\HEAP)} \land s \neq t \>\land \\
&\forall n \in \HEAP \>\bullet\> \PATH{}{\GRAPH}{s}{n} \land \PATH{}{\GRAPH}{n}{t} \label{eq:hammock}\\
\end{aligned}
\end{equation}
\begin{equation}
\begin{aligned}
\mo{ACYCLIC_{\GRAPH}(\HEAP)} \>\> \longeq \>\>& \HEAP \subseteq \GRAPH \land \forall x,y\in\HEAP \>\bullet\> \lnot\EDGE{}{\GRAPH}{x}{y} \lor \lnot\PATH{}{\GRAPH}{y}{x}
\label{eq:acyclic}
\end{aligned}
\end{equation}
\noindent The essential property of the example in~\figref{nonConvexFrameCode} is that the client method \emph{creates} heap edges inside the footprint of the callee (and not just in the frame of the call), whereas the callee \emph{destroys} some of the paths in its footprint.

We start reasoning about \coderef{joinAndModify} in state~0, with the footprint being comprised of two (disjoint) hammocks, \FRAME{} and \HEAP{}, where $s_1$ and $s_2$ are their sources and $t_1$ and $t_2$ are their sinks, resp. The first two operations are field updates, resulting in state~1. They join the two hammocks into one by creating exactly two edges: $(s_1, s_2)$ and $(t_2, t_1)$. Hence, there must exist at least two distinct paths from $s_1$ to $t_1$ in state~1: one path through the nested hammock, \HEAP, and one inside \FRAME. Note that this makes the subheap $\FRAME$ \emph{non-convex} in $\GRAPH$, even though $\HEAP$ is still relatively convex in $\GRAPH$. The last operation in \coderef{joinAndModify} is a method call with a (relatively convex) footprint, \HEAP, which results in state~2. The callee method, \coderef{disconnectAll}, destroys all heap paths \emph{inside} its footprint (first conjunct of the postcondition), while preserving all of its exit points (second conjunct of the postcondition). We omit the callee's implementation because the problem that we are about to explain occurs exclusively at the call site.

\begin{figure*}[t!]
\centering
\begin{subfigure}[t]{0.5\textwidth}
\centering
\phantomsection\label{lst:joinAndModify}
\begin{lstlisting}[escapeinside={(*@}{@*)}]
method joinAndModify($\GRAPH$: Graph,
      $\FRAME$: Graph, $s_1$: Node, $t_1$: Node,
      $\HEAP$: Graph, $s_2$: Node, $t_2$: Node)
  requires $\GRAPH = \FRAME\DISJUNION\HEAP\>\land$
    $\mo{HAMMOCK_{\GRAPH}(\FRAME, \mathit{s}_1, \mathit{t}_1)} \land \mo{HAMMOCK_{\GRAPH}(\HEAP, \mathit{s}_2, \mathit{t}_2)}$
    $\land \> s_1\mo{.left} = \NULLLIT \land t_2\mo{.right = \NULLLIT}$
  ensures $\PATH{}{\GRAPH}{s_1}{t_1}$
{
  /* state $0$ */ $s_1$.left := $s_2$; $t_2$.right := $t_1$
  /* state $1$ */ disconnectAll($\HEAP$)
  /* state $2$ */ }
\end{lstlisting}
\centering
\end{subfigure}%
\begin{subfigure}[t]{0.5\textwidth}
\centering
\phantomsection\label{lst:disconnectAll}
\begin{lstlisting}[escapeinside={(*@}{@*)}]
method disconnectAll($\GRAPH$: Graph)
  ensures $\left(\forall x,y\in\GRAPH \>\bullet\> \PATH{}{\GRAPH}{x}{y} \Leftrightarrow x = y\right) \land$
    $\forall x\in\GRAPH, y\notin\GRAPH \>\bullet\> \EDGE{}{\GRAPH}{x}{y} \Leftrightarrow \EDGE{0}{\GRAPH}{x}{y}$
\end{lstlisting}
\caption*{The method \coderef{disconnectAll} destroys all non-trivial paths inside $\HEAP$ (exemplifying a possible destructive update to the heap structure).}
\end{subfigure}
\caption{The method \coderef{joinAndModify} first attaches the hammock $\HEAP$ to the hammock $\FRAME$, creating a larger hammock, and then calls the method \coderef{disconnectAll}, creating a frame that is non-convex in $\GRAPH$. Verification of the postcondition is challenging, as it requires localizing reachability in the frame, $\FRAME$, of the call to \coderef{disconnectAll}. \figref{nonConvexFrameDiagram} illustrates a typical run of \coderef{joinAndModify}.}
\label{fig:nonConvexFrameCode}
\end{figure*}


The postcondition of \coderef{joinAndModify} says that there \emph{still} exists a \GRAPH-local path $s_1\ldots t_1$ in state~2. Intuitively, this claim should hold, as these two nodes \emph{were}, before the call to \coderef{disconnectAll}, reachable via at least one $\FRAME$-local path that could not have been destroyed as a result of the method call (because $\FRAME$ is the frame of that call). However, our path-partitioning formulas~\eqref{pathPartitioning} do not capture that such a frame-local path definitely existed; we learn from cases~(iv)--(v) of \eqref{pathPartitioning} only the \emph{disjunction} describing that at least one of the two paths from $s_1$ to $t_1$, labeled ``iv'' and ``v'' in~\figref{pathDistribution}, must have existed before the call, but we do not know which. Since the call is known to destroy the paths corresponding to one disjunct, we cannot deduce $\PATH{2}{\GRAPH}{s_1}{t_1}$ after the call unless we can precisely derive frame-local reachability. 

\subsubsection*{Localizing reachability in the frame of a relatively convex footprint.} \figref{sigmaTauSplit} demonstrates the general problem of localizing reachability information in the frame of a method call. Consider a method call with a relatively convex footprint $\HEAP$ and the frame $\FRAME$; the client's footprint $\GRAPH$ is their disjoint union $\GRAPH = \FRAME \DISJUNION \HEAP$. Our path partitioning formulas~\eqref{pathPartitioning} allow us to precisely define \emph{$\HEAP$-local} reachability based solely on $\GRAPH$-local reachability. As demonstrated by our~\coderef{joinAndModify} example of~\figref{nonConvexFrameCode}, we additionally need a complementary formula that would precisely define \emph{$\FRAME$-local} reachability (again, based solely on $\GRAPH$-local reachability). In other words, we are looking for a first-order formula over the relation \PATHSYMB{} (with the first parameter fixed to $\GRAPH$) that, for a given pair of nodes $x,y\in\FRAME$, precisely defines the existence of an $\FRAME$-local path $x \ldots y$. Fortunately, such an \emph{in-frame reachability localization formula} exists if $\FRAME$ is a \emph{frame of a relatively convex footprint} $\HEAP$ (even if $\FRAME$ itself is non-convex in $\GRAPH$):

\begin{equation}
\begin{alignedat}{5}
  \forall x,y\in\FRAME \>\> \bullet& \>\>       &\big(\forall z\in\HEAP \>\bullet\> \lnot \PATH{}{\GRAPH}{x}{z}& \lor \lnot \PATH{}{\GRAPH}{z}{y}\big) \implies \big(\PATH{}{\FRAME}{x}{y} \Leftrightarrow \PATH{}{\GRAPH}{x}{y}\big) \\
  \forall x,y\in\FRAME \>\> \bullet& \>\> &\big(\exists z\in\HEAP \>\bullet\> \PATH{}{\GRAPH}{x}{z}& \land  \PATH{}{\GRAPH}{z}{y}\big) \implies \Big(\PATH{}{\FRAME}{x}{y} \Leftrightarrow \exists \sigma,\tau\in\FRAME \>\bullet\> \\
              &&&\>\>\>\>\>\>\>\>\>\>\>\>\>\>\>\>\>\>\>\>\PATH{}{\GRAPH}{x}{\sigma} \land \EDGE{}{\GRAPH}{\sigma}{\tau} \land \PATH{}{\GRAPH}{\tau}{y} \land \\
              &&&\big(\exists z_1 \in \HEAP.\> \PATH{}{\GRAPH}{\sigma}{z_1}\big) \land \lnot \big(\exists z_2 \in \HEAP.\> \PATH{}{\GRAPH}{\tau}{z_2}\big)\Big)
\end{alignedat}
\label{eq:frameLocalization}
\end{equation}

We explain and justify~\eqref{frameLocalization} using the diagram of~\figref{sigmaTauSplit}. Generally, since $\GRAPH = \FRAME \DISJUNION \HEAP$, we can case split on whether there exists a path $x \ldots y$ that goes through $\HEAP$, allowing us to obtain the required $\FRAME$-local reachability formula. The first formula above covers the case in which such a path \emph{does not} exist; thus, the following must hold: $\forall z\in\HEAP.~\lnot\PATH{}{\GRAPH}{x}{z} \lor \lnot\PATH{}{\GRAPH}{z}{y}$ (which trivially holds for all $x,y \in \FRAME$ in the special case when $\FRAME$ \emph{is} convex in $\GRAPH$). The second formula above says that, if there exists a path through $\HEAP$ (the upper kind of path in~\figref{sigmaTauSplit}), it must pass through some node $z \in \HEAP$; hence, the following condition must hold: $\exists z\in\HEAP.~\PATH{}{\GRAPH}{x}{z} \land  \PATH{}{\GRAPH}{z}{y}$. Under this condition, we must define the existence of an $\FRAME$-local path that \emph{also} connects $x \ldots y$.
%
\begin{figure}[t]
\begin{center}
	\begin{tikzpicture}[>=latex,xscale=0.5,yscale=0.5]
		
		\def\xo{ 0.0}
		\def\xa{ 2.0}
		\def\xb{ 4.0}
		\def\xc{ 6.0}
		\def\xd{ 8.0}
		\def\xe{10.0}
		\def\xf{12.0}
		\def\xg{14.0}
		\def\xh{16.0}

		\def\deltax{ 0.75}
		
		\def\yo{ 5.5}
		\def\ya{ 4.5}
		\def\yb{ 3.5}
		\def\yc{ 2.5}
		\def\yd{ 1.5}
		\def\ye{ 0.5}

		\def\deltay{ 0.5}
		
		\def\p{1.2}              
		\def\d{0.15}             
		\def\l{{(\xa-\p)}}       
		\def\r{{(\xh+\p)}}       
		\def\t{{(\ya+\p)}}       
		\def\b{{(\yd-\p)}}       
		\def\ml{(0.5*(\xb+\xc))} 
		\def\mr{(0.5*(\xf+\xg))} 
		\def\mv{(0.5*(\yc+\yd))} 
		
		\draw [frame] (\xo,\ya+\deltay) -- (\xb-\deltax*0.5,\ya+\deltay) -- (\xb-\deltax*0.5,\yc-\deltay) -- (\xf+\deltax*0.5,\yc-\deltay) -- (\xf+\deltax*0.5,\ya+\deltay)
		-- (\xh,\ya+\deltay) -- (\xh,\ye-\deltay*0.5) -- (\xo,\ye-\deltay*0.5) -- cycle;

		\draw [footprint] (\xb,\yc) rectangle (\xf,\ya+\deltay);
		
		
		\node [fvertex,lbll=$x$] (x) at (\xa,\yb) {};
		\node [fvertex,lblr=$y$] (y) at (\xg,\yb) {};
		\node [vertex,lbl=$z$] (z) at (\xc,\yb+\deltay) {};

		\node [fvertex,lbld=$\sigma$] (sigma) at ({\xd-\deltax},{\yd-\deltay}) {};
		\node [vertex,lbll=$z_1$] (z1) at ({\xd-\deltax},{\yb-\deltay}) {};

		\node [fvertex,lbld=$\tau$] (tau) at ({\xd+\deltax},{\yd-\deltay}) {};
		\node [vertex, draw=none] (z2) at ({\xd+\deltax},{\yb-\deltay}) {};

		\node [vertex, draw=none] (ttt) at ({\xb+\deltax},{\yb-\deltay}) {};

		\draw [path] (x) -- (z);

		\draw [path] (z) -- (y);

		\draw [fpath] (x) to ({\xa+\deltax},{\yc}) to ({\xb},{\yd}) -- (sigma);
		\draw [fedge] (sigma) -- (tau); 
		\draw [fpath] (tau) to ({\xf},{\yd}) to ({\xg-\deltax},{\yc}) -- (y);

		\draw [path] (sigma) -- (z1);
		\draw [path] (tau) --  node[draw, midway, -, black, sloped, cross out, line width=.2ex, minimum width=1.5ex, minimum height=1ex, anchor=center]{} (z2);
		
		\draw [path] ({\xb},{\yd}) -- (ttt);

		\node [framelbl,anchor=south] at ({\xg+\deltax-0.5},{\yb+\deltay}) {frame (\FRAME)};
		\node [footprintlbl,anchor=south] at ({\xe+0.5*\deltax-0.5},{\ya-\deltay}) {footprint (\HEAP)};

	\end{tikzpicture}
\end{center}
\caption{Paths starting and ending in the frame of a relatively convex footprint may either be local to the frame or go through the footprint. Relative convexity of the footprint guarantees that there must be at most one entry and one exit point \emph{per path}. Note that $x$ and $\sigma$, as well as $\tau$ and $y$ may possibly alias each other, but $x \neq y$ and $\sigma \neq \tau$ are guaranteed.}
\label{fig:sigmaTauSplit}
\end{figure}
%
The key idea that we exploit to justify the second formula in~\eqref{frameLocalization} is to use our relative convexity assumption to justify a \emph{three-way} split of the (hypothetical) $\FRAME$-local path $x \ldots y$ into three segments: a path $\PATH{}{\FRAME}{x}{\sigma}$, an edge $\EDGE{}{\FRAME}{\sigma}{\tau}$, and a path $\PATH{}{\FRAME}{\tau}{y}$ (the lower kind of path in~\figref{sigmaTauSplit}). Furthermore, we choose $(\sigma,\tau)$ such that 
$\sigma$ is \emph{the last node that reaches} $\HEAP$ ($\exists z_1 \in \HEAP.\> \PATH{}{\GRAPH}{\sigma}{z_1}$) and $\tau$ is \emph{the first node that does not reach} $\HEAP$ ($\lnot \exists z_2 \in \HEAP.\> \PATH{}{\GRAPH}{\tau}{z_2}$). Under our assumptions about $x$ and $y$, this requirement can always be satisfied because the footprint of the callee, $\HEAP$, is reachable from (at least) the node $x$ and is unreachable from (at least) the node $y$.

We summarize the conditions under which the predicates defining the three-way split of our hypothetical path $x \ldots (\sigma, \tau) \ldots y$ can be rewritten with $\GRAPH$ instead of $\FRAME$, without losing precision: 
\begin{itemize}
  \item The footprint of the call, $\HEAP$, is convex in the client's footprint, $\GRAPH$. 
  \item Both nodes $x$ and $y$ are in the frame of the call, $\FRAME$. 
  \item We picked $\sigma$ that reaches $\HEAP$ and $\tau$ that does not reach $\HEAP$ s.t. $(\sigma,\tau)$ is on the path $x \ldots y$. 
\end{itemize}
For the first predicate, we need to argue by contradiction: suppose that $\PATH{}{\GRAPH}{x}{\sigma}$ were the case, but $\PATH{}{\FRAME}{x}{\sigma}$ were not (the opposite implication is direct, since $\FRAME \subseteq \GRAPH$). Then, the path from $x$ to $\sigma$ must visit the callee's footprint, $\HEAP$. However, by construction, $\sigma$ is known to have a path to some node in the callee's footprint, and this violates the assumption that this footprint is relatively convex. Hence, we get $\PATH{}{\FRAME}{x}{\sigma} = \PATH{}{\GRAPH}{x}{\sigma}$. The second predicate is easiest: a single edge between two nodes in the frame ($\FRAME$) can only depend on the frame itself; therefore, we get $\EDGE{}{\FRAME}{\sigma}{\tau} = \EDGE{}{\GRAPH}{\sigma}{\tau}$. The third predicate expresses the existence of a path $\tau \ldots y$; since we have picked $\tau$ s.t. it does not reach the footprint, such a path exists in this case exactly when it exists in the frame, giving $\PATH{}{\FRAME}{\tau}{y} = \PATH{}{\GRAPH}{\tau}{y}$. Thus, our construction of $\sigma$ and $\tau$, along with our relative convexity property for method footprints, allows us to justify the formulation in \eqref{frameLocalization}. These formulas now provide the missing ingredient for our technique that complements our path-partitioning formulas of~\eqref{pathPartitioning}.
\qed
%
\begin{figure}[t]
\begin{center}
	\begin{tikzpicture}[>=latex,xscale=0.5,yscale=0.5]

	\def\delta{0.2}
	
	\node [hvertex,lbll=$s_2$] (s2) at (2,5.5) {}; 
	\node [hvertex] (a) at (3.5,6.5-\delta) {}; 
	\node [hvertex] (b) at (3.5,4.5+\delta) {}; 
	\node [hvertex] (c) at (5,5.5) {}; 
	\node [hvertex] (d) at (6.5,6.5-\delta) {}; 
	\node [hvertex] (e) at (6.5,4.5+\delta) {}; 
	\node [hvertex,lblr=$t_2$] (t2) at (8,5.5) {}; 

	\node [fvertex,lbll=$s_1$] (s1) at (1,2) {}; 
	\node [fvertex,lbld=$s_1$\mo{.right}~~~~] (f) at (3,2) {}; 
	\node [fvertex] (h) at (5,3) {}; 
	\node [fvertex] (i) at (5,1) {}; 
	\node [fvertex] (k) at (7,3) {}; 
	\node [fvertex] (l) at (7,1) {}; 
	\node [fvertex,lblr=$t_1$] (t1) at (9,2) {}; 

	\draw [hedge] (s2) -- (a);
	\draw [hedge] (a) -- (c);
	\draw [hedge] (s2) -- (b);
	\draw [hedge] (b) -- (c);
	\draw [hedge] (c) -- (d);
	\draw [hedge] (c) -- (e);
	\draw [hedge] (d) -- (t2);
	\draw [hedge] (e) -- (t2);
	\draw [fedge,densely dashed] (s1) -- (s2);
	\draw [hedge,densely dashed] (t2) -- (t1);
	\draw [fedge] (s1) -- (f);
	\draw [fedge] (h) -- (k);
	\draw [fedge] (k) -- (t1);

	\draw [fedge] (l) -- (t1);
	\draw [fedge] (f) -- (i);
	\draw [fedge] (f) -- (h);
	\draw [fedge] (i) -- (l);
	\draw [fedge] (i) -- (k);
	\draw [fedge] (h) -- (l);

	\draw [frame] (-0.5,0+0.5) rectangle (11.5,3.5);
	\draw [footprint] (-0.5,4+0.25) rectangle (11.5,6.75);

	\node [framelbl,anchor=south] at (10,0.5) {frame (\FRAME)};
	\node [footprintlbl,anchor=south] at (9.5,5.75) {footprint (\HEAP)};

	\end{tikzpicture}
\end{center}
\caption{An example scenario of running the method \coderef{joinAndModify}. In state~0, only the solid edges exist. In state~1, the client has created two new edges: $(s_1, s_2)$ and $(t_2, t_1)$. In state~2, the callee has destroyed all solid-red edges, but there still exists a path $s_1 \ldots t_1$ via solid-blue edges. However, we cannot deduce the existence of this path using just the path partitioning formulas alone due to the disjunction in case~(iv)--(v) of \eqref{pathPartitioning} that does not allow us to distinguish whether \emph{all} paths between $s_1$ and $t_1$ were passing through the footprint in state~1. Therefore, recovering this bit after the call to \coderef{disconnectAll} requires precise localization of reachability information in the frame~\eqref{frameLocalization}.}
\label{fig:nonConvexFrameDiagram}
\end{figure}
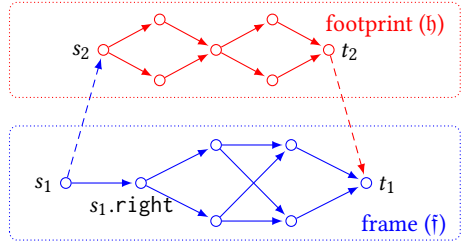
%
\subsubsection*{Revisiting the problematic scenario.}
We return to our \coderef{joinAndModify} method, and show that we can now verify the last conjunct of its postcondition. Previously, we were unable to verify $\PATH{2}{\GRAPH}{s_1}{t_1}$ after the call to \coderef{disconnectAll}, while intuitively, a $\GRAPH$-local path $s_1 \ldots t_1$ exists in state~2, because the method call could not have destroyed the existing \emph{frame}-local path $s_1 \ldots t_1$ that existed in state~1 (\figref{nonConvexFrameDiagram}). Thus, if we could deduce $\PATH{1}{\FRAME}{s_1}{t_1}$ (before the call to \coderef{disconnectAll}), we would obtain our proof goal. This is now possible using the second equation from~\eqref{frameLocalization}: instantiating $s_1$ for $x$ and $t_1$ for $y$, we deduce the hypothesis of the implication, since before the call to \coderef{disconnectAll} we can deduce that paths from $s_1$ to $t_1$ exist passing through the footprint. To deduce $\PATH{1}{\FRAME}{s_1}{t_1}$ from our formula, we need to obtain the following property (recall that $\GRAPH = \FRAME \DISJUNION \HEAP$): 
\[
\exists \sigma,\tau\in\FRAME.~\PATH{}{\GRAPH}{s_1}{\sigma} \land \EDGE{}{\GRAPH}{\sigma}{\tau} \land \PATH{}{\GRAPH}{\tau}{t_1} \land \big(\exists z_1 \in \HEAP.\> \PATH{}{\GRAPH}{\sigma}{z_1}\big) \land \lnot \big(\exists z_2 \in \HEAP.\> \PATH{}{\GRAPH}{\tau}{z_2}\big)
\]
The existentially-quantified pair $(\sigma, \tau)$ can be witnessed by $(s_1, s_1\mo{.right})$. From this, and our hammock properties~\eqref{hammock}, all conditions above follow directly, allowing us to deduce our intermediate proof goals, $\PATH{1,2}{\FRAME}{s_1}{t_1}$, and use the last case of \eqref{pathPartitioning} to deduce the ultimate proof goal, $\PATH{2}{\GRAPH}{s_1}{t_1}$.
\qed

Together with~\eqref{frameLocalization}, the ingredients of our technique presented in~\figref{infFlow} empower \emph{completely general, precise} reasoning about reachability in the presence of method calls with relatively convex footprints. Note that, in examples where stronger properties are known about a method's footprint, our formulas from \eqref{frameLocalization} reduce to much simpler criteria. In particular, if a method operates on a \emph{closed} data structure (no paths leave the footprint), we can always apply the first of our formulas; the full expressiveness of our conditions is required only in the presence of potential paths crossing the callee's footprint (\eg{}~\figref{nonConvexFrameDiagram}). Our technique is complete, provided that callee postconditions specify sufficient information about reachability within their footprint. However, even in examples where this information is incomplete, our technique is applicable and provides useful information at the call site, for instance, by deducing  which frame-local paths will be preserved across a method call. It is the restriction to method calls with relatively convex footprints which enables us to express appropriate formulas to preserve this information; without this restriction, we would not be able to precisely define the existence of frame-local paths exclusively via coarser reachability information. Finally, we note on the efficiency of formulas~\eqref{frameLocalization}: in our encoding (demonstrated in~\figref{viperencoding} and validated in~\secref{evaluation}), we supply appropriate triggers for the universal quantifiers and Skolemize the existential quantifiers.

This completes our treatment of acyclic graphs and method calls; the latter is the most complex part of our technique, and applies equally to the cyclic case, which we tackle in the next section. 

\section{Reachability in Cyclic Structures}
\label{sec:ZOPs}
In the previous section, we presented our technique for enabling modular reasoning about heap reachability in combination with first-order separation logic. The presented technique operates under two key restrictions: (1) that method footprints are always relatively convex in their client's footprint and (2) that all footprints used contain acyclic graphs. These are two \emph{independent} criteria, which our technique checks where necessary. Restriction (1) alone enables our handling of method calls. In this section, we show that we can adapt our technique to a particular setting in which restriction (2) is dropped: that of general \emph{0--1-path graphs}. A graph is called a 0--1-path graph (hereafter, ZOPG) if there exists \emph{at most one} (non-trivial) path (modulo cycles) between all pairs of nodes in the graph; for instance, $\{(a,b),(b,c),(c,a),(c,d)\}$ is a ZOPG, but $\{(a,b),(b,a),(a,c),(c,a)\}$ is not since there are \emph{two} distinct (non-trivial) paths from $a$ to itself~\cite{Dong1995IncrementalAD}. Although this notion does not permit \emph{arbitrary} cyclic graphs, the technique presented in this section allows us to adapt our work to reason about reachability in the presence of potentially-cyclic lists in the heap or more-complex data structures consisting of these, including, for example: trees where the children of each node are stored in a cyclic list (\eg{}~using Java LinkedList), generalized tree-like structures in which some nodes consist of rings, and the ring representation of heap-ordered trees~\cite{Fredman1986ThePH}. Therefore, the ZOPG class is an important generalization of (potentially-cyclic) singly-linked lists, which is the class handled in the closest prior work~\cite{ItzhakyPOPL14}. 


Extending our technique to ZOPGs requires a new way of handling direct field updates (\secref{zopgFieldUpdates}); our handling in \secref{DAGs} depended on acyclicity, and a way to retain that certain graphs in the program \emph{are} ZOPGs; modifying a ZOPG by adding an edge could violate the ZOPG invariant (\secref{qualifiers}). Note that our requirement of relatively convex footprints (\defref{convexity}) is again crucial, enabling an efficient solution of the latter problem. 

\subsection{Field Updates in ZOPGs}
\label{sec:zopgFieldUpdates}

To support direct field updates, we adapt prior work~\cite{Dong1995IncrementalAD} that shows how to precisely update a more-refined reachability relation called \DEPSYMB{} for ZOPGs. There are few changes in our adaptation: our \DEPSYMB{} relation is compatible with the \emph{reflexive} reachability relation (\PATHSYMB{}), whereas Dong~and~Su work with irreflexive reachability, and we parameterize our \DEPSYMB{} relation with two extra parameters, $F$ and $\GRAPH$, supporting separation-logic reasoning, as we did for \PATHSYMB{} in~\secref{localReachability}. The predicate $\DEPF{}{\GRAPH}{x}{y}{u}{v}{F}$ expresses the existence of a (non-trivial) path of field references from $x$ to $y$ such that all objects on the path (except possibly $y$) are in $\GRAPH$, all fields are in $F$, and the path \emph{depends on} the edge $(u,v)$. Intuitively, this means that removing $(u,v)$ from the graph will destroy the path $x\ldots y$ (which is the unique path from $x$ to $y$ in a ZOPG). We will omit the parameter $F$ when it is clear from the context. Note that $\DEP{}{\GRAPH}{x}{y}{u}{v} \Rightarrow u \neq v$ since an edge $(v,v)$ cannot be a dependency of any path: deleting such an edge would not affect reachability. Note also that $\DEP{}{\GRAPH}{x}{y}{u}{v} \Rightarrow x \neq y$ since a trivial path $x \ldots x$ does not depend on any edges. 

Although precisely updating the classical reachability relation \PATHSYMB{} in potentially-cyclic graphs after destructive heap operations is beyond first-order logic (and cannot be efficiently automated), the information about the \DEPSYMB{} relation \emph{can} be updated precisely and efficiently after such destructive operations~\cite{Dong1995IncrementalAD}. For example, if the edge $(\texttt{s},\texttt{t})$ is deleted by executing the statement \lstinline{s.adj := null} in a method with footprint $\GRAPH$, then the new relation, \DEPSYMB{}, can be simply expressed via the old relation \DEPSYMB{0} as follows: 
\begin{equation}
\forall x,y,u,v \>\bullet\> \DEPF{}{\GRAPH}{x}{y}{u}{v}{F} \iff \DEPF{0}{\GRAPH}{x}{y}{u}{v}{F} \land \lnot \DEPF{0}{\GRAPH}{x}{y}{\mo s}{\mo t}{F}
\label{eq:updateDep}
\end{equation}
For fixed $F$ and $\GRAPH$, the intuition for \eqref{updateDep} is this: $(x, y, u, v)$ is in the new relation \emph{iff} it was in the old relation and the deleted edge $(\mo{s}, \mo{t})$ was not a dependency of the path $x \ldots y$ before the update. 

Precisely updating \DEPSYMB{} after an operation that only creates an edge (\eg{}~by executing the statement \code{s.adj := t}) is also possible, provided one additionally checks that the newly-created edge does not violate the ZOPG invariant; we describe how this check is enforced in~\secref{qualifiers}. As before, a general field update entails removing and then adding an edge (see~\appref{zopgDepUpdates}). Our treatment of the \DEPSYMB{} relation is similar to the treatment of the \PATHSYMB{} relation described in~\secref{fieldupdates}: since the mathematical definitions of these relations are beyond first-order logic, we provide the verifier with a \emph{partial} axiomatization (see~\appref{ZOP-rules}). We rewrite each field update with a call to an internal \code{updateZOPG} method with the same footprint $\GRAPH$ as for the current method; the postconditions of \code{updateZOPG} make the \DEPSYMB{} update formulas (\eg{}~\eqref{updateDep}) available to the SMT solver. 

A technical difference between our reachability relation $\PATHSYMB{}$ and the $\DEPSYMB{}$ relation is that the latter carries richer information (in particular, knowledge of every edge on which each path depends). Conversely, it seems unlikely that having to enumerate all edge facts in a graph would be suitable for a method specification; the abstraction provided by $\PATHSYMB{}$ is typically desirable. Thus, we do not provide $\DEPSYMB{}$ as a primitive in our specifications, and instead provide a means of converting between information in one relation and the other, while losing as little information as possible. Our conversion rules are based on the following main axiom: 
\begin{equation}\label{eq:TC_and_DEP}
\forall \HEAP,x,y \>\bullet\> \PATHF{}{\HEAP}{x}{y}{F} \land x \neq y \iff \exists u,v \> \bullet \> \DEPF{}{\HEAP}{x}{y}{u}{v}{F}
\end{equation}
Unlike the update formulas that are emitted for concrete method footprints, our conversion axioms (\eg{}~\eqref{TC_and_DEP}) \emph{quantify} over the footprint ($\HEAP$); as before, we carefully select the triggers for these axioms to guide the SMT solver's quantifier instantiation procedure. 

In general, formula~\eqref{TC_and_DEP} does not capture full information in principle expressible with the $\DEPSYMB{}$ relation; intuitively, this is because a single path $x \ldots y$ (described by the LHS) may depend on multiple edges, all of which match the RHS existential quantifier. To partially mitigate this fact, we augment our axiomatization with a number of additional properties. For instance, one can easily prove the following formula (an axiom in our technique) about ZOPGs, providing (for fixed $F$ and $\GRAPH$) some quadruples which \emph{do not} belong to the \DEPSYMB{} relation: 
\begin{equation}
\forall \HEAP,u,v,w \>\bullet\> \lnot \DEPF{}{\HEAP}{v}{w}{u}{v}{F}
\label{eq:axDepHead}
\end{equation}
Note that if $v = w$, $\lnot\DEP{}{\HEAP}{v}{v}{u}{v}$ holds because a trivial path $v\ldots v$ does not depend on any edges. Assume $v \neq w$. There can be at most one (cycle-free) path from $v$ to $w$ in a ZOPG. If there are no paths from $v$ to $w$, then we get $\lnot\DEP{}{\HEAP}{v}{w}{u}{v}$ from~\eqref{TC_and_DEP}. Otherwise, the edge $(u,v)$ is not part of the (cycle-free) path $v\ldots w$ and cannot be one of its dependencies. 
\qed

Our ZOPG axiomatization is based on a set of formulas like~\eqref{axDepHead} that, together with~\eqref{TC_and_DEP}, help reasoning about the \DEPSYMB{} relation (we provide the full axiomatization in~\appref{ZOP-rules}). Equipped with this conversion between relations $\PATHSYMB{}$ and $ \DEPSYMB{}$, precise reachability information is preserved in all cases that we have observed. This is interesting because the \DEPSYMB{} relation carries more information than the transitive relation \PATHSYMB{}, so (for fixed $F$ and $\GRAPH$) not all quadruples $(x,y,u,v)$ in \DEPSYMB{} can be extracted precisely from \PATHSYMB{}, even if all pairs $(x, y)$ in \PATHSYMB{} are known. Intuitively, these missing quadruples appear not to be needed in practice because the overall proof goals are phrased in terms of just \PATHSYMB{} (and not \DEPSYMB{}). 

We illustrate how reachability information is preserved throughout the transformations between $\PATHSYMB{}$ and $ \DEPSYMB{}$ with a concrete example. Consider the following Hoare triple that describes a heap update in a ZOPG with footprint $\GRAPH$ and a single reference field \code{next}: %
\HoareTriple{%
    {\mo x}, {\mo y} \in \GRAPH \land \mo{x.next} = \mo{y} \>\land \\
    \forall n,m\in\GRAPH \>\bullet\> \PATH{0}{\GRAPH}{n}{m}
}{%
    \code{x.next := null}
}{%
    \forall m\in\GRAPH \>\bullet\> \PATH{}{\GRAPH}{\mo{y}}{m}
}
We can justify the postcondition assertion as follows. Consider an arbitrary node $m \in \GRAPH$. If $m = \mo y$, then we trivially get $\PATH{}{\GRAPH}{\mo y}{\mo y}$. Otherwise, we assume $m \neq \mo{y}$, and the remaining proof obligation is $\PATH{}{\GRAPH}{\mo y}{m}$; to justify this, we need to exploit information from the pre-state. Since by \eqref{TC_and_DEP}, we can reduce the current proof obligation to $\DEP{}{\GRAPH}{\mo y}{m}{\mo y}{\mo{y.next}}$, we can instantiate the \DEPSYMB{} update formula \eqref{updateDep}, obtaining two pre-state conditions: $\DEP{0}{\GRAPH}{\mo y}{m}{\mo y}{\mo{y.next}}$ and $\lnot\DEP{0}{\GRAPH}{\mo y}{m}{\mo x}{\mo y}$. The former is justified by the precondition quantifier (providing $\PATH{0}{\GRAPH}{\mo y}{m}$) and the main conversion axiom \eqref{TC_and_DEP}, whereas the latter can be obtained directly from the additional conversion axiom \eqref{axDepHead}. 

This example, as well as our evaluation (\secref{evaluation}), show that necessary reachability information can be fully recovered after the following steps: first, conversion from \PATHSYMB{} to \DEPSYMB{}, second, application of update formulas for the \DEPSYMB{} relation, and third, conversion from \DEPSYMB{} to \PATHSYMB{}. We plan to investigate as future work the extent to which this approach is always precise for preserving reachability information of this kind.

\subsection{Preservation of the ZOPG Invariant}
\label{sec:qualifiers}

To justify the handling of field updates from the previous subsection, we require knowledge that the graph being updated is a ZOPG\@. Since this fact can be violated by changes to the heap, an important question is how we can know if the ZOPG invariant holds. It can be expressed in first-order logic with the combination of edge and path predicates as follows:
\begin{equation}
\begin{aligned}
\mo{ZOPG(\HEAP)} \;\longeq\; &\big(\forall x_1, x_2, a, b \in \HEAP, y \>\>\bullet\>\> (x_1 \neq x_2 \lor a \neq b) \land \PATH{}{\HEAP}{x_1}{x_2} \land \PATH{}{\HEAP}{x_2}{x_1} \>\land \\
&\qquad\EDGE{}{\HEAP}{x_1}{a} \land \lnot\PATH{}{\HEAP}{a}{x_1} \>\land \\
&\qquad\EDGE{}{\HEAP}{x_2}{b} \land \lnot\PATH{}{\HEAP}{b}{x_2} \implies \lnot \PATH{}{\HEAP}{a}{y} \lor \lnot \PATH{}{\HEAP}{b}{y}\big) \>\>\land \\
&\forall x, a, b \in \HEAP \>\>\bullet\>\>
a \neq x \land b \neq x \land
\EDGE{}{\HEAP}{x}{a} \land \PATH{}{\HEAP}{a}{x} \> \land \\ 
&\>\>\>\>\>\>\>\>\>\>\>\>\>\>\>\>\>\>\>\>\>\>\>\>\>\>\>\>\>\>\>\>\>\>\>\>\>\>\>\>\>\>\>\>\>\>\>\>\>\>\>\>\>\>\>\>\>\>\EDGE{}{\HEAP}{x}{b} \land \PATH{}{\HEAP}{b}{x} \implies a = b
\end{aligned}
\label{eq:ZOPG}
\end{equation}

The first conjunct of the formula expresses a situation in which two (potentially aliasing) nodes $x_1$ and $x_2$ are on the same strongly-connected component (SCC), and two edges (starting in $x_1$ and $x_2$) that are \emph{different}---at least by source or target---end in nodes $a$ and $b$, resp., outside of the SCC ($a$ and $b$ may alias unless $x_1=x_2$). In such a case, it is forbidden that any node $y$ is reachable from both $a$ and $b$ (this would form two different paths from the SCC to $y$). The second conjunct restricts the structure of SCCs themselves: no two \emph{different} edges may leave the node $x$ and stay within the same SCC.

Intuitively, formula \eqref{ZOPG} is hard to automate because it uses a non-trivial combination of edge and reachability predicates. Establishing \mo{ZOPG(\HEAP)} would require, for example, the information about all path splits in $\HEAP$, \ie{}~all nodes $x\in\HEAP$ s.t. $\exists a,b\in\HEAP \>\bullet\> a \neq b \land \EDGE{}{\HEAP}{x}{a} \land \EDGE{}{\HEAP}{x}{b}$. Such details ultimately require specifications to enumerate edges in the graph, which is impractical, and breaks the abstraction that reachability specifications grant. Even if the full information about the edge relation were present, establishing \mo{ZOPG(\HEAP)} would require an induction proof that is beyond the power of modern SMT solvers. Instead of checking this invariant from scratch, we design a mechanism for checking that the ZOPG invariant is \emph{preserved} across changes to the heap.

\begin{figure}[t]
\begin{lstlisting}
$\phantomsection\label{lst:testZopgObligations}$
method testZopgObligations($\GRAPH$: Zopg, $R$: Graph, $r$: Node, $u$: Node)
  requires $\{u\} \DISJUNION R \subseteq \GRAPH \land \mo{CLOSED}(\{u\}) \land \mo{RING}_{\GRAPH}(R) \land r \in R \>\land$
           $\forall x\in \GRAPH,y \>\>\bullet\>\> \PATH{}{\GRAPH}{x}{y} \land \PATH{}{\GRAPH}{x}{u} \implies \lnot \PATH{}{\GRAPH}{r}{y}$ // (Pre)
{
    var $\HEAP$: Zopg := $\{ u\} \DISJUNION R$ // define new ghost parameter
    ringInsert($\HEAP$, $R$, $r$, $u$)
}$\phantomsection\label{lst:ringInsert}$
method ringInsert($\GRAPH$: Zopg, $R$: Graph,  $r$: Node, $u$: Node)
  requires $\GRAPH = \{ u \} \DISJUNION R \land \mo{CLOSED}(\{u\}) \land \mo{RING}_{\GRAPH}(R) \land r \in R$
  ensures $\mo{RING}_{\GRAPH}(\GRAPH) \land \big(\forall n\notin\GRAPH \>\>\bullet\>\> \PATH{}{\GRAPH}{u}{n} \Leftrightarrow \PATH{0}{\GRAPH}{r}{n} \big) \>\land $
          $\forall x\in\GRAPH,y \>\>\bullet\>\> x \neq u \land y \neq u \implies \big( \PATH{}{\GRAPH}{x}{y} \Leftrightarrow \PATH{0}{\GRAPH}{x}{y} \big)$
{
    u.next := r.next
    r.next := u
}
\end{lstlisting}
\begin{equation}
\begin{aligned}
\mo{RING}_{\GRAPH}(\HEAP) &\>\>\longeq\>\> \mo{FUNCTIONAL}(\HEAP) \land \mo{UNSHARED}(\HEAP) \land \mo{SCC}_{\GRAPH}(\HEAP) \\
\mo{SCC}_{\GRAPH}(\HEAP) &\>\>\longeq\>\> \forall x,y \in \HEAP \>\bullet\> \PATH{}{\GRAPH}{x}{y} \\
\mo{FUNCTIONAL}(\HEAP) &\>\>\longeq\>\> \forall a,b,c \in \HEAP \>\bullet\> \EDGE{}{\HEAP}{a}{b} \land \EDGE{}{\HEAP}{a}{c} \implies b = c \\
\mo{UNSHARED}(\HEAP) &\>\>\longeq\>\> \forall a,b,c \in \HEAP \>\bullet\> \EDGE{}{\HEAP}{a}{c} \land \EDGE{}{\HEAP}{b}{c} \implies a = b
\end{aligned}
\label{eq:ring}
\end{equation}
\caption{An example client with a ZOPG footprint. For simplicity, the method \coderef{testZopgObligations} has no postconditions. In order to verify it, one must nonetheless prove that the ZOPG invariant is maintained after the call to \coderef{ringInsert}. The definition of \mo{RING} is given in \eqref{ring}. Note the different meaning of footprint parameters written in subscripts vs. those written in parentheses (\eg{}~$\GRAPH$ and $\HEAP$, resp., in the definition of $\mo{SCC}_{\GRAPH}(\HEAP))$: the former are used as arguments for the \EDGESYMB{} and \PATHSYMB{} predicates, whereas the latter are used for restricting the domain of quantification.}
\label{fig:testZopgObligations}
\end{figure}

\subsubsection*{Extending the specification language for potentially-cyclic footprints.}

As a first step, we introduce an additional annotation in our specification language, so that we can label certain method footprints as ZOPGs. In addition to general graphs (whose structure is only constrained by other specifications), such as \code{$\HEAP$: Graph}, we allow the footprints of some methods to be more specifically marked as ZOPGs, using the syntax \code{$\GRAPH$: Zopg}. For method footprints declared this way, we will explain the additional proof obligations necessary to check that we \emph{maintain} the ZOPG invariant. In particular, a method with footprint \code{$\GRAPH$: Zopg} can be translated to a method with footprint \code{$\GRAPH$: Graph} with additional \emph{ZOPG proof obligations}. 

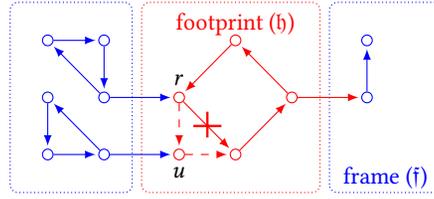
\begin{figure}[t]
\usetikzlibrary{shapes}

\begin{center}
	\begin{tikzpicture}[>=latex,xscale=0.5,yscale=0.5]
		\def\r{1.5}
		\def\xa{-5.0}
		\def\xb{-3.5}
		\def\xc{ 3.5}
		\def\ya{ 1.5}
		\def\yb{ 0.0}
		\def\yc{-1.5}
		
		\def\p{1.25}               
		\def\d{0.25}              
		\def\la{{(\xa-\p)}}       
		\def\ra{{(\xc+\p)}}       
		\def\ta{{(\ya+\p)}}       
		\def\ba{{(\yc-\p)}}       
		\def\lb{-2.5}             
		\def\rb{ 2.5}             
		\def\tb{{(\ya+\p-\d)}}
		\def\bb{{(\yc-\p+\d)}}
		
		\node [fvertex] (a1) at (\xa,\ya) {};
		\node [fvertex] (a2) at (\xb,\ya) {};
		\node [fvertex] (a3) at (\xa,\yb) {};
		\node [fvertex] (a4) at (\xb,\yb) {};
		\node [fvertex] (a5) at (\xa,\yc) {};
		\node [fvertex] (a6) at (\xb,\yc) {};
		
		\node [fvertex] (b1) at (\xc,\ya) {};
		\node [fvertex] (b2) at (\xc,\yb) {};
		
		\node [hvertex] (c1) at (  0:\r) {};
		\node [hvertex] (c2) at ( 90:\r) {};
		\node [hvertex,lbl=$r$] (c3) at (180:\r) {};
		\node [hvertex] (c4) at (270:\r) {};
		\node [hvertex,lbld=$u$] (x) at (-\r,-\r) {};
		
		\draw [frame] (-6.0,\bb) rectangle (\lb-\d,\tb);
		\draw [frame] (\rb,\bb) rectangle (5.5,\tb);
		\draw [footprint] (\lb,\bb) rectangle (\rb-\d,\tb);
	
		\node [framelbl,anchor=south] at (4.0,\bb) {frame ($\FRAME$)};
		\node [footprintlbl,anchor=north] at (0,\tb) {footprint ($\HEAP$)};
		
		\draw [fedge] (a1) -- (a2);
		\draw [fedge] (a2) -- (a4);
		\draw [fedge] (a4) -- (a1);
		\draw [fedge] (a4) -- (c3);
		\draw [fedge] (a6) -- (a3);
		\draw [fedge] (a3) -- (a5);
		\draw [fedge] (a5) -- (a6);
		\draw [fedge] (a6) -- (x);
		
		\draw [hedge] (c1) -- (c2);
		\draw [hedge] (c2) -- (c3);
		\draw [hedge] (c3) -- node[draw, midway, -, red, sloped, cross out, line width=.2ex, minimum width=1.5ex, minimum height=1ex, anchor=center]{} (c4);
		\draw [hedge] (c4) -- (c1);
		\draw [red,edge,dashed] (c3) -- (x);
		\draw [red,edge,dashed] (x) -- (c4);
		
		\draw [hedge] (c1) -- (b2);
		\draw [fedge] (b2) -- (b1);
	\end{tikzpicture}
\end{center}
	\caption{A typical scenario of running \coderef{testZopgObligations}. $r$ is an arbitrary node of the ring, and $u$ is added to the ring after the call to \coderef{ringInsert}. The diagram demonstrates a data structure with nodes that can have two reference fields (for simplicity, the implementation of \coderef{ringInsert} shows a single field, \code{next}). The second conjunct in the client's precondition says that no (non-trivial) paths may originate from $u$ (there may be paths ending in $u$). The footprint may have both entry and exit points, but (as required by the last conjunct in the client's precondition) each connected component of the frame may have \emph{at most one} entry or exit point into the footprint; otherwise, the ZOPG invariant would be violated by the call.}
  \label{fig:blubb}
\end{figure}

We illustrate the generation of ZOPG proof obligations based on the example in~\figref{testZopgObligations}. The client, \coderef{testZopgObligations}, operates on a ZOPG $\GRAPH$ that includes two disjoint parts: a ring $R$ and a (closed) singleton graph consisting of just one node, $u$. The extra node $r$ denotes an arbitrary node of the ring. The only operation performed by the client is a call to \coderef{ringInsert}. To verify that $\GRAPH$ remains a ZOPG by the end of \coderef{testZopgObligations}, we need to check that the callee does not create alternative paths---not just in its footprint, $\HEAP$ (which is guaranteed to remain a ZOPG, as the methods with footprints marked by \code{Zopg} are locally checked to preserve this property), but also in the larger subheap, $\GRAPH$. The callee \coderef{ringInsert}, operates on a ZOPG that \emph{equals} the union of two disjoint parts: a closed singleton graph $u$ and a ring $R$ (these two parts must be mutually-unreachable in the pre-state). The callee attaches $u$ to the ring $R$, resulting in a larger ring, $u \DISJUNION R$. The callee's postcondition says that (in the post-state) its entire footprint is a ring (thus, all pairs of footprint nodes are mutually reachable), and precisely defines its local reachability. Local paths that end \emph{outside} of the callee's footprint (\ie{}~\emph{outgoing paths}) are defined by the last two conjuncts: the former says that all exit points reachable from the ring in the pre-state are exactly the exit points reachable from $u$ in the post-state, whereas the latter preserves all outgoing paths of the initial ring, \ie{}~\emph{all} exit points of the footprint in the pre-state where $\{u\}$ was closed. \figref{blubb} illustrates the client's footprint in a state after \coderef{ringInsert} has executed. 

\subsubsection*{Maintaining the ZOPG invariant after a field update.} The knowledge that a subheap was a ZOPG in the pre-state of an operation helps checking that that subheap is still a ZOPG in the post-state, as we show next. We translate a general field update \code{u.next := v} in a method with the footprint \code{$\GRAPH$: Zopg} to \code{u.next := null; u.next := v}, where (assuming \code{v} is not null) the first update deletes an edge and the second one creates a new edge. Deleting edges does not alter the graph class of $\GRAPH$. However, a newly added edge may create an alternative path between some nodes of the graph. Concretely, new paths will be created between all pairs of nodes $(x,y)$ s.t.~there exist two paths: $x\ldots \mo{u}$ and $\mo{v}\ldots y$. Therefore, we get the following soundness criterion (emitted as a proof obligation before the second update) for a field update in a ZOPG:
\begin{equation}
  \mo{u} \neq \mo{v} \implies \forall x \in \GRAPH, y \>\bullet\> \PATH{}{\GRAPH}{x}{\mo u} \land \PATH{}{\GRAPH}{\mo v}{y} \implies \lnot \PATH{}{\GRAPH}{x}{y}
\label{eq:ZopgFieldUpdatePO}
\end{equation}
Note that $y$ may be outside of the current method's footprint because a $\GRAPH$-local path may leave $\GRAPH$ (\emph{iff} its last edge leaves that subheap). The formula \eqref{ZopgFieldUpdatePO} is much simpler than, \eg{}~\eqref{ZOPG} because it is (a) an \emph{incremental} condition (we used the knowledge that the subheap \emph{was} a ZOPG before the update; otherwise, we would need to consider alternative paths other than those introduced by the new creation) and (b) the operation is a field update (hence, only one edge has been added to the graph). Keeping track of graph classes in the presence of method calls is more involved, but the idea (a) is again helpful for tackling this problem.

\subsubsection*{Maintaining the ZOPG invariant after a method call.} A method call may violate the ZOPG invariant at call site even if the footprint of the call remains a ZOPG (a condition which is checked locally for the callee). The condition that a method call \emph{does not} violate the ZOPG invariant at call site is generally as hard to check as the formula \eqref{ZOPG} itself. Fortunately, this condition can be drastically simplified if the footprints or the callee and the client are \hyperref[def:convexity]{\emph{relatively convex}}.

\begin{figure}[t!]
\begin{center}
	\begin{tikzpicture}[>=latex,xscale=0.5,yscale=0.5]
		\def\xa{0.0}
		\def\xb{1.5}
		\def\xc{3.5}
		\def\xd{5.0}
		
		\def\ya{2.0}
		\def\yb{1.0}
		\def\yc{0.0}

		\def\dta{0.5}
		
		\begin{scope}[shift={(0,0)}]
			\node [fvertex,lbl=$x$] (x) at (\xa,\yb) {};
			\node [hvertex,lbld=$a$] (a) at (\xc,\ya) {};
			\node [hvertex,lbl=$b$] (b) at (\xc,\yc) {};
			\node [hvertex,lbl=$y$] (y) at (\xd,\yb) {};
			
			\draw [fpath] (x) -- (a);
			\draw [fpath] (x) -- (b);
			\draw [hpath] (a) -- (y);
			\draw [hpath] (b) -- (y);

			\draw [frame]     ({-\dta},{-0.5-\dta})    rectangle (2.5-0.5*\dta,2+\dta);
			\draw [footprint] ({2.5+0.5*\dta},{-0.5-\dta}) rectangle (5+\dta,2+\dta);

			\node [framelbl,anchor=south]     at (0.5,-1) {frame};
			\node [footprintlbl,anchor=south] at (4.125,-1) {footprint};

			\node [lbl,anchor=south] at (2.5,-2.5) {$(\alpha)$};

		\end{scope}
		
		\begin{scope}[shift={(7,0)}]
			\node [hvertex,lbl=$x$]  (x) at (\xa,\yb) {};
			\node [fvertex,lbld=$a$] (a) at (\xc,\ya) {};
			\node [fvertex,lbl=$b$]  (b) at (\xc,\yc) {};
			\node [fvertex,lbl=$y$]  (y) at (\xd,\yb) {};
			
			\draw [hpath] (x) -- (a);
			\draw [hpath] (x) -- (b);
			\draw [fpath] (a) -- (y);
			\draw [fpath] (b) -- (y);

			\draw [footprint]     ({-\dta},{-0.5-\dta})    rectangle (2.5-0.5*\dta,2+\dta);
			\draw [frame] ({2.5+0.5*\dta},{-0.5-\dta}) rectangle (5+\dta,2+\dta);

			\node [footprintlbl,anchor=south]     at (0.875,-1) {footprint};
			\node [framelbl,anchor=south] at (4-0.25,-1) {frame};

			\node [lbl,anchor=south] at (2.5,-2.5) {$(\beta)$};
		\end{scope}

		\begin{scope}[shift={(14,0)}]
			\node [fvertex,lbl=$x$] (x) at (\xa,\yb) {};
			\node [hvertex,lbld=$a$] (a) at (\xb,\ya) {};
			\node [fvertex,lbld=$b$] (b) at (\xc,\ya) {};
			\node [fvertex,lbl=$y$] (y) at (\xd,\yb) {};
			
			\draw [fpath] (x) -- (a);
			\draw [hpath] (a) -- (b);
			\draw [fpath] (b) -- (y);
			\draw [fpath] (x) -- (\xb-0.75,\yc) -- (\xc,\yc) -- (y);

			\draw [frame] (-\dta,-0.5-\dta) -- (\xd+\dta,-0.5-\dta) -- (\xd+\dta,2+\dta) -- (\xc-1.5*\dta,2+\dta) -- (\xc-1.5*\dta,\yb-\dta)
			-- (0.75-0.5*\dta,\yb-\dta) -- (0.75-0.5*\dta,2+\dta) -- (-\dta,2+\dta) -- cycle;
			\draw [footprint] ({\xb-\dta},{\yb}) rectangle (\xc-2.5*\dta,2+\dta);

			\node [framelbl,anchor=south] at (0.5,-1) {frame};

			\node [lbl,anchor=south] at (2.5,-2.5) {$(\gamma)$};

		\end{scope}

		\begin{scope}[shift={(21,0)}]
			\node [fvertex,lbl=$x$] (x) at (\xa,\yb) {};
			\node [hvertex,lbld=$a$] (a) at (\xb,\ya) {};
			\node [hvertex,lbl=$b$] (b) at (\xb,\yc) {};
			\node [fvertex,lbld=$c$] (c) at (\xc,\ya) {};
			\node [fvertex,lbl=$d$] (d) at (\xc,\yc) {};
			\node [fvertex,lbl=$y$] (y) at (\xd,\yb) {};
			
			\draw [fpath] (x) -- (a);
			\draw [fpath] (x) -- (b);
			\draw [hpath] (a) -- (c);
			\draw [hpath] (b) -- (d);
			\draw [fpath] (c) -- (y);
			\draw [fpath] (d) -- (y);

			\draw [frame]     ({-\dta},{-0.5-\dta})    rectangle (0.75-0.5*\dta,2+\dta);
			\draw [footprint] ({\xb-\dta},{-0.5-\dta}) rectangle (\xc-2.5*\dta,2+\dta);
			\draw [frame]     ({\xc-1.5*\dta},{-0.5-\dta})    rectangle (\xd+\dta,2+\dta);

			\node [framelbl,anchor=south]     at (0,-1) {fr.};
			\node [footprintlbl,anchor=south]     at (1.5,-1) {ft.};
			\node [framelbl,anchor=south] at (3.75,-1) {frame};

			\node [lbl,anchor=south] at (2.5,-2.5) {$(\delta)$};

		\end{scope}
	\end{tikzpicture}
\end{center}
\vspace{-4mm}
	\caption{The four configurations that violate the ZOPG invariant after a method call with a relatively convex ZOPG footprint (in red) and a ZOPG frame (in blue).}
	\label{fig:badZopgConfigs}
\end{figure}
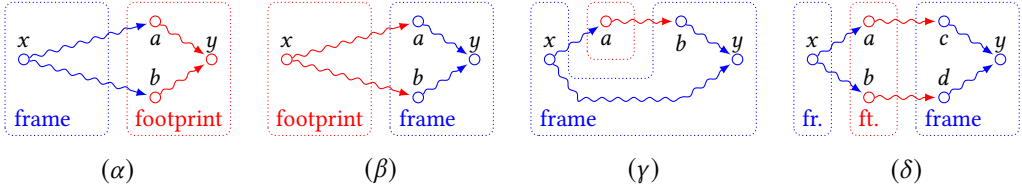

We proceed as follows. First, we enumerate the ways in which a method call that preserves the ZOPG invariant on its own footprint, could potentially \emph{violate} that invariant for its client's footprint. In particular, this must be by the creation of at least one new path. A call to a method with a \emph{convex} footprint may result in one of the four \emph{bad heap configurations} (violating the ZOPG invariant) depicted in \figref{badZopgConfigs}. Second, we conjoin the \emph{negated} formulas \eqref{badZopgConfigAlpha}, \eqref{badZopgConfigBeta}, \eqref{badZopgConfigGamma}, \eqref{badZopgConfigDelta} characterizing these four bad configurations, comprising an efficient criterion for preserving the ZOPG invariant. Checking this criterion can be easily automated: unlike formula~\eqref{ZOPG}, our criterion requires no information about the edge relation whatsoever. Our technique encodes this criterion as a proof obligation for the client. Finally, we sketch a proof of completeness for the four cases in \figref{badZopgConfigs}. 

The bad configuration in \figref{badZopgConfigs}~($\alpha$) corresponds to a scenario in which the method call has created an alternative path from $x$ to $y$, where the former node does not belong to the callee's footprint. We can describe this configuration via the following formula: 
\begin{equation}
\begin{aligned}
\exists x\in\FRAME,a,b\in\HEAP,y\notin\FRAME
\>\bullet\> a \neq b\>\land\>&\PATH{0}{\FRAME}{x}{a} \land \PATH{0}{\FRAME}{x}{b} \\
        \>\land\>&\PATH{}{\HEAP}{a}{y} \land \lnot \PATH{0}{\HEAP}{a}{y} \land \PATH{}{\HEAP}{b}{y}
\end{aligned}
\label{eq:badZopgConfigAlpha}
\end{equation}
The symbols \PATHSYMB{0} and \PATHSYMB{} denote the reachability relation before and after the method call; $\HEAP$ is the callee's footprint; $\FRAME$ is the frame of the call. We evaluate the first two reachability predicates in the old state because frame-local reachability is not affected by the call. 
The information about the last three predicates comes from the postcondition of the callee\footnote{It is also possible to get the information about the old reachability relation from a modified version of \eqref{badZopgConfigAlpha} where $\lnot \PATH{0}{\HEAP}{a}{y}$ is dropped and all other path predicates are evaluated in the pre-state.}. We assume \WLOG{} that $a\ldots y$ has been \emph{newly created} by the call (whereas $b\ldots y$ may have existed before the call). Both paths could not have existed before the call, as that would contradict our assumption that $\GRAPH$ was a ZOPG. 

Returning to our example of \figref{testZopgObligations}, we observe that the precondition of \coderef{testZopgObligations} is strong enough to prevent the bad configuration ($\alpha$) after the call to \code{ringInsert};  
we prove this by contradiction. Assume that, while preserving its local ZOPG invariant, the call results in the bad configuration ($\alpha$) for some $x\in\FRAME,a,b\in\HEAP,y\notin\FRAME$; thus, we learn the conjuncts (say, \#1 to \#6) from the body of~\eqref{badZopgConfigAlpha}. Note that $a$ and $b$ must be distinct (due to \#1) and cannot \emph{both} be in $R$ (due to \#4 and \#6; otherwise, there would be alternative paths, violating the ZOPG invariant of the callee's footprint in the post-state). We draw the contradiction by instantiating the last conjunct, \code{(Pre)}, of the precondition of \coderef{testZopgObligations}: $\forall x\in \GRAPH,y.\>\PATH{}{\GRAPH}{x}{y} \land \PATH{}{\GRAPH}{x}{u} \Rightarrow \lnot \PATH{}{\GRAPH}{r}{y}$. With our path partitioning formulas~\eqref{pathPartitioning}, \#2 and \#3 imply $\PATH{0}{\GRAPH}{x}{a}$ and $\PATH{0}{\GRAPH}{x}{b}$, resp. Together, \#4 and \#5 express that $a \ldots y$ is a \emph{newly created path}; hence either $a=u$ or $y=u$ (see~\figref{blubb}). If $a=u$, $y \neq u$, then $b \in R$; we draw the contradiction by instantiating \code{(Pre)} with $x,b$ for $x,y$. Otherwise, $a \neq u$, $y = u = b$, then $a \in R$; we draw the contradiction by instantiating \code{(Pre)} with $x,a$ for $x,y$. 
\qed

Similarly, we can describe the bad configuration in \figref{badZopgConfigs}~($\beta$) using the following formula: 
\begin{equation}
\begin{aligned}
\exists x\in\HEAP,a,b\in\FRAME,y\notin\HEAP
\>\bullet\> a \neq b \>\land\> & \PATH{0}{\FRAME}{a}{y} \land \PATH{0}{\FRAME}{a}{y} \\
                     \>\land\> & \PATH{}{\HEAP}{x}{a} \land \lnot \PATH{0}{\HEAP}{x}{a} \land \PATH{}{\HEAP}{x}{b}
\end{aligned}
\label{eq:badZopgConfigBeta}
\end{equation}
In this configuration, the source of the alternative paths falls into the callee footprint, and their end into the frame; this results in alternative paths $x\ldots a\ldots y$ and $x\ldots b\ldots y$. In our example of \figref{testZopgObligations}, the new outgoing paths that \coderef{ringInsert} creates originate in $u$; all other outgoing paths also existed \emph{before} the call (due to the last postcondition). In order to avoid the bad configuration ($\beta$), \coderef{testZopgObligations} requires that \emph{no paths} may originate in the attached node $u$. Thus, any \emph{new outgoing path} must pass through $R$ before it reaches the callee's footprint. Since \coderef{ringInsert} preserves the paths that start in $R$ and end in the frame (due to its last postcondition), and we assumed that the callee's footprint is a ZOPG before and after the call, the last three conjuncts in \eqref{badZopgConfigBeta} cannot be satisfied. Hence, our specification is strong enough to prevent ($\beta$). 
\qed

The scenario depicted in \figref{badZopgConfigs}~($\gamma$) illustrates that \emph{any new path} $a\ldots b$ created by the method call, combined with suitable frame paths, may violate the ZOPG invariant: 
\begin{equation}
\begin{aligned}
\exists x,b\in\FRAME,a\in\HEAP,y\notin\HEAP
\>\bullet\>
\PATH{0}{\FRAME}{x}{y} \land
\PATH{0}{\FRAME}{x}{a} \land \PATH{0}{\FRAME}{b}{y} \land
\PATH{}{\HEAP}{a}{b} \land \lnot\PATH{0}{\HEAP}{a}{b}.
\end{aligned}
\label{eq:badZopgConfigGamma}
\end{equation}
In order to avoid the bad configuration $(\gamma)$, we must ensure that an arbitrary frame node $x$ that reaches the footprint node $a$ does not reach any of the frame nodes (\eg{}~$y$) that will be reachable from $a$ after the call. In our example of \figref{testZopgObligations}, the precondition of \code{testZopgObligations} is strong enough to prevent ($\gamma$) after the call to \code{ringInsert}. The nodes $x$ and $y$ in the last conjunct of this precondition can be thought of as those in \eqref{badZopgConfigGamma} and \figref{badZopgConfigs} ($\gamma$); the condition rules out the possibility that the effect of the call will connect up such alternative path. 
\qed

The most subtle bad configuration is \figref{badZopgConfigs}~($\delta$), where both alternative paths $x\ldots y$ go via the footprint of the method call. This heap configuration can be expressed via the following formula: 
\begin{equation}
\begin{aligned}
\exists x,c,d\in\FRAME,a,b\in\HEAP,y\notin\HEAP
\>\bullet\> a \neq b \land c \neq d \>\land\> &\PATH{0}{\FRAME}{x}{a} \land \PATH{0}{\FRAME}{x}{b} \land \PATH{0}{\FRAME}{c}{y}\land \PATH{0}{\FRAME}{d}{y}\\
                                    \>\land\> &\PATH{}{\HEAP}{a}{c}\land \lnot \PATH{0}{\HEAP}{a}{c} \land \PATH{}{\HEAP}{b}{d}.
\end{aligned}
\label{eq:badZopgConfigDelta}
\end{equation}
This configuration can be realized when $a$ and $b$ are \emph{mutually unreachable in both states} (otherwise, the configuration is covered by ($\alpha$) and ($\beta$)). This configuration cannot occur in the post-state of our example because after the method call $u$ is attached to the ring. 
\qed

\subsubsection*{Completeness proof sketch.} To derive the four cases in \figref{badZopgConfigs}, consider a ZOPG subheap $\GRAPH$ comprised of the ZOPG frame $\FRAME$ and the (relatively convex) ZOPG footprint $\HEAP$ of a method call. Assume that the method call creates at least one new path s.t. the ZOPG invariant of its footprint is maintained while the ZOPG invariant of the (larger) client's footprint is violated. Consider as well two nodes $x$ and $y$ that are connected (in the state after the call) by \emph{multiple} (at least two) $\GRAPH$-local paths $x \ldots y$. 
We assume that $x$ and $y$ are both in $\GRAPH$; if $y$ is outside $\GRAPH$, we first apply \eqref{crossBorderPaths}, providing some node $u\in\GRAPH$ s.t. $\PATH{}{\GRAPH}{x}{u}\land\EDGE{}{\GRAPH}{u}{y}$; we then continue the argument for $x \ldots u$ instead of $x \ldots y$. 

Multiple paths $x \ldots y$ may not be entirely inside just one of the two subheaps $\FRAME$ or $\HEAP$ because that would violate our assumption that these are ZOPG subheaps. Therefore, \emph{at least one of these paths must cross the border between $\FRAME$ and $\HEAP$}. We proceed with a case analysis based on the distribution of the nodes $x$ and $y$ between (disjoint) subheaps $\HEAP$ and $\FRAME$: 
\begin{itemize}
\item The case $x,y\in\HEAP$ cannot be realized because a path starting in $x$ may leave $\HEAP$ just once and \emph{may never come back} to reach $y$ (otherwise our convexity assumption would be violated). 
\item If $x\in\HEAP, y\in\FRAME$, then again, the paths starting in $x$ may leave $\HEAP$ just once and may never come back due to $\HEAP \CONVEXIN \GRAPH$. Since two different paths starting in $x$ may not merge in $\HEAP$ (otherwise, alternative paths will exist within $\HEAP$, contradicting our assumption that it is a ZOPG), such paths must reach \emph{two different frame nodes} $a,b\in\FRAME$, creating alternative paths of the form $x \ldots a \ldots y$ and $x \ldots b \ldots y$, as covered by case ($\beta$) of \figref{badZopgConfigs}. 
\item If $x\in\FRAME, y\in\HEAP$, then no path that starts in the frame node $x$ can enter the footprint more than once due to $\HEAP \CONVEXIN \GRAPH$. Next, since in this case these paths must end in $y$, they cannot leave the (relatively convex) footprint $\HEAP$ at all. Finally, these paths \emph{may not merge until at least one of them enters $\HEAP$} (otherwise, alternative paths will exist within $\FRAME$, contradicting our assumption that it is a ZOPG). This gives us \emph{two different footprint nodes} $a,b\in\HEAP$, creating alternative paths of the form $x \ldots a \ldots y$ and $x \ldots b \ldots y$, as covered by case ($\alpha$) of \figref{badZopgConfigs}. 
\item In the most subtle case of $x,y\in\FRAME$, each pair of alternative paths of the form $x \ldots y$ is s.t. either \emph{just one} or \emph{each of the two} alternative paths \emph{enters and exits} the footprint exactly once, as covered by cases ($\gamma$) and ($\delta$) of \figref{badZopgConfigs}, resp. 
\qed
\end{itemize}

The simplicity of our formulas \eqref{badZopgConfigAlpha}, \eqref{badZopgConfigBeta}, \eqref{badZopgConfigGamma}, \eqref{badZopgConfigDelta} is due to the fact that, in our technique, footprints of method calls must be relatively convex, limiting the number of bad configurations to just four. The bad configurations that we have identified are helpful for deriving weakest preconditions for method calls that operate over ZOPGs, like in our \code{testZopgObligations} example. In combination with local heap updates (for which \eqref{ZopgFieldUpdatePO} is the efficient ZOPG preservation criterion), we have explained how our technique is generalized for modular reasoning about ZOPGs. 


\section{Evaluation}
\label{sec:evaluation}
\label{sec:impl}
We have evaluated our technique on a variety of challenging example programs taken from the literature, illustrating our technique for different classes of graphs and data structures (including the running examples of closely-related work). 

\begin{table}[t]
\setlength{\tabcolsep}{4pt}
\newcommand{\h}[1]{\multicolumn{1}{c}{#1}}
\newcommand{\hb}[1]{\multicolumn{1}{c}{#1}}
\begin{center}
\scalebox{0.9}{
\begin{tabular}{@{}lllcccrl@{}}\toprule
  \thead{Example}                           & \thead{Variant}& \thead{Class}& \outDegreeSymb & \inDegreeSymb & \CONVEXIN & \thead{Time} & \thead{Notes} \\
\midrule
Merge (\figref{runningExample})             & Tree  &\acyclic& \tick    &      & \tick                   & 16.1 & Path-partitioning, \\
                                            & DAG   &\acyclic& \tick    &\tick & \tick                   & 14.5 & Unbounded cut-points  \\
                                            & Fail 1 &\acyclic& \tick    &\tick & \tick                   & 13.2 & Bug in code  \\
                                            & Fail 2 &\acyclic& \tick    &\tick & \tick                   & 33.9 & Bug in spec.  \\
\hdashline
Left-Child-                                 & Tree, add sibl.&\acyclic& \tick    &      & \tick                   & 10.5 & Encodes n-ary tree as binary\\
Right-Sibling                               & Tree, add child&\acyclic& \tick    &      & \tick                   & 15.0 &  \ditto  \\
                                            & DAG, add sibl. &\acyclic& \tick    &\tick & \tick                   & 10.1 & Unbounded cut-points \\
                                            & DAG, add child &\acyclic& \tick    &\tick & \tick                   & 17.1 & \ditto \\
\hdashline
Harris List                                 & Original &\acyclic&         & \tick &       & 14.5 & From \cite{Krishna2017GoWT} \\
\hdashline
Acyclic List                                & Reverse  &\acyclic&         &       &       & 7.9  & From~\cite{SagivLMCS09} \\
                                            & Append   &\acyclic&         &       &       & 6.9  &  \ditto \\
\hdashline
Ring-Insert:                                & Sorted   &\zeroOnePG& \tick & \tick &       & 87.2  & Functional spec. \\
Impl.                  & Anywhere &\zeroOnePG& \tick & \tick &       & 10.1   & \ditto \\
\hdashline
Ring-Insert:                                & Closed $\{u\} $ &\zeroOnePG& \tick & \tick & \tick & 11.5  &  Non-convex frame, \\
Client (\figref{testZopgObligations})       & Open $\{u\} $   &\zeroOnePG& \tick & \tick & \tick & 10.8  &  \zeroOnePG{} obligations \\
                                            & Fail 1 &\zeroOnePG & \tick & \tick & \tick & 12.4  & Failure due to ($\beta$) \\
                                            & Fail 2 &\zeroOnePG& \tick & \tick & \tick & 10.7 & Failure due to ($\alpha$), ($\gamma$) \\
\bottomrule
\end{tabular}%
}
\end{center}
\caption{Experimental results. We indicate example features via \tick{} where \outDegreeSymb{} and \inDegreeSymb{} denote examples with \emph{greater-than-one outdegree} and \emph{with sharing}, resp.; \CONVEXIN{} means \emph{convex framing}.}
\label{tab:experiments}
\end{table}

\subsection{Experimental Setup}

We encoded each example by-hand into the Viper verification language~\cite{mueller-2015-viper}: an intermediate verification language designed for expressing heap-based verification problems, and with native support for separation logic reasoning. Although manual, our encoding of each example was performed methodically, simulating the translation that a front-end verification tool could perform. Each example consists of a common set of background definitions and axioms, along with a translation of the code of the example, statement by statement, according to the technique presented in \secref{DAGs} and \secref{ZOPs}. For instance, a source-level method call is encoded with additional \code{assume} and \code{assert} statements before and after the call which enable reachability framing on relatively convex method footprints, as defined in \secref{nonconvexframe}. 

The background definitions common to our examples are organized in separately-included library files, and we make heavy use of Viper's macros to improve the readability of our encoded examples. Our examples are verified with Viper's standard Boogie-based~\cite{ThisIsBoogie2} verifier, which uses the Z3 SMT solver~\cite{Z3solver} for checking verification conditions. We indicate Viper's run time for each example in \tabref{experiments}. The experiments were performed on a laptop running macOS, with a 2.8 GHz Intel Core i7 CPU, with Z3 version 4.8.5 - 64 bit. The Viper files used in our experiments are available as artifact of this paper~\cite{ArtifactCitation}. 

An important practical issue arising in the successful use of SMT-based verification tools is controlling the instantiation of quantifiers; our technique employs a large number of quantified formulas, and we have carefully selected appropriate \emph{triggers} \cite{Simplify,MoskalEtAl,SMTLIB,Z3solver} for these, guided by the intended situations in which these formulas are relevant; for the rich reachability properties expressed by our technique, such triggers are essential for performance. Since our source-level specifications can also contain quantified formulas, we require these to be annotated with appropriate triggers (for simple cases, Viper can also infer appropriate choices if omitted).

\subsection{Experiments}

\tabref{experiments} gives an overview of our experiments. The ``Merge'' example is our first running example of \figref{runningExample}, in variants with both tree and DAG structures for the underlying graphs (obtaining the DAG variant simply requires dropping the tree requirements throughout; no other changes are necessary). ``Left-Child-Right-Sibling'' is a technique for encoding trees with arbitrary multiplicities using only two fields (representing a list of children at each node), as employed in binomial heaps \cite{Cormen09}, and recently proposed as a verification challenge \cite{Mueller18}. We again show a DAG variant (directly obtained by removing tree requirements), and verify adding sibling and child structures. As with the running example, these are non-functional graphs with (in the DAG case) sharing and requiring our convex framing to frame reachability across sub-calls; to our knowledge, they are beyond reach for all existing automated graph-verification techniques. 

The ``Fail'' variants of Merge are buggy, with the bug being (1) negation of the branch condition in the body of \code{merge} and (2) missing \code{merge}'s last precondition. We have observed that the failure time does not diverge from the time of a successful verification attempt. This is important in practice if a program's implementation and specification are developed iteratively, with multiple invocations of the verifier.


Lev-Ami \etal{} verify reachability for linked-list reverse and append methods \cite{SagivLMCS09}; the recent Flows framework \cite{Krishna2017GoWT} uses the Harris List as running example. In both cases, we prove the same invariants and reachability specifications, simply encoded in our language. In the latter case, we use two reachability relations based on different edges. 

``Ring-Insert'' is a series of six 0--1-path graph examples. We wrote two variants of the Ring-Insert method. ``Sorted'' is an implementation that traverses a sorted ring and inserts a newly allocated node into the right place. We can prove both reachability (the ring remains a ring) and sortedness; our connection to separation-logic reasoning makes layering additional functional specifications of this kind straightforward. ``Anywhere'' is the version discussed in~\secref{ZOPs}, where the insertion happens at an arbitrary point in the ring. We also verified two types of clients of Ring-Insert. ``Closed $\{u\}$'' is the example of \figref{testZopgObligations}, where the attached node does not have outgoing paths, whereas ``Open $\{u\}$'' permits the attached node to be both reachable from the frame and have outgoing paths. The latter requires a more subtle precondition to satisfy the 0--1-path preservation criteria. In the final two cases, we show that our technique allows us to automatically identify the type of bad configurations that may violate the 0--1-path invariant in cases where the heap is under-constraint before a method call~\secref{qualifiers}. 

\subsection{Results}







Our experiments show that reachability properties are amenable to SMT-based verification for a broad class of heap-manipulating programs. In particular, we have observed that our technique is well-suited for this task despite heavy usage of quantified formulas. While developing the specifications, we have experienced that our technique helps the programmer to better understand the subtle effects of heap operations on data structure invariants. Even with good tool support, writing consistent preconditions and postconditions requires particular craftsmanship, especially for recursive methods, like \code{merge}. Additionally, SMT-based verification with quantifiers requires the programmer to annotate the specifications with triggers. 





%
%
%
%

\section{Related Work}
\label{sec:related-work}

\begin{table}[t]
\begin{center}
\scalebox{0.9}{
\begin{tabular}{@{}llp{8pt}p{8pt}p{8pt}l@{}}\toprule
\thead{Data structure}                  & \thead{Class}                & \hspace*{-2pt}\thead{Itz.} & \hspace*{-2pt}\thead{Gr.}                   & \hspace*{-2pt}\thead{Nv.} & \thead{Author}                   \\
\midrule                                                                                                                                                                    
Arbitrary linked-list structures        & \zeroOnePG                   & \tick                &                               &             & --                               \\
2-opt, 3-opt, etc.                      & \zeroOnePG                   &                      &                               & \tick       & \citet{twoOptPaper,threeOptPaper} \\
Hierarchical rings                      & \zeroOnePG                   &                      &                               & \tick       & \citet{Fredman1986ThePH}          \\
The priority inheritance protocol       & \zeroOnePG                   &                      &                               & \tick       & \citet{pipPaper}                  \\
Trees encoded via Java’s LinkedList     & \zeroOnePG                   &                      &                               & \tick       & Sun, Oracle                      \\
Union-Find                              & \zeroOnePG/\acyclic          & \tick                &                               &             & \citet{unionFindPaper}            \\
Trees                                   & \zeroOnePG/\acyclic          &                      & \tick                         &             & --                               \\
Binary Decision Diagrams                & \multicolumn{1}{r}{\acyclic} &                      &                               & \tick       & \citet{BDDsInitialPaper,Akers78}  \\
General DAGs (DFG, VCS, etc.)           & \multicolumn{1}{r}{\acyclic} &                      &                               & \tick       & --                               \\
\bottomrule
\end{tabular}%
}
\end{center}
\caption{Supported data structure categories. \textbf{``Itz.''} denotes~\cite{ItzhakyPOPL14}; \textbf{``Gr.''} denotes GRASShopper~\cite{Piskac2014GRASShopperC}; \textbf{``Nv.''} indicates whether, to our knowledge, our technique enables automated verification of modularly specified reachability properties for the first time.}
\label{tab:supportedStructures}
\end{table}

Most work on separation logic focuses on data structures with limited sharing, with some notable exceptions. Iterated separating conjunction has been used to verify the Schorr-Waite graph marking algorithm~\cite{Yang01anexample}, but without any tool support or automation. Recent work on Flows~\cite{Krishna2017GoWT} allows one to prove the \emph{preservation} of a rich variety of graph invariants including reachability properties, but requires fixpoint computations that are hard to automate. Methods can operate on a subgraph; under the condition that \emph{interfaces}~\cite{Krishna2017GoWT} of these subgraphs are preserved, a view on the caller's graph can be reconstructed. They make no convexity restriction, but the interface preservation conditions rule out the possibility of method calls adding or removing paths between nodes in subgraph boundaries. By contrast, our reachability framing technique explicitly enables such side-effectful methods, and the reconstruction of appropriate changes in the caller's footprint. For instance, in our running example of \figref{merge}, new reachability relations are first \emph{established} (by creating an edge from \code{link} to the root of \code{rdag}) and then \emph{propagated} (by the enclosing method calls) to the larger context (the entirety of the client’s footprint).

We adapted the precise transitive-closure update formulas from~\citet{Dong1995IncrementalAD} to program heaps and separation logic, rather than mathematical graphs. Their work also inspired our \DEPSYMB{} relation; however, our version of the \DEPSYMB{} relation is compatible with the \emph{reflexive} reachability relation and is used only in the internal encoding, whereas theirs is exposed to programmers. Reachability has been integrated into separation logic before (\eg{} in GRASShopper~\cite{Piskac2014GRASShopperC}), but only in a limited way that supports lists and trees but not heap structures with sharing.

Our work was inspired by \citet{ItzhakyCAV13,ItzhakyPOPL14}. Their verification technique allows one to prove reachability properties in various forms of list data structures. A focus of their work is to obtain decidable proof obligations. We sacrificed decidability in favor of supporting arbitrary acyclic graphs (with bounded out-degree) as well as 0--1-path graphs; our evaluation shows that we nevertheless achieve good automation. In contrast to Itzhaky et~al., we integrated our work into separation logic, which allows us to verify concurrent programs and to reason about reachability and other properties in a uniform way. Moreover, we do not restrict method footprints in the number of entry and exit points or the number of SCCs in them. \tabref{supportedStructures} summarizes the expressiveness of our technique and compares it with closely-related work.

\section{Conclusions}
\label{sec:conclusion}

We presented a specification and verification technique that allows one to reason about heap reachability properties modularly. The technique is integrated into separation logic and, thus, benefits immediately from the plurality of techniques and tools in this area. The key challenge of this integration is to specify reachability locally, within the footprint of a method. We solved this challenge by specifying reachability relatively to a given heap fragment and introducing a novel form of reachability framing to extend reachability properties in the footprint of a callee method to the larger footprint of the caller. Even though reasoning about general reachability properties is difficult to automate, the proof obligations required by our technique are amenable to SMT solvers, which we demonstrate in our experiments.

As future work, we plan to extend our technique to graphs with unbounded outdegree. This can be done by using a generalized version of iterated separating conjunction~\cite{QuantifiedPermissions} that can specify permissions to sets of resources. Another direction of future work is to adapt our technique to separation logic with fractional permissions~\cite{Boyland2003CheckingIW} to distinguish read and write access, especially in concurrent settings. We also plan to investigate the extent to which our approach to cyclic graphs is always precise. Finally, we are planning to implement a front-end verification tool that will simplify the process of writing modular reachability specifications.

\begin{acks}
We are grateful to the anonymous referees for their thoughtful comments. We also thank 
Uri~Juhasz,
Marco~Eilers,
Gishor~Sivanrupan,
Jérôme~Dohrau,
Sviatlana-Maryia~Zdobnikava, \\
Alexandra-Olimpia~Bugariu,
Siddharth~Krishna,
Felix~Wolf,
Vytautas~Astrauskas,
Federico~Poli,
Martin~Clochard,
K.~Rustan~M.~Leino,
and Shachar~Itzhaky for their help.
This work was funded in part by SNF under project 200021-156980.
\end{acks}

\bibliography{all}


\begin{thebibliography}{31}


\ifx \showCODEN    \undefined \def \showCODEN     #1{\unskip}     \fi
\ifx \showDOI      \undefined \def \showDOI       #1{#1}\fi
\ifx \showISBNx    \undefined \def \showISBNx     #1{\unskip}     \fi
\ifx \showISBNxiii \undefined \def \showISBNxiii  #1{\unskip}     \fi
\ifx \showISSN     \undefined \def \showISSN      #1{\unskip}     \fi
\ifx \showLCCN     \undefined \def \showLCCN      #1{\unskip}     \fi
\ifx \shownote     \undefined \def \shownote      #1{#1}          \fi
\ifx \showarticletitle \undefined \def \showarticletitle #1{#1}   \fi
\ifx \showURL      \undefined \def \showURL       {\relax}        \fi
\providecommand\bibfield[2]{#2}
\providecommand\bibinfo[2]{#2}
\providecommand\natexlab[1]{#1}
\providecommand\showeprint[2][]{arXiv:#2}

\bibitem[\protect\citeauthoryear{{Akers Jr.}}{{Akers Jr.}}{1978}]%
        {Akers78}
\bibfield{author}{\bibinfo{person}{Sheldon~B. {Akers Jr.}}}
  \bibinfo{year}{1978}\natexlab{}.
\newblock \showarticletitle{Binary Decision Diagrams}.
\newblock \bibinfo{journal}{\emph{IEEE Trans. Comput.}} \bibinfo{volume}{27},
  \bibinfo{number}{6} (\bibinfo{year}{1978}), \bibinfo{pages}{509--516}.
\newblock


\bibitem[\protect\citeauthoryear{Barrett, Fontaine, and Tinelli}{Barrett
  et~al\mbox{.}}{2017}]%
        {SMTLIB}
\bibfield{author}{\bibinfo{person}{Clark Barrett}, \bibinfo{person}{Pascal
  Fontaine}, {and} \bibinfo{person}{Cesare Tinelli}.}
  \bibinfo{year}{2017}\natexlab{}.
\newblock \bibinfo{booktitle}{\emph{{The SMT-LIB Standard: Version 2.6}}}.
\newblock \bibinfo{type}{{T}echnical {R}eport}.
  \bibinfo{institution}{Department of Computer Science, The University of
  Iowa}.
\newblock
\newblock
\shownote{Available at {\tt www.SMT-LIB.org}.}


\bibitem[\protect\citeauthoryear{Boyland}{Boyland}{2003}]%
        {Boyland2003CheckingIW}
\bibfield{author}{\bibinfo{person}{John~Tang Boyland}.}
  \bibinfo{year}{2003}\natexlab{}.
\newblock \showarticletitle{Checking Interference with Fractional Permissions}.
  In \bibinfo{booktitle}{\emph{{SAS}}} \emph{(\bibinfo{series}{Lecture Notes in
  Computer Science})}, Vol.~\bibinfo{volume}{2694}.
  \bibinfo{publisher}{Springer}, \bibinfo{pages}{55--72}.
\newblock


\bibitem[\protect\citeauthoryear{Cormen, Leiserson, Rivest, and Stein}{Cormen
  et~al\mbox{.}}{2009}]%
        {Cormen09}
\bibfield{author}{\bibinfo{person}{Thomas~H. Cormen},
  \bibinfo{person}{Charles~E. Leiserson}, \bibinfo{person}{Ronald~L. Rivest},
  {and} \bibinfo{person}{Clifford Stein}.} \bibinfo{year}{2009}\natexlab{}.
\newblock \bibinfo{booktitle}{\emph{Introduction to Algorithms, 3rd Edition}}.
\newblock \bibinfo{publisher}{{MIT} Press}.
\newblock
\showISBNx{978-0-262-03384-8}


\bibitem[\protect\citeauthoryear{Croes}{Croes}{1958}]%
        {twoOptPaper}
\bibfield{author}{\bibinfo{person}{G.~A. Croes}.}
  \bibinfo{year}{1958}\natexlab{}.
\newblock \showarticletitle{A Method for Solving Traveling-Salesman Problems}.
\newblock \bibinfo{journal}{\emph{Operations Research}} \bibinfo{volume}{6},
  \bibinfo{number}{6} (\bibinfo{year}{1958}), \bibinfo{pages}{791--812}.
\newblock


\bibitem[\protect\citeauthoryear{de~Moura and Bj{\o}rner}{de~Moura and
  Bj{\o}rner}{2008}]%
        {Z3solver}
\bibfield{author}{\bibinfo{person}{Leonardo~Mendon{\c{c}}a de Moura} {and}
  \bibinfo{person}{Nikolaj Bj{\o}rner}.} \bibinfo{year}{2008}\natexlab{}.
\newblock \showarticletitle{{Z3:} An Efficient {SMT} Solver}. In
  \bibinfo{booktitle}{\emph{{TACAS}}} \emph{(\bibinfo{series}{Lecture Notes in
  Computer Science})}, \bibfield{editor}{\bibinfo{person}{C.~R. Ramakrishnan}
  {and} \bibinfo{person}{Jakob Rehof}} (Eds.), Vol.~\bibinfo{volume}{4963}.
  \bibinfo{publisher}{Springer}, \bibinfo{pages}{337--340}.
\newblock


\bibitem[\protect\citeauthoryear{Detlefs, Nelson, and Saxe}{Detlefs
  et~al\mbox{.}}{2005}]%
        {Simplify}
\bibfield{author}{\bibinfo{person}{David Detlefs}, \bibinfo{person}{Greg
  Nelson}, {and} \bibinfo{person}{James~B. Saxe}.}
  \bibinfo{year}{2005}\natexlab{}.
\newblock \showarticletitle{Simplify: A Theorem Prover for Program Checking}.
\newblock \bibinfo{journal}{\emph{Journal of the {ACM}}} \bibinfo{volume}{52},
  \bibinfo{number}{3} (\bibinfo{year}{2005}), \bibinfo{pages}{365--473}.
\newblock


\bibitem[\protect\citeauthoryear{Dong and Su}{Dong and Su}{1995}]%
        {Dong1995IncrementalAD}
\bibfield{author}{\bibinfo{person}{Guozhu Dong} {and} \bibinfo{person}{Jianwen
  Su}.} \bibinfo{year}{1995}\natexlab{}.
\newblock \showarticletitle{Incremental and Decremental Evaluation of
  Transitive Closure by First-Order Queries}.
\newblock \bibinfo{journal}{\emph{Inf. Comput.}}  \bibinfo{volume}{120}
  (\bibinfo{year}{1995}), \bibinfo{pages}{101--106}.
\newblock


\bibitem[\protect\citeauthoryear{Fredman, Sedgewick, Sleator, and
  Tarjan}{Fredman et~al\mbox{.}}{1986}]%
        {Fredman1986ThePH}
\bibfield{author}{\bibinfo{person}{Michael~L. Fredman}, \bibinfo{person}{Robert
  Sedgewick}, \bibinfo{person}{Daniel~Dominic Sleator}, {and}
  \bibinfo{person}{Robert~E. Tarjan}.} \bibinfo{year}{1986}\natexlab{}.
\newblock \showarticletitle{The pairing heap: {A} new form of self-adjusting
  heap}.
\newblock \bibinfo{journal}{\emph{Algorithmica}}  \bibinfo{volume}{1}
  (\bibinfo{year}{1986}), \bibinfo{pages}{111--129}.
\newblock


\bibitem[\protect\citeauthoryear{Itzhaky, Banerjee, Immerman, Lahav, Nanevski,
  and Sagiv}{Itzhaky et~al\mbox{.}}{2014}]%
        {ItzhakyPOPL14}
\bibfield{author}{\bibinfo{person}{Shachar Itzhaky}, \bibinfo{person}{Anindya
  Banerjee}, \bibinfo{person}{Neil Immerman}, \bibinfo{person}{Ori Lahav},
  \bibinfo{person}{Aleksandar Nanevski}, {and} \bibinfo{person}{Mooly Sagiv}.}
  \bibinfo{year}{2014}\natexlab{}.
\newblock \showarticletitle{Modular Reasoning About Heap Paths via Effectively
  Propositional Formulas}. In \bibinfo{booktitle}{\emph{{POPL}}},
  \bibfield{editor}{\bibinfo{person}{Suresh Jagannathan} {and}
  \bibinfo{person}{Peter Sewell}} (Eds.). \bibinfo{publisher}{ACM},
  \bibinfo{pages}{385--396}.
\newblock
\showISBNx{978-1-4503-2544-8}


\bibitem[\protect\citeauthoryear{Itzhaky, Banerjee, Immerman, Nanevski, and
  Sagiv}{Itzhaky et~al\mbox{.}}{2013}]%
        {ItzhakyCAV13}
\bibfield{author}{\bibinfo{person}{Shachar Itzhaky}, \bibinfo{person}{Anindya
  Banerjee}, \bibinfo{person}{Neil Immerman}, \bibinfo{person}{Aleksandar
  Nanevski}, {and} \bibinfo{person}{Mooly Sagiv}.}
  \bibinfo{year}{2013}\natexlab{}.
\newblock \showarticletitle{Effectively-Propositional Reasoning about
  Reachability in Linked Data Structures}. In
  \bibinfo{booktitle}{\emph{{CAV}}}, \bibfield{editor}{\bibinfo{person}{Natasha
  Sharygina} {and} \bibinfo{person}{Helmut Veith}} (Eds.).
  \bibinfo{publisher}{Springer Berlin Heidelberg}, \bibinfo{pages}{756--772}.
\newblock


\bibitem[\protect\citeauthoryear{Krishna, Shasha, and Wies}{Krishna
  et~al\mbox{.}}{2018}]%
        {Krishna2017GoWT}
\bibfield{author}{\bibinfo{person}{Siddharth Krishna},
  \bibinfo{person}{Dennis~E. Shasha}, {and} \bibinfo{person}{Thomas Wies}.}
  \bibinfo{year}{2018}\natexlab{}.
\newblock \showarticletitle{Go with the flow: compositional abstractions for
  concurrent data structures}.
\newblock \bibinfo{journal}{\emph{{PACMPL}}} \bibinfo{volume}{2},
  \bibinfo{number}{{POPL}} (\bibinfo{year}{2018}),
  \bibinfo{pages}{37:1--37:31}.
\newblock


\bibitem[\protect\citeauthoryear{{Lee}}{{Lee}}{1959}]%
        {BDDsInitialPaper}
\bibfield{author}{\bibinfo{person}{C.~Y. {Lee}}.}
  \bibinfo{year}{1959}\natexlab{}.
\newblock \showarticletitle{Representation of switching circuits by
  binary-decision programs}.
\newblock \bibinfo{journal}{\emph{The Bell System Technical Journal}}
  \bibinfo{volume}{38}, \bibinfo{number}{4} (\bibinfo{year}{1959}),
  \bibinfo{pages}{985--999}.
\newblock


\bibitem[\protect\citeauthoryear{Leino}{Leino}{2008}]%
        {ThisIsBoogie2}
\bibfield{author}{\bibinfo{person}{K~Rustan~M Leino}.}
  \bibinfo{year}{2008}\natexlab{}.
\newblock \bibinfo{title}{This is Boogie 2}.  (\bibinfo{year}{2008}).
\newblock
\urldef\tempurl%
\url{https://www.microsoft.com/en-us/research/publication/this-is-boogie-2-2/}
\showURL{%
\tempurl}


\bibitem[\protect\citeauthoryear{Leino and Monahan}{Leino and Monahan}{2009}]%
        {limitedfunctions}
\bibfield{author}{\bibinfo{person}{K.~Rustan~M. Leino} {and}
  \bibinfo{person}{Rosemary Monahan}.} \bibinfo{year}{2009}\natexlab{}.
\newblock \showarticletitle{Reasoning about comprehensions with first-order
  {S}{M}{T} solvers}. In \bibinfo{booktitle}{\emph{{SAC}}},
  \bibfield{editor}{\bibinfo{person}{Sung~Y. Shin} {and}
  \bibinfo{person}{Sascha Ossowski}} (Eds.). \bibinfo{publisher}{ACM},
  \bibinfo{pages}{615--622}.
\newblock


\bibitem[\protect\citeauthoryear{Lev{-}Ami, Immerman, Reps, Sagiv, Srivastava,
  and Yorsh}{Lev{-}Ami et~al\mbox{.}}{2009}]%
        {SagivLMCS09}
\bibfield{author}{\bibinfo{person}{Tal Lev{-}Ami}, \bibinfo{person}{Neil
  Immerman}, \bibinfo{person}{Thomas~W. Reps}, \bibinfo{person}{Mooly Sagiv},
  \bibinfo{person}{Siddharth Srivastava}, {and} \bibinfo{person}{Greta Yorsh}.}
  \bibinfo{year}{2009}\natexlab{}.
\newblock \showarticletitle{Simulating reachability using first-order logic
  with applications to verification of linked data structures}.
\newblock \bibinfo{journal}{\emph{Logical Methods in Computer Science}}
  \bibinfo{volume}{5}, \bibinfo{number}{2} (\bibinfo{year}{2009}).
\newblock


\bibitem[\protect\citeauthoryear{Lin}{Lin}{1965}]%
        {threeOptPaper}
\bibfield{author}{\bibinfo{person}{Shen Lin}.} \bibinfo{year}{1965}\natexlab{}.
\newblock \showarticletitle{Computer solutions of the traveling salesman
  problem}.
\newblock \bibinfo{journal}{\emph{The Bell System Technical Journal}}
  \bibinfo{volume}{44}, \bibinfo{number}{10} (\bibinfo{year}{1965}),
  \bibinfo{pages}{2245--2269}.
\newblock


\bibitem[\protect\citeauthoryear{Moskal}{Moskal}{2009}]%
        {MoskalEtAl}
\bibfield{author}{\bibinfo{person}{Michal Moskal}.}
  \bibinfo{year}{2009}\natexlab{}.
\newblock \showarticletitle{Programming with triggers}.
\newblock \bibinfo{journal}{\emph{ACM International Conference Proceeding
  Series}} (\bibinfo{year}{2009}).
\newblock


\bibitem[\protect\citeauthoryear{M{\"u}ller}{M{\"u}ller}{2018}]%
        {Mueller18}
\bibfield{author}{\bibinfo{person}{Peter M{\"u}ller}.}
  \bibinfo{year}{2018}\natexlab{}.
\newblock \showarticletitle{The Binomial Heap Verification Challenge in
  {V}iper}.
\newblock In \bibinfo{booktitle}{\emph{Principled Software Development}},
  \bibfield{editor}{\bibinfo{person}{Peter M{\"u}ller} {and}
  \bibinfo{person}{Ina Schaefer}} (Eds.). \bibinfo{publisher}{Springer-Verlag},
  \bibinfo{pages}{203--219}.
\newblock


\bibitem[\protect\citeauthoryear{M{\"{u}}ller, Schwerhoff, and
  Summers}{M{\"{u}}ller et~al\mbox{.}}{2016a}]%
        {QuantifiedPermissions}
\bibfield{author}{\bibinfo{person}{Peter M{\"{u}}ller}, \bibinfo{person}{Malte
  Schwerhoff}, {and} \bibinfo{person}{Alexander~J. Summers}.}
  \bibinfo{year}{2016}\natexlab{a}.
\newblock \showarticletitle{Automatic Verification of Iterated Separating
  Conjunctions using Symbolic Execution}. In \bibinfo{booktitle}{\emph{{CAV}}}
  \emph{(\bibinfo{series}{LNCS})}, \bibfield{editor}{\bibinfo{person}{Swarat
  Chaudhuri} {and} \bibinfo{person}{Azadeh Farzan}} (Eds.),
  Vol.~\bibinfo{volume}{9779}. \bibinfo{publisher}{Springer-Verlag},
  \bibinfo{pages}{405--425}.
\newblock


\bibitem[\protect\citeauthoryear{M{\"{u}}ller, Schwerhoff, and
  Summers}{M{\"{u}}ller et~al\mbox{.}}{2016b}]%
        {mueller-2015-viper}
\bibfield{author}{\bibinfo{person}{Peter M{\"{u}}ller}, \bibinfo{person}{Malte
  Schwerhoff}, {and} \bibinfo{person}{Alexander~J. Summers}.}
  \bibinfo{year}{2016}\natexlab{b}.
\newblock \showarticletitle{Viper: A Verification Infrastructure for
  Permission-Based Reasoning}. In \bibinfo{booktitle}{\emph{{VMCAI}}}
  \emph{(\bibinfo{series}{LNCS})}, \bibfield{editor}{\bibinfo{person}{Barbara
  Jobstmann} {and} \bibinfo{person}{K.~Rustan~M. Leino}} (Eds.),
  Vol.~\bibinfo{volume}{9583}. \bibinfo{publisher}{Springer-Verlag},
  \bibinfo{pages}{41--62}.
\newblock


\bibitem[\protect\citeauthoryear{Parkinson and Bierman}{Parkinson and
  Bierman}{2005}]%
        {Parkinson05}
\bibfield{author}{\bibinfo{person}{Matthew~J. Parkinson} {and}
  \bibinfo{person}{Gavin~M. Bierman}.} \bibinfo{year}{2005}\natexlab{}.
\newblock \showarticletitle{Separation logic and abstraction}. In
  \bibinfo{booktitle}{\emph{{POPL}}}, \bibfield{editor}{\bibinfo{person}{Jens
  Palsberg} {and} \bibinfo{person}{Mart{\'{\i}}n Abadi}} (Eds.).
  \bibinfo{publisher}{{ACM}}, \bibinfo{pages}{247--258}.
\newblock


\bibitem[\protect\citeauthoryear{Parkinson and Summers}{Parkinson and
  Summers}{2011}]%
        {ParkinsonSummers12}
\bibfield{author}{\bibinfo{person}{Matthew~J. Parkinson} {and}
  \bibinfo{person}{Alexander~J. Summers}.} \bibinfo{year}{2011}\natexlab{}.
\newblock \showarticletitle{The Relationship between Separation Logic and
  Implicit Dynamic Frames}. In \bibinfo{booktitle}{\emph{{ESOP}}},
  \bibfield{editor}{\bibinfo{person}{Gilles Barthe}} (Ed.).
  \bibinfo{pages}{439--458}.
\newblock


\bibitem[\protect\citeauthoryear{Piskac, Wies, and Zufferey}{Piskac
  et~al\mbox{.}}{2014}]%
        {Piskac2014GRASShopperC}
\bibfield{author}{\bibinfo{person}{Ruzica Piskac}, \bibinfo{person}{Thomas
  Wies}, {and} \bibinfo{person}{Damien Zufferey}.}
  \bibinfo{year}{2014}\natexlab{}.
\newblock \showarticletitle{GRASShopper - Complete Heap Verification with Mixed
  Specifications}. In \bibinfo{booktitle}{\emph{{TACAS}}}
  \emph{(\bibinfo{series}{Lecture Notes in Computer Science})},
  \bibfield{editor}{\bibinfo{person}{Erika {\'{A}}brah{\'{a}}m} {and}
  \bibinfo{person}{Klaus Havelund}} (Eds.), Vol.~\bibinfo{volume}{8413}.
  \bibinfo{publisher}{Springer}, \bibinfo{pages}{124--139}.
\newblock


\bibitem[\protect\citeauthoryear{Reynolds}{Reynolds}{2002}]%
        {reynolds2002}
\bibfield{author}{\bibinfo{person}{John~C. Reynolds}.}
  \bibinfo{year}{2002}\natexlab{}.
\newblock \showarticletitle{Separation Logic: A Logic for Shared Mutable Data
  Structures}. In \bibinfo{booktitle}{\emph{{IEEE} Symposium on Logic in
  Computer Science}} \emph{(\bibinfo{series}{LICS '02})}.
  \bibinfo{publisher}{IEEE Computer Society}, \bibinfo{pages}{55--74}.
\newblock
\showISBNx{0-7695-1483-9}


\bibitem[\protect\citeauthoryear{Sha, Rajkumar, and Lehoczky}{Sha
  et~al\mbox{.}}{1990}]%
        {pipPaper}
\bibfield{author}{\bibinfo{person}{Lui Sha}, \bibinfo{person}{Ragunathan
  Rajkumar}, {and} \bibinfo{person}{John~P. Lehoczky}.}
  \bibinfo{year}{1990}\natexlab{}.
\newblock \showarticletitle{Priority inheritance protocols: an approach to
  real-time synchronization}.
\newblock \bibinfo{journal}{\emph{IEEE Trans. Comput.}} \bibinfo{volume}{39},
  \bibinfo{number}{9} (\bibinfo{year}{1990}), \bibinfo{pages}{1175--1185}.
\newblock


\bibitem[\protect\citeauthoryear{Smans, Jacobs, and Piessens}{Smans
  et~al\mbox{.}}{2012}]%
        {ImpDynFrames}
\bibfield{author}{\bibinfo{person}{Jan Smans}, \bibinfo{person}{Bart Jacobs},
  {and} \bibinfo{person}{Frank Piessens}.} \bibinfo{year}{2012}\natexlab{}.
\newblock \showarticletitle{Implicit Dynamic Frames}.
\newblock \bibinfo{journal}{\emph{{ACM} Transactions on Programming Languages
  and Systems}} \bibinfo{volume}{34}, \bibinfo{number}{1}, Article
  \bibinfo{articleno}{2} (\bibinfo{year}{2012}), \bibinfo{numpages}{58}~pages.
\newblock


\bibitem[\protect\citeauthoryear{Tarjan}{Tarjan}{1975}]%
        {unionFindPaper}
\bibfield{author}{\bibinfo{person}{Robert~Endre Tarjan}.}
  \bibinfo{year}{1975}\natexlab{}.
\newblock \showarticletitle{Efficiency of a Good But Not Linear Set Union
  Algorithm}.
\newblock \bibinfo{journal}{\emph{Journal of the {ACM}}} \bibinfo{volume}{22},
  \bibinfo{number}{2} (\bibinfo{year}{1975}), \bibinfo{pages}{215--225}.
\newblock


\bibitem[\protect\citeauthoryear{Ter-Gabrielyan, Summers, and
  M{\"u}ller}{Ter-Gabrielyan et~al\mbox{.}}{2019}]%
        {ArtifactCitation}
\bibfield{author}{\bibinfo{person}{Arshavir Ter-Gabrielyan},
  \bibinfo{person}{Alexander~J. Summers}, {and} \bibinfo{person}{Peter
  M{\"u}ller}.} \bibinfo{year}{2019}\natexlab{}.
\newblock \bibinfo{title}{{Modular Verification of Heap Reachability Properties
  in Separation Logic (Artifact)}}.
\newblock
\newblock
\urldef\tempurl%
\url{https://doi.org/10.5281/zenodo.3367478}
\showDOI{\tempurl}


\bibitem[\protect\citeauthoryear{Yang}{Yang}{2001a}]%
        {Yang01anexample}
\bibfield{author}{\bibinfo{person}{Hongseok Yang}.}
  \bibinfo{year}{2001}\natexlab{a}.
\newblock \showarticletitle{An example of local reasoning in {BI} pointer
  logic: the {S}chorr-{W}aite graph marking algorithm}. In
  \bibinfo{booktitle}{\emph{Proceedings of the {SPACE} Workshop}}.
\newblock


\bibitem[\protect\citeauthoryear{Yang}{Yang}{2001b}]%
        {YangPhD}
\bibfield{author}{\bibinfo{person}{Hongseok Yang}.}
  \bibinfo{year}{2001}\natexlab{b}.
\newblock \emph{\bibinfo{title}{Local Reasoning for Stateful Programs}}.
\newblock \bibinfo{thesistype}{Ph.D. Dissertation}.
\newblock Advisor(s) Uday S. Reddy.
\newblock
\showISBNx{0-493-35008-X}


\end{thebibliography}
\newpage

\ifTechReport
\def\includeAppendices{1}
\fi

\ifdefined\includeAppendices
  \begin{subappendices}
  \appendix

\section{Precise Update Formulas for the Reachability Query in Acyclic Structures}
\label{sec:acyclicTCupdates}

\newcommand{\inDifferentSCCs}{\mo{NOT\_IN\_SAME\_SCC}}
\newcommand{\dagLinkTcUpdate}{\mo{LINK}^\mathit{F}}
\newcommand{\dagUnlinkTcUpdate}{\mo{UNLINK}^\mathit{F}}
\newcommand{\linkEdgeUpdate}{\mo{G\_PLUS\_DELTA}}
\newcommand{\unlinkEdgeUpdate}{\mo{G\_MINUS\_DELTA}}
\newcommand{\tcgPlusDelta}{\mo{TC\_G\_PLUS\_DELTA}}
\newcommand{\tcgMinusDelta}{\mo{TC\_G\_MINUS\_DELTA}}
\newcommand{\snaps}{\mo{snap}^\mathit{F}}

\newcommand{\fname}{\ensuremath{\mathit{f}}}
\newcommand{\ename}{\ensuremath{\mathit{e}}}

\newcommand{\memspecs}[2]{\mo{MEMORY\_SPECS}_{\fname}^{F}(#1, #2)}

In this section, we present the general encoding of field updates in acyclic graphs, as introduced in~\secref{DAGs}. Our goal is to rewrite operations of the form \code{from.$\fname$ := to} in such a way that would introduce reachability update formulas~\cite{Dong1995IncrementalAD} to the verifier. We denote the footprint of the method enclosing this field update as $\GRAPH$. We assume that this methods implementation and specification mentions reference fields from the set $F$; in particular, $\fname \in F$. 


Since general field updates may result in a deletion and then a creation of two different heap edges, we rewrite them, \WLOG{}, as sequence of two (conceptually simpler) operations: 

\phantomsection\label{lst:updateDAG}
\begin{lstlisting}
define updateDAG$^{F}_{\fname}$($\GRAPH$: Graph, from: Node, to: Node) { 
  // state 1
  if ( to != from.$\fname$ ) {
    if ( $\mo{from.}\fname \neq \NULLLIT $ ) {
      assert $\ACYCLIC{\GRAPH}$
      unlinkDAG$^{F}_{\fname}$($\GRAPH$, from)
    } // state 2
    if ( $\mo{to} \neq \NULLLIT$ ) {
      linkDAG$^{F}_{\fname}$($\GRAPH$, from, to)
      assert $\ACYCLIC{\GRAPH}$
}} // state 3 }
\end{lstlisting}

If \code{to == from.$\fname$}, then state~1 and state~3 are identical; therefore, neither the field value, nor the reachability relation should be updated. Otherwise, we proceed with two (conditional) operations (we postpone the discussion of the explicit assertions until the end of this section). The first operation, called \coderef{unlinkDAG}, \emph{removes exactly one heap edge} by assigning $\NULLLIT$ to $\mo{node.}\fname$; note that the pre- and post-states of this operation (state~1 and state~2, resp.) are identical if the value of the considered field is already $\NULLLIT$; hence, the first \code{if} statement. The second operation, called \coderef{linkDAG}, \emph{requires} that its last argument, \code{to} (\ie{} the destination of the created edge) is non-null; the operation then \emph{creates exactly one heap edge} by assigning \code{to} to $\mo{node.}\fname$; note that the pre- and post-states of this operation (state~2 and state~3, resp.) are identical if the value of the considered field is already \code{to}; hence, the second \code{if} statement. 

The memory specifications in our \coderef{unlinkDAG} and \coderef{linkDAG} operations express access permissions to all reference fields from $F$ of all objects in the current method's footprint ($\GRAPH$). In logics that differentiate \emph{read} and \emph{write} permissions (\eg{}~\cite{Boyland2003CheckingIW}), we require write permissions only to the single reference field that is \emph{modified} by this operation; that is, concretely, \code{node.$\fname$} and \code{from.$\fname$} in the contracts of \coderef{unlinkDAG} and \coderef{linkDAG}, resp. For all other reference fields, we require read permissions, because we are only interested in the existing values of those fields (to be able to define our $\GRAPH$-local reachability relation). These memory specifications can be realized via the iterated separating conjunction~\cite{QuantifiedPermissions} available in Viper~\cite{mueller-2015-viper}. 

\newpage

\phantomsection\label{lst:unlinkDAG}
\begin{lstlisting}
method unlinkDAG$^{F}_{\fname}$($\GRAPH$: Graph, node: Node) 
  requires $\memspecs{\GRAPH}{\mo{node}}$
  requires $\mo{node} \in \GRAPH \land \fname \in F \land \mo{node.}\fname \neq \NULLLIT$ 
  ensures $\memspecs{\GRAPH}{\mo{node}}$
  ensures $\mo{node.}\fname = \NULLLIT$
  ensures $\big( \bigwedge\nolimits_{\ename \in F, \ename \neq \fname} \OLDEXP{\mo{node.}\ename} \neq \OLDEXP{\mo{node.}\fname} \big) \implies \dagUnlinkTcUpdate(\GRAPH, \mo{node}, \OLDEXP{\mo{node.}\fname})$
\end{lstlisting}

\phantomsection\label{lst:linkDAG}
\begin{lstlisting}
method linkDAG$^{F}_{\fname}$($\GRAPH$: Graph, from: Node, to: Node) 
  requires $\memspecs{\GRAPH}{\mo{from}}$ 
  requires $\mo{from} \in \GRAPH \land \fname \in F \land \mo{from.}\fname = \NULLLIT \land \mo{to} \neq \NULLLIT$ 
  ensures $\memspecs{\GRAPH}{\mo{from}}$ 
  ensures $\mo{from.}\fname = \mo{to}$
  ensures $\big( \bigwedge\nolimits_{\ename \in F, \ename \neq \fname} \OLDEXP{\mo{from.}\ename} \neq \mo{to} \big) \implies \dagLinkTcUpdate(\GRAPH, \mo{from}, \mo{to})$
\end{lstlisting}




\begin{equation*}
\begin{aligned}
\dagLinkTcUpdate(\GRAPH, \alpha, \beta) \>\longeq\>   &\linkEdgeUpdate(\OLDEXP{\snaps(\GRAPH)}, \snaps(\GRAPH), \alpha, \beta) \land \\
                                                      &\tcgMinusDelta(\OLDEXP{\snaps(\GRAPH)}, \snaps(\GRAPH), \alpha, \beta) \land \\
                                                      &\tcgPlusDelta(\OLDEXP{\snaps(\GRAPH)}, \snaps(\GRAPH), \alpha, \beta), \\
\dagUnlinkTcUpdate(\GRAPH, \alpha, \beta) \>\longeq\> &\unlinkEdgeUpdate(\snaps(\GRAPH), \OLDEXP{\snaps(\GRAPH)}, \alpha, \beta) \land \\
                                                      &\tcgMinusDelta(\snaps(\GRAPH), \OLDEXP{\snaps(\GRAPH)}, \alpha, \beta) \land \\
                                                      &\tcgPlusDelta(\snaps(\GRAPH), \OLDEXP{\snaps(\GRAPH)}, \alpha, \beta), \\ 
  \linkEdgeUpdate(g, G, \alpha, \beta) \>\longeq\> & \forall x, y \>\bullet\> (x,y)\in G \>\>{=\joinrel=}\>\> (x,y)\in g \>\lor\> x=\alpha \land y=\beta, \\
\unlinkEdgeUpdate(g, G, \alpha, \beta) \>\longeq\> & \forall x, y \>\bullet\> (x,y)\in g \>\>{=\joinrel=}\>\> (x,y)\in G \>\land\> (x\neq\alpha \lor y\neq\beta), \\
\tcgPlusDelta(g, G, \alpha, \beta) \>\longeq\>  & \forall x, y \>\bullet\> \hat{P}(G,x,y) \>\>{=\joinrel=}\>\> \hat{P}(g,x,y) \lor \hat{P}(g,x,\alpha) \land \hat{P}(g,\beta,y), \\ 
\tcgMinusDelta(g, G, \alpha, \beta) \>\longeq\> & \big(\forall x, y \>\bullet\> \lnot \hat{P}(G,x,\alpha) \lor \lnot \hat{P}(G,\beta,y) \implies \hat{P}(g,x,y) \>{=\joinrel=}\> \hat{P}(G,x,y) \big) \\
                                                &\land \forall x, y \>\bullet\> \hat{P}(G,x,\alpha) \land \hat{P}(G,\beta,y) \implies \\ 
                                                & \qquad\qquad\qquad\qquad\hat{P}(g,x,y) \>{=\joinrel=}\> \mo{case}_{i}(G, x, \alpha, \beta, y) \lor \\
                                                & \qquad\qquad\qquad\qquad\qquad\qquad\quad\>\> \mo{case}_{ii}(G, x, \alpha, \beta, y) \lor \\
                                                & \qquad\qquad\qquad\qquad\qquad\qquad\quad\>\> \mo{case}_{iii}(G, x, \alpha, \beta, y), \\
\mo{case}_{i}(G, x, \alpha, \beta, y)  \>\longeq\> & \exists u \>\bullet\> u \neq \alpha \land u \neq \beta \land \hat{P}(G, \alpha, u) \land \hat{P}(G, u, \beta), \\
\mo{case}_{ii}(G, x, \alpha, \beta, y) \>\longeq\> & \exists u \>\bullet\> \hat{P}(G, x, u) \land \hat{P}(G, u, y) \>\> \land \\ 
                                                &\big( \lnot \hat{P}(G, u, \alpha) \land \lnot \hat{P}(G, \alpha, u) \lor \lnot \hat{P}(G, u, \beta) \land \lnot \hat{P}(G, \beta, u) \big), \\
\mo{case}_{iii}(G, x, \alpha, \beta, y) \>\longeq\> & \exists u,v \>\bullet\> (u \neq \alpha \lor v \neq \beta) \land \hat{E}(G, u, v) \>\> \land \\ 
                                                &\hat{P}(G, x, u) \land \hat{P}(G, u, \alpha) \land \hat{P}(G, \beta, v) \land \hat{P}(G, v, y).
\end{aligned}
\label{eq:updateTcInDag}
\end{equation*}

Our reachability update formulas, called $\dagLinkTcUpdate$ and $\dagUnlinkTcUpdate$, are three-fold. First, we update the edge relation, \EDGESYMB{}, using the macros called $\unlinkEdgeUpdate$ and $\linkEdgeUpdate$. Second, we \emph{directly update} the transitive path relation, \PATHSYMB{}, using the macros $\tcgMinusDelta$ and $\tcgPlusDelta$ for $\dagUnlinkTcUpdate$ and $\dagLinkTcUpdate$, resp. These correspond to the decremental and incremental transitive closure update formulas by~\citet{Dong1995IncrementalAD}, with two exceptions: (1) our macros use ternary reachability and edge predicates and (2) our reachability relation is reflexive, whereas \citet{Dong1995IncrementalAD} consider an irreflexive transitive closure relation. The direct update formulas provide a canonical form of the reachability relation in the new state (\eg{}~$\PATHSYMB{}~\Leftrightarrow~(\ldots)$) based on the reachability relation in the old state (\eg{}~$(\ldots)~\Leftrightarrow~Q[\PATHSYMB{0}]$, where $Q[R]$ is a first-order formula over $R$). Finally, we exploit the fact that both \coderef{unlinkDAG} and \coderef{linkDAG} relate two neighboring states: a state in which the heap graph has exactly one fewer edges and a state in which the heap graph has exactly one extra edge. Therefore, for each of the operations, we can also use the \emph{indirect update} of the reachability relation: $\tcgPlusDelta$ and $\tcgMinusDelta$ for $\dagUnlinkTcUpdate$ and $\dagLinkTcUpdate$, resp. 

Remarkably, the formulation in $\tcgMinusDelta$ is more complex than in $\tcgPlusDelta$. This is due to the problem of recovering reachability after destructive updates (\cf{}~\figref{atomicUpdateDiagram}). The formal justifications for both formulations are given in~\cite{Dong1995IncrementalAD}. 

Since the reachability update formulas described above are defined for DAGs, the assertions in~\coderef{updateDAG} ensure soundness of our technique for field updates. The first assertion checks that the heap graph represented by $\GRAPH$ is acyclic \emph{before} the first operation, \coderef{unlinkDAG}. The second assertion checks that the acyclic invariant is \emph{preserved} after the second operation, \coderef{linkDAG}. 

\section{Precise Update Formulas for the Reachability Query in 0--1-Path Structures}
\label{sec:zopgDepUpdates}

\newcommand{\noAltPathsViaEdge}{\mo{NO\_ALT\_PATHS\_VIA\_EDGE}}
\newcommand{\zopgLinkTcUpdate}{\mo{ZOPG\_LINK}^\mathit{F}}
\newcommand{\zopgUnlinkTcUpdate}{\mo{ZOPG\_UNLINK}^\mathit{F}}
\newcommand{\depgPlusDelta}{\mo{DEP\_G\_PLUS\_DELTA}}
\newcommand{\depgMinusDelta}{\mo{DEP\_G\_MINUS\_DELTA}}
\newcommand{\psiFormula}{\Psi}

In this section, we present the general encoding of field updates for ZOPGs, as introduced in~\secref{ZOPs}. We assume that the reader understand the previous case of DAGs discussed in~\appref{acyclicTCupdates}; similarly, our goal now is to rewrite operations of the form \code{from.$\fname$ := to} in such a way that would introduce precise update formulas to the verifier, except this time we are interested in the auxiliary \DEPSYMB{} relation for ZOPGs (and not \PATHSYMB{}). As before, we denote the footprint of the method enclosing this field update as $\GRAPH$. We assume that this methods implementation and specification mentions reference fields from the set $F$; in particular, $\fname \in F$.

\begin{lstlisting}[escapeinside={(*@}{@*)}]
define updateZOPG$^{F}_{\fname}$($\GRAPH$: Zopg, from: Node, to: Node) {
  // state 1
  if ( to != from.$\fname$ ) {
    if ( from.$\fname$ != $\NULLLIT$ ) { 
      assert $\noAltPathsViaEdge(\snaps(\GRAPH), \mo{from}, \mo{to})$
      unlinkZOPG$^{F}_{\fname}$($\GRAPH$, from) 
    } // state 2
    if ( to != null ) { 
      linkZOPG$^{F}_{\fname}$($\GRAPH$, from, to)                    
      assert $\noAltPathsViaEdge(\snaps(\GRAPH), \mo{from}, \mo{to})$ 
}} // state 3 } 

\end{lstlisting}

The assertions in the code above are the main difference from the DAG case so far. The macro used in these assertions directly corresponds to the~\eqref{ZopgFieldUpdatePO}. 

\begin{equation*}
  \noAltPathsViaEdge(g, \alpha, \beta) \>\longeq\> \forall x,y \>\bullet\> \hat{P}(g, x, \alpha) \land \hat{P}(g, \beta, y) 
    \implies \lnot \hat{P}(g, \alpha, \beta)
\end{equation*}
\newpage
\begin{lstlisting}
method unlinkZOPG$^{F}_{\fname}$($\GRAPH$: Graph, node: Node) 
  requires $\memspecs{\GRAPH}{\mo{node}}$
  requires $\mo{node} \in \GRAPH \land \fname \in F \land \mo{node.}\fname \neq \NULLLIT$ 
  ensures $\memspecs{\GRAPH}{\mo{node}}$
  ensures $\mo{node.}\fname = \NULLLIT$
  ensures $\big( \bigwedge\nolimits_{\ename \in F, \ename \neq \fname} \OLDEXP{\mo{node.}\ename} \neq \OLDEXP{\mo{node.}\fname} \big) \implies \zopgUnlinkTcUpdate(\GRAPH, \mo{node}, \OLDEXP{\mo{node.}\fname})$
\end{lstlisting}

\begin{lstlisting}
method linkZOPG$^{F}_{\fname}$($\GRAPH$: Graph, from: Node, to: Node) 
  requires $\memspecs{\GRAPH}{\mo{from}}$
  requires $\mo{from} \in \GRAPH \land \fname \in F \land \mo{from.}\fname = \NULLLIT \land \mo{to} \neq \NULLLIT$
  ensures $\memspecs{\GRAPH}{\mo{from}}$
  ensures $\mo{from.}\fname = \mo{to}$
  ensures $\big( \bigwedge\nolimits_{\ename \in F, \ename \neq \fname} \OLDEXP{\mo{from.}\ename} \neq \mo{to} \big) \implies \zopgLinkTcUpdate(\GRAPH, \mo{from}, \mo{to})$
\end{lstlisting}

\begin{equation*}
\begin{aligned}
\zopgLinkTcUpdate(\GRAPH, \alpha, \beta) \>\longeq\>  &\linkEdgeUpdate(\OLDEXP{\snaps(\GRAPH)}, \snaps(\GRAPH), \alpha, \beta) \land \\
                                                        &\depgMinusDelta(\OLDEXP{\snaps(\GRAPH)}, \snaps(\GRAPH), \alpha, \beta) \land \\
                                                        &\depgPlusDelta(\OLDEXP{\snaps(\GRAPH)}, \snaps(\GRAPH), \alpha, \beta), \\
\zopgUnlinkTcUpdate(\GRAPH, \alpha, \beta) \>\longeq\> &\unlinkEdgeUpdate(\snaps(\GRAPH), \OLDEXP{\snaps(\GRAPH)}, \alpha, \beta) \land \\
                                                         &\depgMinusDelta(\snaps(\GRAPH), \OLDEXP{\snaps(\GRAPH)}, \alpha, \beta) \land \\
                                                         &\depgPlusDelta(\snaps(\GRAPH), \OLDEXP{\snaps(\GRAPH)}, \alpha, \beta), \\ 
\depgMinusDelta(g, G, \alpha, \beta) \>\longeq\>  & \forall x,y,u,v \>\bullet\>  \longhat{DEP}(g,x,y) \>\>{=\joinrel=}\>\> \longhat{DEP}(G,x,y,u,v) \>\land \\
                                                  & \quad\quad\quad\quad\quad\quad\quad\quad\quad\quad\quad\>\>\> \lnot \longhat{DEP}(G,x,y,\alpha,\beta), \\ 
\depgPlusDelta(g, G, \alpha, \beta) \>\longeq\> & \alpha \neq \beta \land \forall x,y,u,v \>\bullet\> u \neq v \land x \neq y \implies \\
                                                & \Big( \longhat{DEP}(G,x,y,u,v) \>\>{=\joinrel=}\>\> \longhat{DEP}(g,x,y,u,v) \>\lor \\
                                                & \quad\quad\quad\quad\quad\quad\quad\quad\> (x=u=\alpha \land y=v=\beta) \>\lor \\
                                                & \quad\quad\quad\quad\quad\quad\quad\quad\> \big( \exists w,z \>\>\>\bullet\> \psiFormula(g,\>\alpha,\beta,\>x,y,\>u,v,\>w,z) \big) \>\lor \\ 
                                                & \quad\quad\quad\quad\quad\quad\quad\quad\> \big( \exists u',v'     \>\bullet\> \psiFormula(g,\>\alpha,\beta,\>x,y,\>u',v',\>u,v) \big) \>\lor \\ 
                                                & u=\alpha \land v=\beta \land \exists u',v',w,z \>\bullet\> \psiFormula(g,\>\alpha,\beta,\>x,y,\>u',v',\>w,z) \Big), \\
\psiFormula(g,\alpha,\beta,x,y,U,V,w,z) \>\longeq\> & \big( \forall u'',v'' \>\bullet\> \lnot \longhat{DEP}(g,x,y,u'',v'') \big) \land \\
                                             & \big( \longhat{DEP}(g,x,\alpha,u,v) \land x\neq\alpha \lor x=u=\alpha \land v=\beta \big) \\
                                             & \big( \longhat{DEP}(g,\beta,y,w,z)  \land y\neq\beta  \lor \alpha=w \land \beta=z=y \big). 
\end{aligned}
\label{eq:updateDepInDag}
\end{equation*}

The structure of $\zopgUnlinkTcUpdate$ and $\zopgLinkTcUpdate$ is conceptually the same as in their DAG counterparts, $\dagUnlinkTcUpdate$ and $\dagLinkTcUpdate$. The essential difference here are the update formulas which, for the \DEPSYMB{} relation, are more complex in the case of $\depgPlusDelta$ than $\depgMinusDelta$. Our \DEPSYMB{} update formulas are adapted from~\cite{Dong1995IncrementalAD} with few modifications, as explained in~\secref{ZOPs}.



\section{Conversion Rules for the DEP Relation}
\label{sec:ZOP-rules}

The auxiliary \DEPSYMB{} relation enables precise reasoning about reachability in 0--1-path graphs. However, the formal definition of \DEPSYMB{}\footnote{Note that here we omit the first parameter (\ie{}~the sub-heap) in all relation symbols as it is implicitly universally quantified and the same in all formulas.} is beyond first-order logic: 
\begin{align}
\label{eq:metaDep}
\begin{split}
\forall x, y, u, v \;\;\bullet\;\; \BDEP{x}{y}{u}{v} \longeq& \>\> \BEDGE{u}{v} \land u \neq v \>\land \\
                                   & \>\>\BPATH{x}{u} \land \BPATH{v}{y} \>\land \\
                                   & \>\>\lnot \mathtt{TC}[ \EDGESYMB{} \setminus \{ (u, v) \} ](x, y)
\end{split}
\tag{MetaDefinitionDEP}
\end{align}
\noindent Here $\mathtt{TC}[ R ]$ is the reflexive, transitive closure of the binary relation $R$ and \EDGESYMB{} is the edge relation. The last conjunct says that removing the edge $(u, v)$ results in the destruction of the path $x..y$. 

Our 0--1-path graph axiomatization includes a number of formulas---that can be derived from \eqref{metaDep}---enabling conversions between reachability information in terms of \PATHSYMB{} vs. \DEPSYMB{}. The core of this axiomatization consists of the following formulas: 
\begin{align}
\forall x, y, u, v \;\;\bullet\;\; & \BPATH{x}{y} \land x \neq y \iff \exists u,v \> \bullet \> \BDEP{x}{y}{u}{v} \tag{PToDep} \\
\forall x, y, u, v \;\;\bullet\;\; & \BDEP{x}{y}{u}{v} \implies x \neq y \land u \neq v \land \BPATH{x}{y} \land \BEDGE{u}{v} \tag{DepToP} \\
\forall x, y \;\;\bullet\;\; & \BDEP{x}{y}{x}{y} \iff \BEDGE{x}{y} \land x \neq y \tag{SimplePath}
\end{align}

Given the knowledge that the graph is a 0--1-path graph, we derive the remaining part of our axiomatization: 
\begin{align}
\forall x, y, n \;\;\bullet\;\; & \lnot\BDEP{y}{n}{x}{y} \tag{UnrollFromHead} \\
\forall x, y, n \;\;\bullet\;\; & \lnot\BDEP{n}{x}{x}{y} \tag{UnrollFromTail} \\ 
\forall x,y,v \;\;\bullet\;\;   & x \neq y \land x \neq v \land \tag*{} \\
                                &\lnot\BPATH{v}{x} \land \BEDGE{x}{v} \land \BPATH{v}{y} \implies \BDEP{x}{y}{x}{v} 
\tag{HeadTriangleImposable}\label{eq:HeadTriangleImposable} \\
\forall x,y,u \;\;\bullet\;\;   & x \neq y \land u \neq y \land \tag*{} \\
                                &\lnot\BPATH{y}{u} \land \BPATH{x}{u} \land \BEDGE{u}{y} \implies \BDEP{x}{y}{u}{y} 
\tag{TailTriangleImposable} \\
\forall \nu, n, \mu, \sigma \;\;\bullet\;\; & \nu \neq n \land n \neq \sigma \land \nu \neq \mu \land \mu \neq \sigma \land \tag*{} \\ 
                                            &\BEDGE{\nu}{\sigma} \land \BEDGE{\mu}{\sigma} \implies \lnot \BDEP{\nu}{n}{\mu}{\sigma} 
\tag{Slingshot} \\
\forall x, y, u, v \;\;\bullet\;\; & \lnot\BPATH{x}{u} \lor \lnot\BPATH{v}{y} \implies \lnot\BDEP{x}{y}{u}{v} \tag{CoalignedEdgeAndPath}
\end{align}

Some of the formulas above can be written more precisely by using additional quantification, but it seems that these more-precise formulas can only be applied in an inductive proof. For example, we can strengthen the formula \eqref{HeadTriangleImposable} as follows: 

\begin{align*}
    \forall x,y,v \;\;\bullet\;\; &  x \neq y \land x \neq v \>\land                                                                             &                           \\ 
                                  & (\lnot\BPATH{v}{x} \lor \forall \beta \>\bullet\> \BPATH{x}{\beta} \Rightarrow \lnot\BDEP{v}{y}{x}{\beta}) &                           \\
                                  & \land \BEDGE{x}{v} \land \BPATH{v}{y}                                                                      &\implies \BDEP{x}{y}{x}{v} {}. 
\end{align*}
However, the above formula is difficult to efficiently automate, because in its premise it would require $\DEPSYMB{}$ information about a potentially unbounded number of quadruples, either of the form $(v, y, x, \beta)$ or $(x, y, x, v)$. 

In practice, the imprecision of our 0--1-path graph axiomatization does not cause loss of reachability information, as explained in~\secref{ZOPs}. 


  \end{subappendices}
\fi

\end{document}